\documentclass[
aps,superscriptaddress,
twocolumn,prd,
showpacs,preprintnumbers,nofootinbib,
amsmath,amssymb,
aps,
superscriptaddress]{revtex4-1}
\usepackage{Preamble}
\makeatletter
\AtBeginDocument{\let\includegraphics\saved@includegraphics}

\newcommand{\muB}{\mu_{\rm B}}

\newcommand{\nncor}[1]{{\textcolor{black}{#1}}}

\begin{document}

%\title{Did the gravitational-wave signal GW190521g originate in head-on collision of black holes?}

%\title{Some head blowing title :D}
%\title{Constraining the mass of ultralight vector bosons with GW190521\\
%Estimating the mass of a hypothetical ultralight vector boson compatible with GW190521 \\
%\title{GW190521 as a collision of vector boson stars with mass $\mu_{\rm B}=6.50\times 10^{-13}$ eV\\
\title{Searching for vector boson-star mergers within LIGO-Virgo \\ intermediate-mass black-hole merger candidates}
%Searching for boson-star mergers in the LIGO-Virgo gravitational-wave catalogue}

%% Notice placement of commas and superscripts and use of &
%% in the author list

\author{Juan Calder\'on~Bustillo}
 	\email{juan.calderon.bustillo@gmail.com}
	\affiliation{Instituto Galego de F\'{i}sica de Altas Enerx\'{i}as, Universidade de
Santiago de Compostela, 15782 Santiago de Compostela, Galicia, Spain}
	\affiliation{Department of Physics, The Chinese University of Hong Kong, Shatin, N.T., Hong Kong}
\author{Nicolas Sanchis-Gual}
	\email{nicolas.sanchis@uv.es}
	\affiliation{Departamento de Astronom\'{i}a y Astrof\'{i}sica, Universitat de Val\`{e}ncia,
Dr. Moliner 50, 46100, Burjassot (Val\`{e}ncia), Spain}
	\affiliation{Departamento  de  Matem\'{a}tica  da  Universidade  de  Aveiro  and  Centre  for  Research  and  Development in  Mathematics  and  Applications  (CIDMA),  Campus  de  Santiago,  3810-183  Aveiro,  Portugal}
%	\affiliation{Centro de Astrof\'{i}sica e Gravita\c{c}\~{a}o - CENTRA,
%Departamento de F\'{i}sica, Instituto Superior T\'{e}cnico - IST, Universidade de Lisboa - UL, Avenida Rovisco Pais 1, 1049-001, Portugal}
\author{Samson H. W. Leong}
	\affiliation{Department of Physics, The Chinese University of Hong Kong, Shatin, N.T., Hong Kong}
\author{Koustav Chandra}
	\affiliation{Department of Physics, Indian Institute of Technology Bombay, Powai, Mumbai, Maharashtra 400076, India}
\author{Alejandro Torres-Forn\'e}
	\affiliation{Departamento de Astronom\'{i}a y Astrof\'{i}sica, Universitat de Val\`{e}ncia,
Dr. Moliner 50, 46100, Burjassot (Val\`{e}ncia), Spain}
\affiliation{Observatori Astron\`{o}mic, Universitat de Val\`{e}ncia,
C/ Catedr\'{a}tico Jos\'{e} Beltr\'{a}n 2, 46980, Paterna (Val\`{e}ncia), Spain}
\author{Jos\'e A. Font}
	\affiliation{Departamento de Astronom\'{i}a y Astrof\'{i}sica, Universitat de Val\`{e}ncia,
Dr. Moliner 50, 46100, Burjassot (Val\`{e}ncia), Spain}
	\affiliation{Observatori Astron\`{o}mic, Universitat de Val\`{e}ncia,
C/ Catedr\'{a}tico Jos\'{e} Beltr\'{a}n 2, 46980, Paterna (Val\`{e}ncia), Spain}
\author{Carlos Herdeiro} 
		\affiliation{Departamento  de  Matem\'{a}tica  da  Universidade  de  Aveiro  and  Centre  for  Research  and  Development in  Mathematics  and  Applications  (CIDMA),  Campus  de  Santiago,  3810-183  Aveiro,  Portugal}
\author{Eugen Radu}
	\affiliation{Departamento  de  Matem\'{a}tica  da  Universidade  de  Aveiro  and  Centre  for  Research  and  Development in  Mathematics  and  Applications  (CIDMA),  Campus  de  Santiago,  3810-183  Aveiro,  Portugal}
\author{Isaac C.F. Wong}
	\affiliation{Department of Physics, The Chinese University of Hong Kong, Shatin, N.T., Hong Kong}
\author{Tjonnie G. F. Li}
	\affiliation{Department of Physics, The Chinese University of Hong Kong, Shatin, N.T., Hong Kong}

\begin{abstract}

We present the first systematic search for exotic compact mergers in Advanced LIGO and Virgo events. We compare the short gravitational-wave signals GW190521, GW190426$\_$190642, GW200220$\_$061928 and the trigger 200114$\_$020818 (or \nncor{S200114f}) to a new catalogue of 759 numerical simulations of head-on mergers of horizonless exotic compact objects known as Proca stars, interpreted as self-gravitating lumps of (fuzzy) dark matter sourced by an ultralight (vector) bosonic particle. The Proca-star merger hypothesis is strongly rejected with respect to the black hole merger one by GW190426, weakly rejected by GW200220 and weakly favoured by GW190521 and \nncor{S200114f}. GW190521 and GW200220 yield highly consistent boson masses of $\mu_{\rm B} = 8.69^{+0.61}_{-0.75}\times10^{-13}$ eV and $\mu_{\rm B} = 9.13^{+1.18}_{-1.30}\times10^{-13}$ eV at the $90\%$ credible level. We conduct a preliminary population study of the compact binaries behind these events. Excluding (including) \nncor{S200114f} as a real event, and ignoring boson-mass consistencies across events, we estimate a fraction of Proca-star mergers of $\zeta = 0.27^{+0.43}_{-0.25} \ (0.39^{+0.38}_{-0.33})$. We discuss the impact of boson-mass consistency across events in such estimates. Our results maintain GW190521 as a Proca-star merger candidate and pave the way towards population studies considering exotic compact objects.

\end{abstract}

\maketitle

\section{Introduction}

The gravitational-wave (GW) detectors, Advanced LIGO \cite{AdvancedLIGOREF} and Virgo \cite{TheVirgo:2014hva}, have made the observation of compact binary mergers almost routine. In only 6 years, these have reported $\sim 90$ such observations \cite{GWTC1_PRX,abbott2021gwtc2,abbott2021gwtc3,GWTC2.1} that have provided us with unprecedented knowledge on how black holes (BHs) and neutron stars form and how they populate our Universe~\cite{Populations_GWTC3}. Moreover, these observations have enabled the first tests of General Relativity in the strong-field regime~\cite{GWTC3-TGR} and qualitatively new studies of the Universe at a large scale \cite{H0_nature_lvk,Cosmology_GWTC3,Lensing_LIGO}. All such studies require an accurate identification of the source parameters, which has been possible for most observations owing to a clear initial inspiral stage that allows to identify the parameters of the merging binary. In particular, most of such events have been confidently identified as circular black holes or neutron star mergers (BBHs and BNSs) with negligible orbital eccentricity.%\\ 

The detection of the GW190521 event represented the first departure from such ``canonical'' events \cite{GW190521D, GW190521I}. Owing to the large mass of its source, GW190521 barely displays any pre-merger dynamics, with the vast majority of the signal coming from the final distorted, merged object while it relaxes to its final BH form. In such a situation, there is little information about the parents of the final object, making the inference of their parameters depends strongly on the prior assumptions about them and leading to a variety of interpretations of this event \cite{Olsen2021_IAS_Likelihood,GW190521_Nitz,RomeroShaw2020_ecc_apjl,Gayathri2022_ecc_natastro,Proca,Gamba2022_ecc_natastro}.%\\

First, the LIGO-Virgo-KARGA collaboration (LVK) reported a circular BH merger with mild signatures of orbital precession \cite{GW190521D,GW190521I}. However, Ref \cite{HOC} showed that, for such short signals, orbital precession can be confused with high eccentricity. Consistently, \cite{RomeroShaw2020_ecc_apjl} and \cite{Gayathri2022_ecc_natastro} argued that GW190521 could be interpreted as an eccentric merger. Despite their differences, all the mentioned interpretations lead to two main conclusions. First, the remnant BH has a mass $M_{\rm f} > 100\,M_\odot$, making it the first observation of a compact object in the intermediate-mass BH mass range. Second, \nncor{in most of the above interpretations}, the heavier merging BH shows significant support within so-called pair-instability supernova (PISN) gap, located within the approximate range $\sim [65,130]\,M_\odot$ \footnote{See references \cite{Barkat1967_PISN,Woosley2019,Farmer2019_PISN,Mapelli2020} for possible variations of this range.}, where no BH formation is expected to occur from stellar collapse \cite{Barkat1967_PISN,Woosley2019,Farmer2019_PISN,Mapelli2020}; and nearly null support outside of it \cite{GW190521D,Gayathri2022_ecc_natastro,RomeroShaw2020_ecc_apjl}\footnote{Although see \cite{Belczynski2020} for possible alternative formation of the heaviest merging BH in GW190521 through stellar collapse.}. With these two characteristics, GW190521 provides on the one hand an invaluable clue towards understanding the formation of supermassive BHs via hierarchical merger channels \cite{Volonteri2003,Volonteri2010}. On the other, it poses the challenge of explaining the origin of such merging BHs populating the PISN gap, e.g., invoking hierarchical formation channels~\cite{Kimball2021,Liu2021}. While several other explanations for the origin of such BHs have been proposed \cite{Costa2020,DallAmico2021}, alternative studies have shown that the heavier BH in GW190521 may actually avoid the PISN gap with some probability. For instance, using a population informed prior \cite{straddling} hinted that GW190521 could actually involve one BH above the PISN gap and one below, known as a “straddling binary”. Also, using an alternative mass prior, \cite{GW190521_Nitz} showed that GW190521 could be a high-mass ratio binary. Finally, as the departing point of this work, Ref \cite{Proca} showed that GW190521 is consistent with numerically simulated head-on mergers \cite{sanchis2019head} of horizonless compact objects known as Proca stars~\cite{brito2016proca}. While exotic, this interpretation automatically eliminates the presence of a BH populating the PISN gap while still yielding an IMBH remnant.\\

The third observing run of the Advanced LIGO - Virgo network has delivered more short signals\footnote{We refer to signals that display a small enough number of inspiral cycles in the detector band that we shall be able to fit them with our catalog of waveform templates for head-on Proca-star mergers.} similar to GW190521, namely GW190426$\_$190642 and GW200220$\_$061928 (which we will refer to as GW190426 and GW200220), albeit with a much lower statistical significance \cite{abbott2021gwtc3,GWTC2.1}. In addition, a targeted search for intermediate-mass black holes delivered the intriguing  trigger 200114\_020818 (S200114f in the following) which, while observed with a larger statistical significance than the former two, was not conclusively classified as either a GW or a noise artefact \cite{o3_imbh,abbott2021gwtc3} \footnote{GW190426 and GW200220 have false alarm rates (FAR) larger than 1 per year, or 1/1\,yr \cite{GWTC2.1,abbott2021gwtc3}, while S200114f was associated a FAR of 1/17\,yr \cite{o3_imbh}. For comparison, GW190521 was detected with a FAR of 1/4900\,yr \cite{GW190521D}.}. The morphological characteristics of these signals make them merit further investigation exploring possibilities beyond the BBH paradigm. In this work, we compare all of these events to a catalogue of 759 numerical simulations of Proca-star mergers (PSMs). In particular, we perform model selection on these events between our PSM and a classical BBH model and report the estimated parameters under the PSM model. Finally, we perform a preliminary population study to estimate the fraction of PSMs within the observed set of compact mergers.  \\

\subsection{Proca stars and dark matter}

Bosonic stars are self-gravitating lumps of bosonic fields, first constructed for massive, complex scalar fields in the late 1960s~\cite{Kaup:1968zz,Ruffini:1969qy} and more recently constructed also for massive, complex vector fields~\cite{brito2016proca}. The latter are also known as Proca fields, and thus the corresponding stars have been dubbed Proca stars. These stars can be either spherical and non-rotating~\cite{Herdeiro:2017fhv} or axially-symmetric and spinning~\cite{Herdeiro:2019mbz}. They can  be rather Newtonian but become compact in regions of the parameter space, to the point that their compactness becomes comparable (albeit smaller) to that of BHs. In this case, bosonic stars are an example of exotic compact objects (ECOs) that can mimic some of the phenomenology attributed to BHs (see e.g.~\cite{Proca, Herdeiro:2021lwl}).

From a macroscopic perspective, the simplest bosonic stars are described by free, complex, massive bosonic fields minimally coupled to gravity. Self-interactions can be introduced in the model and can change their properties~\cite{Schunck:2003kk,liebling2017dynamical,Cough_ghost,Coates2022_selfint_pathology,Siemonsen2021}, but are not mandatory for the existence of solutions (and are absent in the models considered here). From a microscopic perspective, they can be interpreted as many-particle states of ultralight bosons. The ultra-lightness requirement for the fundamental bosonic particle guarantees (in the simplest models) that  the bosonic stars achieve masses in the astrophysical BH range.  In particular, ultralight bosons with a particle mass $\mu_{\rm B}$ within $10^{-13}\leq\mu_{\rm B}\leq10^{-10}$~eV, yield stars with maximal masses in the interval $\sim$ 1000 and 1 solar masses, respectively. Such ultralight bosons can be motivated by particle-physics models, from the QCD axion~\cite{Peccei:1977hh}, to the string axiverse~\cite{Arvanitaki2010} and also by simple extensions of the Standard Model of particle physics~\cite{Freitas:2021cfi}. Such ultralight particles could form part, or the whole, of the dark matter budget of the Universe \cite{DarkMatterBoook,EuCAPT_astroparticle}, making bosonic stars only detectable via their gravitational signatures.

%Proca stars belong to a family of theoretical exotic compact objects (ECOs) known as bosonic stars. These are part of wider family of objects known as ``black hole'' mimickers, which lacking the characteristic event horizon of black holes, can reproduce many of their properties.
%ECOs have been proposed, e.g., as dark-matter candidates, often invoking
%, with some of these invoking 
%the existence of hypothetical ultralight (i.e.~sub-eV) bosonic particles. One common candidate is the pseudo-scalar QCD axion,
%particle, 
%but other ultralight bosons arise, e.g., in the string axiverse \cite{Arvanitaki2010}. In particular, vector bosons are also motivated in extensions of the Standard Model of elementary particles and can clump together forming macroscopic entities dubbed bosonic
%boson 
%stars. 

Unlike other ECO models, bosonic stars have a well-established, field-theoretical description. Their dynamics have been extensively studied (see e.g.~\cite{liebling2017dynamical,bezares2017final,palenzuela2017gravitational,sanchis2017numerical,sanchis2019head}). 
%Scalar boson stars and their vector analogues, Proca stars~\cite{brito2016proca,sanchis2017numerical} (PSs), are self-gravitating stationary solutions of the Einstein-(complex, massive) Klein-Gordon~\cite{Schunck:2003kk} and of the Einstein-(complex) Proca~\cite{brito2016proca} systems, respectively. 
The corresponding bosonic fields oscillate at a well-defined frequency $\omega$, which provides a dispersive nature counteracting gravity and determines
%gives 
the mass and compactness of the star. %Unlike other ECOs, 
Moreover, bosonic stars have a precise formation mechanism, which needs no fine-tuning, known as gravitational cooling~\cite{Seidel1994,DiGiovanni2018}. This is consistent with their dynamical robustness, which has been established for spherical boson stars  both perturbatively and non-perturbatively~\cite{liebling2017dynamical}. On the other hand, spinning bosonic stars are more subtle; only recently it was found that in the simplest models they are unstable in the scalar case, but not in the Proca case~\cite{SanchisGual2019,di2020dynamical}.
%While spinning solutions have been obtained for both scalar and vector bosons, the former are unstable against non-axisymmetric perturbations
This motivated considering collisions of spinning Proca stars. In Ref.~\cite{Proca}, 
% Hence, we will focus on the vector case in this work. 
% For non-self-interacting bosonic fields, the maximum possible mass of the corresponding stars is determined  by the boson particle mass $\mu_V$. In Ref.~\cite{Bustillo:2021proca1}, 
it was established that the event GW190521 is consistent with a head-on collision of two Proca stars with $\mu_{\rm B}=8.7\times 10^{-13}$\,eV.\\

We note that alternative searches for signatures of ultra-light bosons in gravitational-wave data have been performed, in particular focusing on the effects that (scalar) boson clouds can produce when surrounding black-holes. On the one hand, these include searches for continuous GW emission arising from super-radiant instability e.g., \cite{boson_LVK_followup,palomba2019direct,Dergachev2020,Sun2020}, which should in principle be detectable by current detectors. On the other hand, such clouds can extract angular momentum from the host black-holes leading to a reduction of its spin, an effect which has also been searched for \cite{Ken_1,Ken_2}. While none of these methods has delivered an actual detection these have been used to place constraints on the possible range of masses of (scalar) ultra-light bosons. Finally, further methods targeting LISA observations have been designed that may establish the existence of ultra-light bosons through a single observation \cite{Hannuksela2019}.

\subsection{Aim and structure of this work}

We perform a systematic analysis of the events GW190521, GW190426 and GW200220 using an expanded catalogue of 759 numerical simulations of head-on mergers of Proca stars (PSMs). In addition, we analyse the trigger S200114f. We compare the incoming detector data to both our catalogue of numerical simulations and to a state-of-the-art waveform model for circular black hole mergers. For the BBH case, we perform a ``canonical'' analysis comparing strain-data to strain-templates. For the case of our numerical simulations, however, we make use of a novel framework that we introduced in~\cite{Psi4} that allows for a comparison of the signal data to the waveform templates for the Newman-Penrose scalar directly outputted by our numerical simulations, commonly denoted as $\psi_4$. 
The rest of this article is organised as follows. %Commenting the following as we have merged it with Sec. II.  %Section \ref{sec:psi4} summarises our new framework to perform GW data analysis using $\psi_4$. 
In section~\ref{sec:setup} we describe our analysis setup, including our waveform models, simulation catalogue and prior choices. In section~\ref{sec:results} we report our parameter estimation and model selection results for all individual events and in section~\ref{sec:pop} we conduct a preliminary population study. Finally, we close with a discussion of the limitations and potential implications of our work.

\section{Analysis set-up}
\label{sec:setup}

For given detector data $d(t)$ and a waveform template model $M_i$ spanning parameters $\theta$, we aim to compute the posterior probability distribution for $\theta_i$ 
\begin{equation}
    p_{M_i}(\theta\,|\,d) = \frac{\pi(\theta)\,{\cal{L}}_{M_i}(d \,|\, \theta)}{{\cal{Z}}_{M_i}}.
\label{eq:bayesian_prob}
\end{equation}

Here, $\pi(\theta)$ denotes the prior probability for the parameters $\theta$, the term ${\cal{L}}_{M_i}(d\,|\,\theta)$ denotes the likelihood of the data $d$ according to the waveform model $M_i$ given parameters $\theta$. This is given by \cite{Finn1992,Cutler1994,Romano2017}
\begin{equation}
    {\cal{L}}(d\,|\,\theta) \propto \exp \bigg{[} - \frac{(d-h(\theta)|d-h(\theta))}{2} \bigg{]},
\label{eq:logl}
\end{equation}
where the operation $(a|b)$ denotes the noise-weighted inner product \cite{Cutler1994} 
\begin{equation}
(a|b) = 4 \times \Re \int_{f_{\rm min}}^{f_{\rm max}} \frac{\tilde{a}(f)\tilde{b}^{*}(f)}{S_n(f)}\, \dd f
\label{eq:nwip}
\end{equation}
with $S_n(f)$ the one-sided power-spectral density of the background noise and $(f_{\rm min},f_{\rm max})$ the lower and upper frequency limits. The term ${\cal{Z}}_{M_{i}}$ denotes the Bayesian evidence for the waveform model $M_{i}$. This is equal to the integral of the numerator of Eq.~\eqref{eq:bayesian_prob} over the explored parameter space $\Theta$, given by  
\begin{equation}
    {\cal{Z}}_{M_{i}}=\int_{\Theta} \pi(\theta) {\cal{L}}_{M_{i}}(d\,|\,\theta) \,\dd\theta.
    \label{eq:evidence}
\end{equation}
Finally, given two waveform models $M_1$ and $M_2$, the relative probability for the data given the models, or relative Bayes Factor ${\cal{B}}^{M_1}_{M_2}$, is given by 
\begin{equation}
{\cal{B}}^{M_1}_{M_2} = \frac{{\cal{Z}}_{M_1}}{{\cal{Z}}_{M_2}}.
\label{eq:bayesfactor}
\end{equation}

\subsection{Data and Waveform models}

We perform Bayesian parameter estimation and model selection on four seconds of publicly available data \cite{GWOSC,GWOSC:GWTC} from the two Advanced LIGO and Virgo detectors around the time of GW190521, GW200220, GW190426 and S200114f. We compare the detector data to numerical-relativity simulations of head-on PSMs \cite{sanchis2019head,sanchis2022impact} and to the state-of-the-art waveform model for circular BBHs \texttt{NRSur7dq4} \cite{NRSur7dq4} implemented in the \texttt{LALSuite} library \cite{lalsuite}. In previous work \cite{Proca} we made use of a catalogue of 96 numerical simulations of PSMs. These were divided into two sets: one is of equal-mass and equal-spin, therefore equal boson-field frequency $\omega$; and the other is an exploratory unequal-mass family. Here we make use of an expanded catalogue of 759 simulations spanning a grid in the frequencies of the two stars $\omega_1/\mu_{\rm B}$ and $\omega_2/\mu_{\rm B}$, which we describe in detail in Appendix~\ref{sec:appendix1}). These simulations include the co-dominant GW emission modes $(\ell,m) = (2,0),(2, \pm 2)$ and the largest sub-dominant modes $(3,\pm 2),(3,\pm 3)$ \footnote{We note that while the mass ratio of our simulations is larger than $m_2/m_1 = 0.65$, we have found a few cases where the amplitude of the $(3,3)$ mode is half of that of the dominant $(2,2)$ and $(2,0)$ modes. We attribute this to the interference effects described in \cite{Sanchis:catalogue}. See also our priors section \nncor{ (Sec.~\ref{sec:bayes_priors})}.}. The \texttt{NRSur7dq4} model is the only existing waveform model directly trained on numerical simulations of circular BBHs including the impact of orbital precession \cite{SXSCatalog}. The model is trained for mass-ratios $q\in[1,4]$ and spin magnitudes $a_1\in[0,0.8]$ but can be extrapolated to values of $q\in[1,6]$  and $a_1\in[0,0.99]$. This model includes all GW modes up to $\ell = 4$. \\

Finally, we note that as in \cite{Proca}, we do not marginalise over detector calibration uncertainties. The reason is that while this would increase the computational cost of already very expensive runs making use of PSM waveforms, such effects are known to be negligible for current detector sensitivities \cite{Vitale2012_calenvs,Payne2020_calenvs,Huang_calenvs}.

\subsection{Data analysis using the Newman-Penrose scalar}

GW data analysis relies on the comparison of the strain data read by the detectors to waveform templates for such strain. We rely on this ``classical'' approach for the case of comparing the data to the strain model \texttt{NRSur7dq4}. Numerical simulations performed by a large collection of numerical relativity codes as, e.g. the Einstein Toolkit \cite{EinsteinToolkit,Loffler:2011ay}, however, do not directly output the GW strain but a quantity known as the  Newman-Penrose scalar, or $\psi_4$, related to the GW strain as $\psi_4(t) = \dv*[2]{h(t)}{t}$ \cite{Newman:1961qr}\footnote{We note that there exist methods to directly extract the GW strain, as the Regge-Wheeler-Zerilli \cite{SXSCatalog,Regge1957,Zerilli1970}, Cauchy Characteristic Extraction \cite{Bishop1996_CCE,Moxon2023_CCE} or Cauchy Characteristic Matching \cite{CCM_Bishop,Ma_CCM} formalisms. Please see \cite{Bishop2016_Summary} and references therein for a discussion of these methods.}. Obtaining the corresponding strain templates therefore requires a double time integration that is subject to well-known potential systematic errors due to spurious low frequencies contaminating the resulting $h(t)$ \cite{Pollney_Reissweig}. These can be especially relevant for highly eccentric mergers for which there is no natural way to diminish these. While we used such strain templates in \cite{Proca}, here we adopt a novel framework presented in~\cite{Psi4} that allows for a comparison of the detector data to the $\psi_4$ templates directly extracted from our numerical simulations, therefore avoiding further systematic errors. To do this, given the discrete detector data strain $d[n]$ of duration $T = M\Delta t$ sampled at frequency $1/\Delta t$ and the corresponding PSD $S_n[k]$, we perform the transformation:
\begin{equation}
\begin{aligned}
       d[n]  &\longrightarrow d_{\Psi_4}[n] \equiv (\delta^{2}d)[n]  \\
       S_n[k] &\longrightarrow S_{n{\Psi_4}}[k]
\end{aligned}
    \label{eq:transformation}
\end{equation}
Above, $\delta^2 d[n]$ represents the second-order finite difference of $d[n]$, given by
\begin{equation}
   \delta^2 d [n] = \frac{d[n+1] - 2d[n] + d[n-1]}{(\Delta t)^{2}},
\end{equation}
and the transformed PSD $S_{n\Psi_4}(f)$ is obtained through
\begin{equation}
    S_{n\Psi_4}[k] = \frac{1}{(\Delta t)^{4}}
    \left(
    6 -
    8 \cos(\frac{2\pi k}{M}) +
    2\cos(\frac{4\pi k}{M})
    \right)
    S[k].
    \label{eq:noise_psd}
\end{equation}
Finally, we replace the typical strain templates $h[n]$ by the $\psi_4[n]$ templates outputted from numerical-relativity simulations \textit{after applying a correction} that accounts for the difference between second derivative and second-order finite differencing. We denote the resulting template by $\Psi_4[n]$. In particular, expressing waveform templates in the frequency domain, we substitute: 
\begin{equation}
       \widetilde{h}[k] \longrightarrow \widetilde{\Psi_4}[k] = K(k\Delta f)\,\widetilde{\psi_4}(k\Delta f), 
    \label{eq:transformation2}
\end{equation}
where
\begin{equation}
 K(k\Delta f) = \frac{1-\cos(2\pi \,k\Delta f\,\Delta t)}{2\pi^{2}(k\Delta f\,\Delta t)^{2}}
    \label{eq:correction_factor}
\end{equation}
and $\Delta t = 1/(M\Delta f)$.\\

Finally, we note that since \texttt{NRSur7dq4} waveform model is trained using numerical simulations that directly extract the GW strain (with no integration process), these are free of such errors. Nevertheless, see \cite{SXSCatalog} for a detailed description of further possible systematic errors

%In order to perform model selection against the BBH scenario, we also compare the data to three models for quasi-circular BBHs, namely \texttt{NRSur7dq4} \cite{NRSur7dq4}, \texttt{IMRPhenomXPHM} and \texttt{IMRPhenomTPHM} trying several mass-ratio priors and ranges. Among these, \texttt{NRSur7dq4} is the only model fully calibrated to numerical simulations of precessing BBHs including higher harmonics. Therefore, consistently with \cite{GW190521D,GW190521I,Proca} we will treat this as our ``preferred'' BBH model. The other two \texttt{IMRPhenom} models rely on different approximations. In particular, these are calibrated to non-precessing simulations and then ``twisted'' to reproduce the effect of precession. While such technique is reliable in the post-newtonian regime, far from merger, it is significantly less reliable during merger and ringdown, which is the observable regime of GW190521. Moreover, sub-dominant modes in \texttt{IMRPhenomXPHM} are not fitted to NR simulations but computed via a scaling technique that is also not appropriate during the merger and ringdown stages. In particular, this can cause the relative phases of the different modes to be wrong, therefore leading to wrong waveform morphologies. Despite these caveats, we will compute two types of evidence for the Proca vs. BBH models: that against \texttt{NRSur7dq4}, denoted by $\log{\cal{B}}^{\text{Proca}}_{{\text{BBH,NRSur7dq4}}}$, and that against the BBH model yielding the largest evidence, denoted by $\log{\cal{B}}^{\text{Proca}}_{{\text{BBH,Best}}}$. 

\subsection{Bayesian Priors}\label{sec:bayes_priors}

\subsubsection{Intrinsic source parameters} 

\subsubsection*{Proca-star mergers: field frequencies, masses and spins} 

In GW data analysis, it is a common practice to place uniform priors on the individual masses of the source. Our discrete PSM catalogue, however, prevents us from imposing such prior. Instead, we exploit the fact that each of our PSM simulations -- for a given mass-ratio and spins -- scales trivially with the total mass, enabling us to place a uniform prior in the total red-shifted mass of the source. In addition, while our simulations do not uniformly cover the space formed by the two bosonic frequencies $\omega_{1,2}/\muB$, we appropriately weight these to impose an uniform prior across the triangle defined by $\omega_{1,2} / \muB \in [0.80,0.93]$, with $\omega_1/\muB  \geq \omega_2/\muB $ (for details, please see Appendix I, which includes a representation of our simulation bank and weights in Fig. \ref{fig:weights}). Finally, we place a prior in the total (redshifted) mass uniform in $M\in[50,500]\,M_\odot$.\\

We note that due to the properties of Proca stars, our prior on $\omega_{1,2}/\muB$ determines those for the spins (in all cases above 1\footnote{Note that, unlike black holes, Proca-stars are not subject to the Cosmic Censorship conjecture that sets the maximum dimensionless spin to 1.}) and the mass ratio \cite{Herdeiro2019_omega_J_plot}. First, we find that the induced mass-ratio prior approximately follows $\pi(q) \propto q$, with $q \in [0.657,1]$. Second, we empirically find that $\omega/\muB$ and the spin magnitude $a$ are \nncor{approximately} related by $a=\exp[0.9579\times (\omega/\muB)^{9.4}]$. This induces non-trivial spin priors %\SL{The offset 0.63 (and the proportionality) is necessary to produce the linear fit. Also, not sure if this remark is needed, but that linear relation is a good fit only for high $q \approx 1$, see the last section (4.2) of the ``supplementary material''.}

\begin{equation}
\begin{aligned}
    \pi(a_1) & \propto \frac{(\log(a_1))^{-1 + 1/\gamma}}{a_1}\,\qty[\log(a_1)^{1/\gamma} - \log(a_{\rm min})^{1/\gamma}] \\
    \pi(a_2) & \propto \frac{(\log(a_2))^{-1 + 1/\gamma}}{a_2}\,\qty[\log(a_{\rm max})^{1/\gamma} - \log(a_2)^{1/\gamma}]
\end{aligned}   
\end{equation}

with $a_{\rm min}=1.1$, $a_{\rm max}=1.6$, $\gamma = 9.4$.\\

The most important consequence of the exponential relation between $a$ and $\omega/\muB$ is that, in principle, results computed under our prior and an analogous one uniform across an equivalent $[a_1,a_2]$ triangle may widely differ. To check this, we re-weighted our posterior probabilities to obtain Bayesian evidences under this new prior. This results in mild increments of the evidence for the PSM model for all events except for S200114f \footnote{The log evidence for S200114f is reduced by $~0.1$ while the rest are increased by values between $0.3$ and $0.8$. In no case these changes lead to qualitatively different conclusions regarding model selection.}. However, we note that our catalogue is too sparsely populated in the large $a_{1,2}$ region, where the likelihood peaks. Therefore we do not think robust conclusions can be extracted, leaving a detailed analysis under a uniform spin prior for future work.

\subsubsection*{Relative phase of Proca-stars} 

Since Proca-stars are described by complex fields, these are not only characterised by the field frequency $\omega_{1,2}/\mu_{B}$ but also by an initial phase $\epsilon_{1,2}(t_0)$ expressed e.g., at the start of our simulations. While $\epsilon$ is rather irrelevant for an isolated star, the relative phase between the two stars at merger $\Delta\epsilon(t_{\rm merger})=\epsilon_{1}(t_{\rm merger})-\epsilon_{2}(t_{\rm merger})$, which is determined by $\omega_{1,2}/\muB$ and $\Delta\epsilon(t_{0})$, causes an interference phenomenon that can have dramatic effects on both the amplitude and frequency content of the emitted waves \cite{Sanchis:catalogue}. However, on the one hand, we only noticed this after the submission of this work, reason why all of our simulations are characterised by $\Delta\epsilon(t_{0})=0$. On the other hand, including this effect in our simulation catalogue requires us to generate many copies of our current one (one for each value of $\Delta\epsilon(t_{0})$, spanning a reasonably dense grid). Since this is computationally extremely expensive, however, we shall leave such investigation for future work. 

The limitation of our catalog to $\Delta\epsilon(t_0)=0$ cases has two main consequences. First, the catalog is clearly sub-optimal, as alternative $\Delta\epsilon(t_0)$ may better fit the studied signals. Second, as we will show in the results sections, this will cause our two-dimensional posteriors on $\omega_{1,2}/\muB$ to be non-smooth, showing instead ``diagonal probability bands'' (or spikes in the 1-dimensional cases) corresponding to regions of similar $\Delta\epsilon(t_{\rm merger})$ (see later in Fig. \ref{fig:omega_posteriors_one}).

\subsubsection*{Initial star separation and momentum}

All of our simulations start with the two stars at rest, separated by a distance of $r\mu=40$ in geometric units (see Appendix II). We note that the choice of initial momentum and separation is somewhat equivalent to that of initial eccentricity and momentum for eccentric compact mergers. While our choice may lack a solid astrophysical motivation \nncor{--in addition to head-on mergers having essentially zero astrophysical probability --}, this is motivated by two main factors: it is simple and leads to conservative results. On the one hand, we are not in a position to choose any particular initial momentum, as the distribution of these among eccentric Proca-star systems with our initial separation (should Proca-stars exist) is clearly unknown. On the other hand, a systematic exploration of this parameter would require a much larger simulation catalogue and greatly increase computational cost. Again, we understand the sub-optimality of our catalogue makes our results rather conservative.\\ 

Finally, our choice of initial star separation is the smallest possible so that spurious ``junk radiation'' present at the start of numerical simulations can be clearly separated from the true GW emission, avoiding it to impact our results. Larger initial separations (as well as non-zero initial momenta) would cause to the stars colliding at larger speeds, producing a louder signal. As we will show later, the intrinsic loudness is critical in model selection, as louder systems are by default be preferred over weaker ones by physically sensible distance priors. Therefore, we understand that choosing the minimum possible initial separation makes our results conservative \nncor{\footnote{This is true when averaging over our entire catalog. We note, however, that given particular values of the star frequencies and initial relative phase, smaller initial separations may indeed lead to a larger signal amplitude at merger, due to the variation of the relative phase at merger.}}

\subsubsection{Black-hole mergers} 

To keep as much consistency as possible with the PSM model, for the BBH case we place the same prior on the total mass. We explore two different priors on the mass ratio: uniform in $Q=m_1/m_2 \geq 1$ and uniform in $q=m_2/m_1 \leq 1$. The motivation to choose both of these priors is that, as shown in e.g.~\cite{GW190521_Nitz,Estells2022_GW190521,21g_ZTF}, certain prior choices can prevent the exploration of high-likelihood regions of the parameter space strongly down-weighted by them. For each of these two priors, we perform two runs respectively using upper (lower) limits for the mass ratio of 4 (1/4) -- within the calibration region of \texttt{NRSur7dq4}) -- and 6 (1/6) to which the model can, in principle, extrapolate. In order to obtain conservative results (i.e., to minimise the evidence in favour of the PSM model) we will always consider the BBH analysis returning the largest Bayesian evidence. Finally, for the spins, we place priors uniform in spin magnitude and isotropic in spin direction. 

% Note that the value of the Hubble parameter we are using is 67.74, from Planck2015 (used by Bilby and astropy), which is quoted in Table 4 of Ade et. al. (Column: TT, TE, EE + lowP + lensing + ext). This citation can be found from the docstring of the Planck15 object.
\subsubsection{Distance} As in \cite{Proca}, we explore two different distance priors. First, we use a standard prior uniform in co-moving volume with $d_{\rm L}\in[10,10000]$\,Mpc, assuming a flat $\Lambda$CDM cosmology with Hubble parameter H$_0=67.74$\,km\,s$^{-1}$\,Mpc$^{-1}$~\cite{Ade:2015xua} \footnote{This prior is known as \texttt{UniformComovingVolume} in the code \texttt{Bilby}~\cite{Isobel-catalogue}, which employs the \texttt{Planck15} cosmology in \texttt{astropy}~\cite{astropyv5}, with non-relativistic matter density $\Omega_{m,0} = 0.3075$, massive neutrino density $\Omega_{\nu,0} = 0.0014$ and dark energy density $\Omega_{\Lambda,0} = 0.6910$~\cite{Ade:2015xua}.}. We note, however, that such a prior does favour intrinsically louder sources -- like BBHs -- that can produce the observed signals from larger distances than weaker sources like our head-on PSMs \nncor{released} from rest at quite close distances. In order to gauge this effect we make use of a rather un-physical prior uniform in distance. In addition, we note that we are essentially observing the final stages of (putative) Proca-star mergers and the final ringing BH. These signals may be reproducible (modulo global amplitude factors) by suitable sets of, intrinsically louder, quasi-circular BH mergers, less eccentric mergers or even just head-on mergers with larger initial momenta than ours, which should yield distances similar to those obtained for BBHs \nncor{\footnote{For instance, GW190521 has been shown to be reproducible by a quasi-circular merger with and without higher-order modes \cite{GW190521I,Capano33}, a dynamical capture with no higher-order modes \cite{Gamba_GW190521} and a Proca-star merger with a slight signature of a $(2,0)$ mode \cite{Bustillo2021}}}. While we do not yet have such numerical simulations at our disposal, we consider the usage of our secondary prior as an attempt to obtain a ballpark evidence that would be obtained with such simulations. 

\subsubsection{Source orientation, sky-location and polarisation} We place standard priors in all of these quantities, namely isotropic in source orientation and sky-location and uniform in signal polarization.\\

Finally, we sample the parameter space in both the BBH and PSM cases using the (publicly available) parallelizable version of the software \texttt{Bilby} \cite{Ashton:2018jfp} known as \texttt{Parallel Bilby} \cite{pbilby} and the nested sampler \texttt{Dynesty} \cite{Dynesty}.

\begin{figure*}[t!]
\begin{center}
\includegraphics[width=0.32\textwidth]{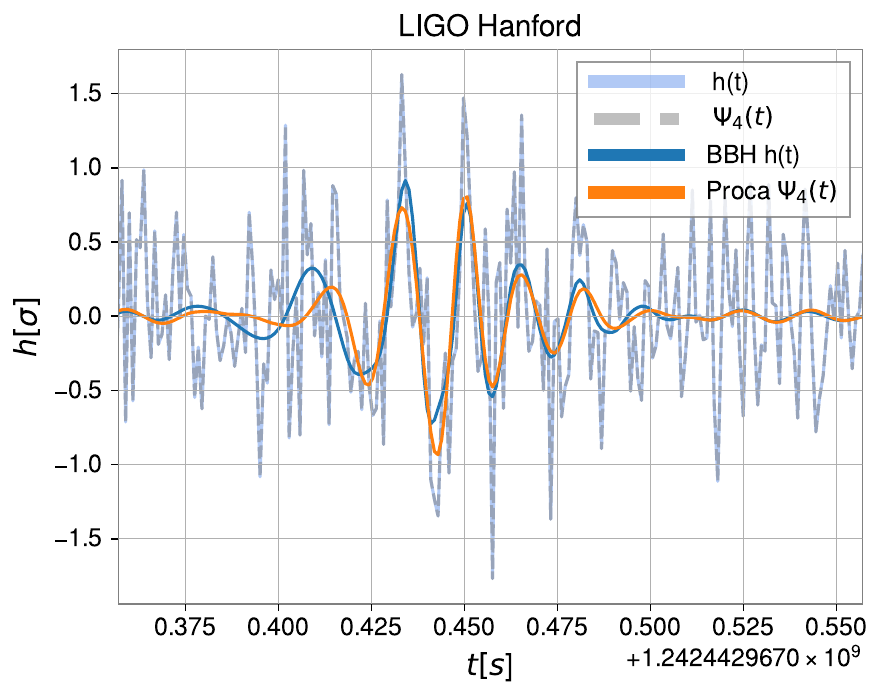}
\includegraphics[width=0.32\textwidth]{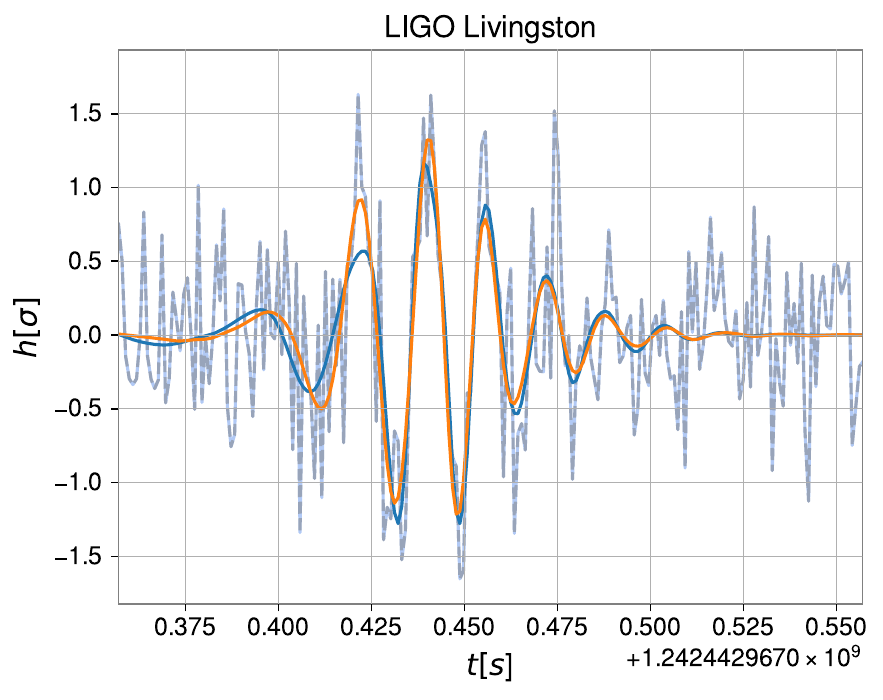}
\includegraphics[width=0.32\textwidth]{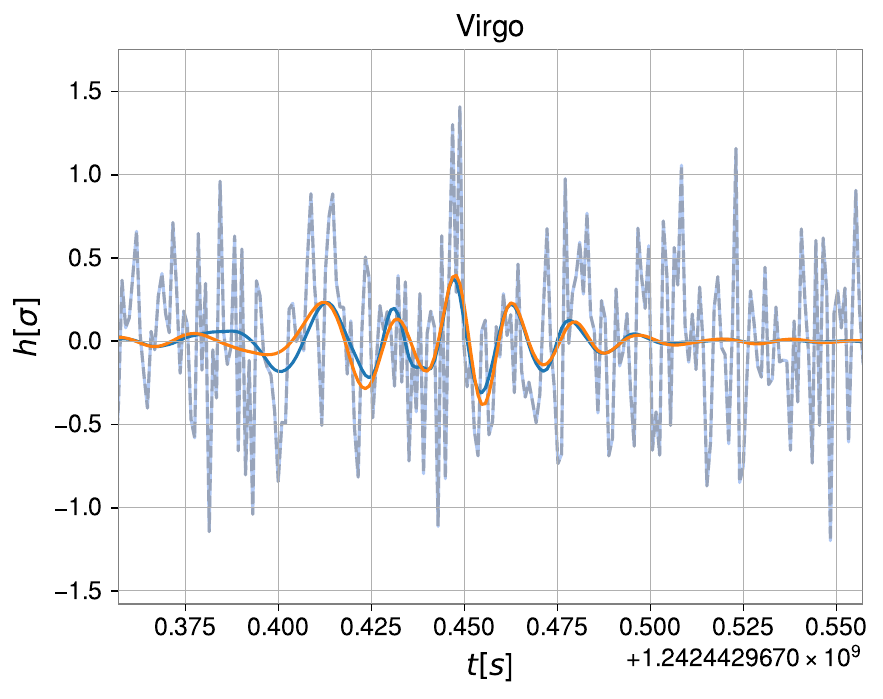}
\caption{\textbf{Whitened $\Psi_4$ and strain time-series around the time of GW190521}, together with the maximum likelihood waveforms returned by the BBH NRSur7dq4 model (blue) and our Proca star head-on merger model (orange). We note that whitened $\Psi_4$ and strain time-series are expected to be indistinguishable since whitened data represents ``the deviation of the data from the expected average background noise'', which should be independent of the way the data is represented.}
\label{fig:21g}
\end{center}
\end{figure*}

\begin{figure*}[t!]
\begin{center}
\includegraphics[width=0.32\textwidth]{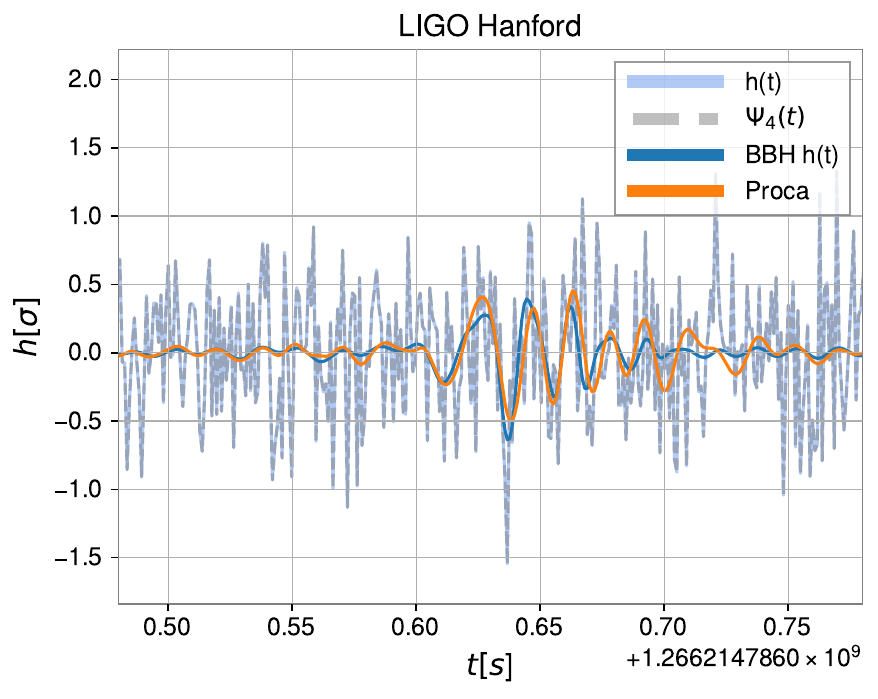}
\includegraphics[width=0.32\textwidth]{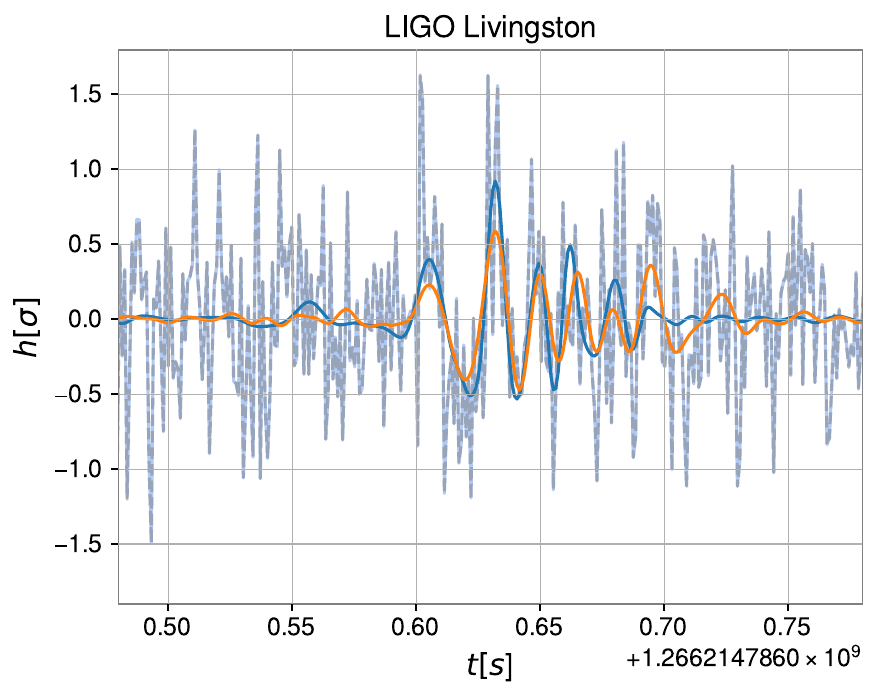}
\includegraphics[width=0.32\textwidth]{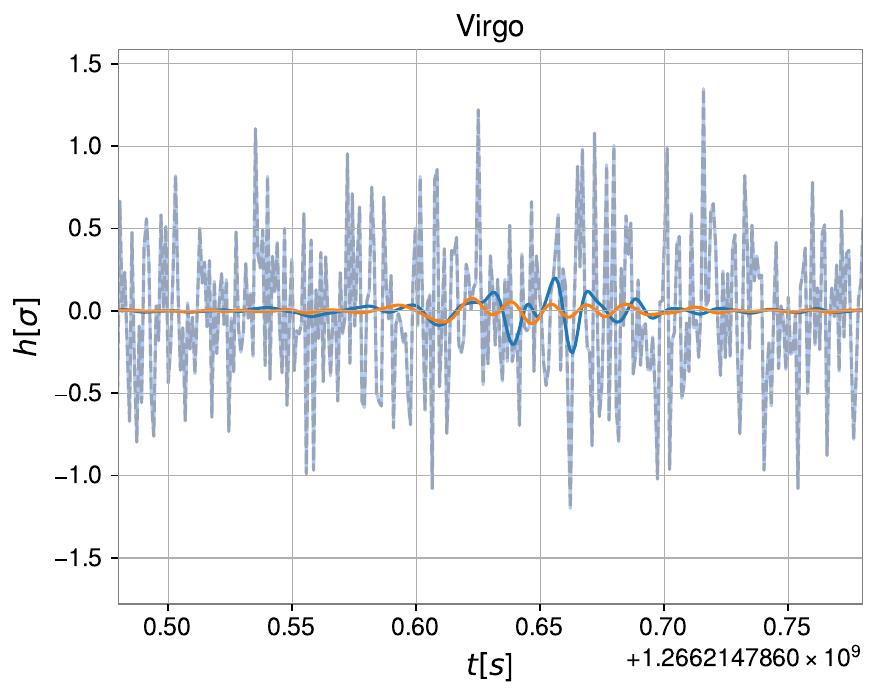}
\caption{\textbf{Whitened $\Psi_4$ and strain time-series around the time of GW200220}, together with the maximum likelihood waveforms returned by the BBH NRSur7dq4 model (blue) and our Proca star head-on merger model (orange).
}
\label{fig:20ad}
\end{center}
\end{figure*}

\begin{figure*}[t!]
\begin{center}
\includegraphics[width=0.32\textwidth]{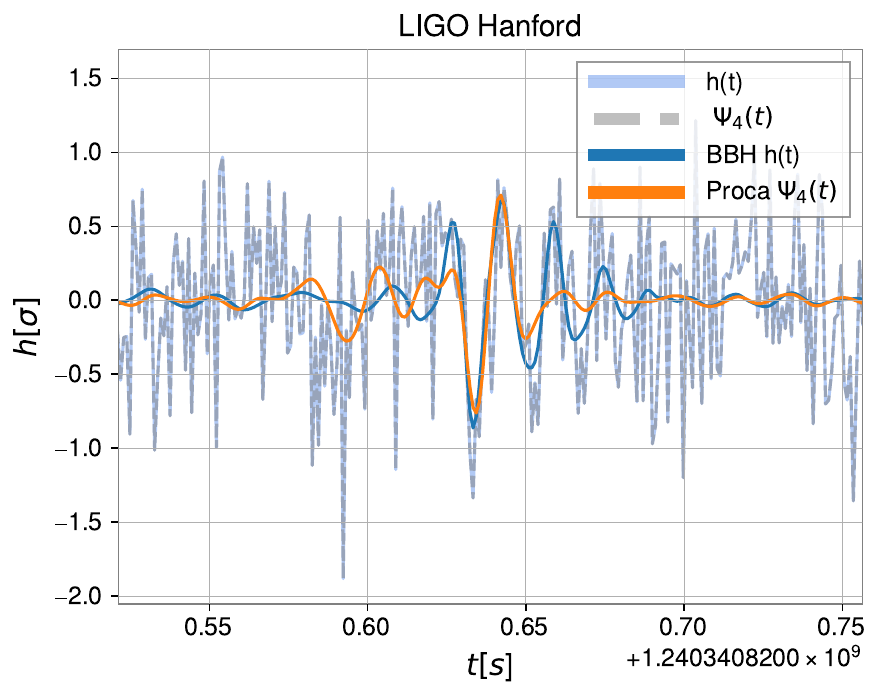}
\includegraphics[width=0.32\textwidth]{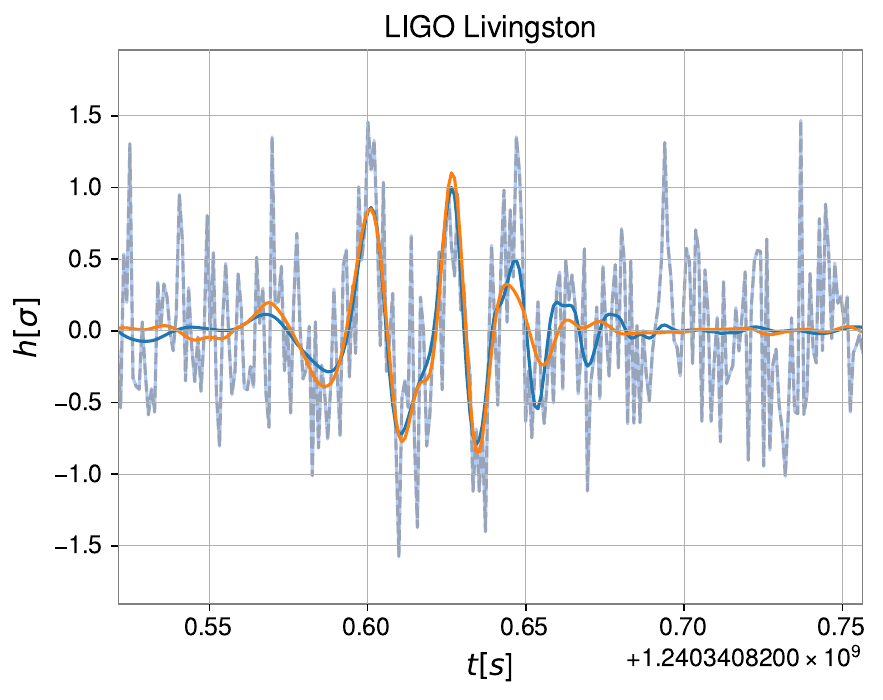}
\caption{\textbf{Whitened $\Psi_4$ and strain time-series around the time of GW190426}, together with the maximum likelihood waveforms returned by the BBH NRSur7dq4 model (blue) and our Proca star head-on merger model (orange).
}
\label{fig:26l}
\end{center}
\end{figure*}

\begin{figure*}[t!]
\begin{center}
\includegraphics[width=0.32\textwidth]{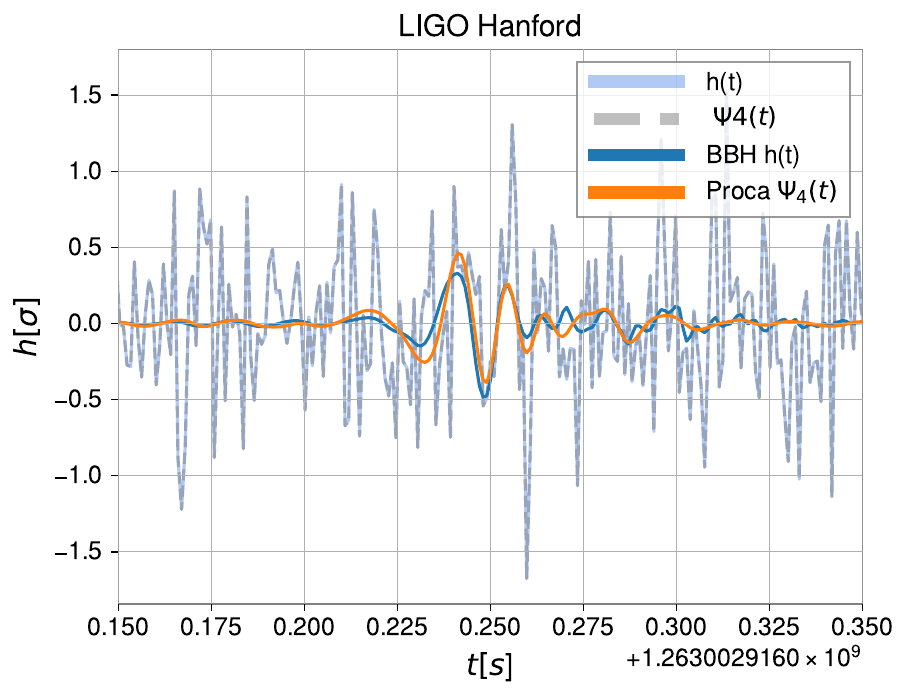}
\includegraphics[width=0.32\textwidth]{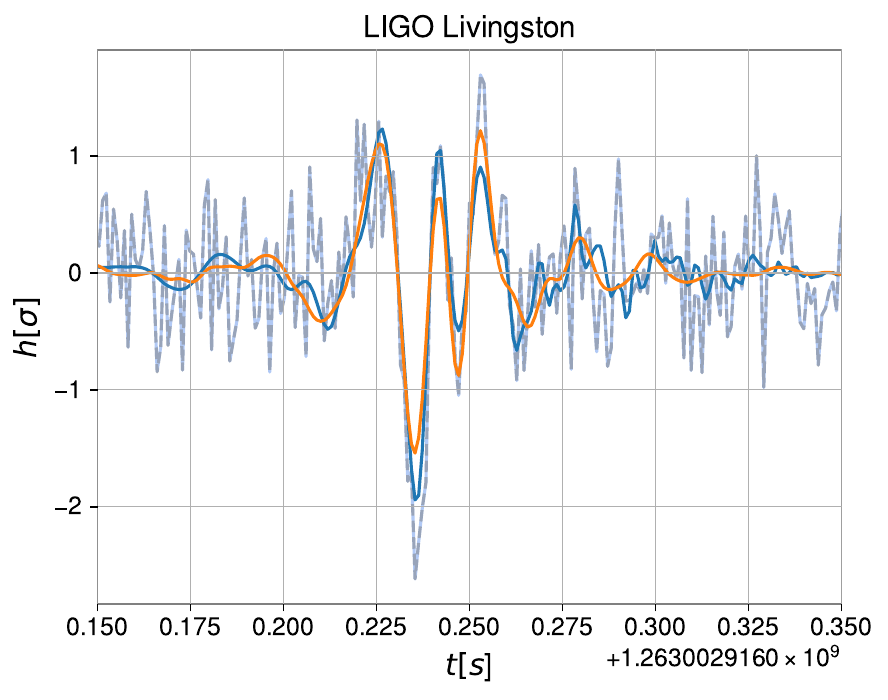}
\includegraphics[width=0.32\textwidth]{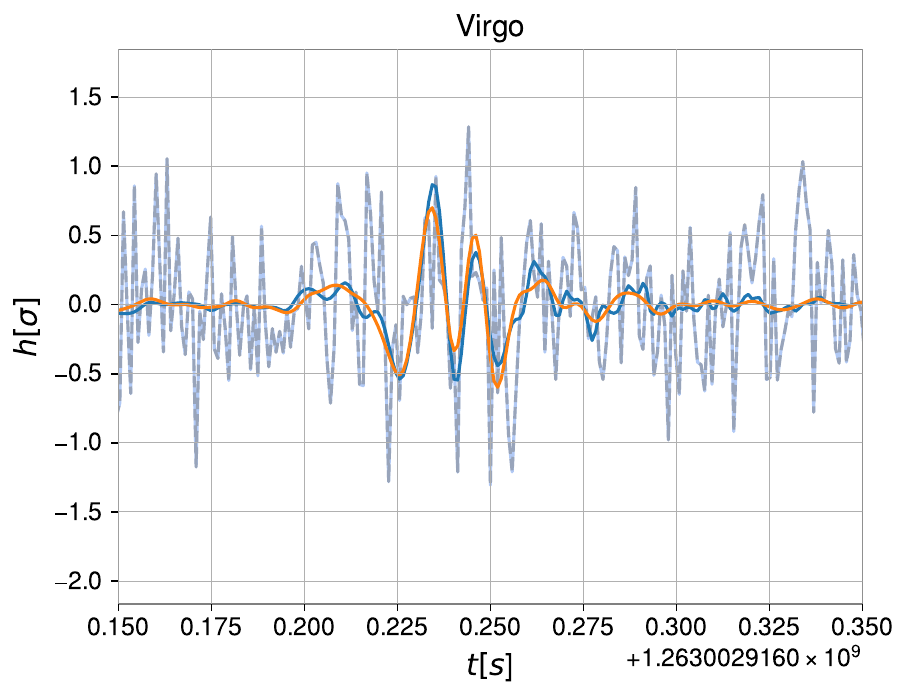}
\caption{\textbf{Whitened $\Psi_4$ and strain time-series around the time of the trigger S200114f}, together with the maximum likelihood waveforms returned by the BBH NRSur7dq4 model (blue) and our Proca star head-on merger model (orange).
}
\label{fig:14f}
\end{center}
\end{figure*}

\section{Results} 
\label{sec:results}

Figs.~\ref{fig:21g}-\ref{fig:26l} show the whitened strain and $\Psi_4$ detector data at times around the four analysed events together with the maximum likelihood templates returned by the BBH and the PSM models. The corresponding signal parameters can be found in Appendix III. Table \ref{tab:logb1} shows the result of our model selection for the events for our two choices of the distance prior. These are labelled by ``V'' (for uniform in co-moving volume) and ``D'' (for uniform in distance). Table \ref{tab:pe} shows our parameter estimates for these events under the PSM scenario. We report median values and symmetric $90\%$ credible intervals. In the following, we first present the result of model selection for individual events to then proceed with a detailed discussion of the properties of each of them.

\subsection{Model Selection} 

Table \ref{tab:logb1} reports natural log Bayes factors, $\log\mathcal{B}$, for the signal $vs.$ noise hypothesis for the events we consider when these are modelled as either BBHs or PSMs. The bottom row reports the relative probability, or Bayes factor, for PSM $vs.$ BBH, ${\cal{B}}^{\text{PSM}}_{\text{BBH}}$. As expected, in all cases the weakness of head-on mergers adds an extra penalty to the PSM model when we use the V prior. For this reason, Bayes factors for the PSM case always grow when we use our D prior while those for the much louder BBH scenario remain almost unchanged. Under the former ``physically realistic'' prior, the PSM merger scenario is mildly favoured by GW190521 and S200114f, with ${\cal{B}}^{\text{PSM}}_{\text{BBH}} \simeq 3.0$ and $7.2$ respectively. Next, the PSM hypothesis is weakly rejected by GW200220, with ${\cal{B}}^{\text{PSM}}_{\text{BBH}} \simeq 0.02$ and strongly rejected by GW190426 with ${\cal{B}}^{\text{PSM}}_{\text{BBH}} \simeq 2\times 10^{-4}$. Using our D prior has somewhat significant consequences. For GW190521, the preference for PSM grows to ${\cal{B}}^{\text{PSM}}_{\text{BBH}} \simeq 40.5$. More spectacularly, for the trigger S200114f we obtain a strong preference for the PSM scenario of ${\cal{B}}^{\text{PSM}}_{\text{BBH}} \simeq 200$ owing to its very small distance estimate of $d_{\rm L} \simeq 150$\,Mpc (see later). Finally, the PSM hypothesis remains strongly rejected for GW190426 with ${\cal{B}}^{\text{PSM}}_{\text{BBH}} \simeq 3\times 10^{-3}$ but very weakly rejected for GW200220, with ${\cal{B}}^{\text{PSM}}_{\text{BBH}} \simeq 0.15$.\\

All in all, for GW190521 we find the same qualitative preference for the PSM model presented in~\cite{Proca} that has a much smaller catalogue. For the other two catalogued events, GW200220 and GW190426, we find mild and strong preferences for the BBH scenario. Finally, the trigger S200114f shows the strongest preference for the PSM scenario. In the following, we analyse in detail these four events, focusing on the parameters we infer under the PSM scenario and, in particular, on potential coincidences in the inferred boson mass $\mu_{\rm B}$ across events.

\begin{table*}
\centering
\begin{center}
%\begin{tabularx}{\columnwidth}{>{\raggedright\arraybackslash}Xrrr}
\renewcommand{\arraystretch}{1.5}
\begin{tabular}{c|ccc|ccc|ccc|ccc}
%\hline 
%\hline
\rule{0pt}{3ex}%
Event & \multicolumn{3}{c}{GW190521}  &  \multicolumn{3}{c}{GW200220} & \multicolumn{3}{c}{GW190426} & \multicolumn{3}{c}{S200114f}  \\
\hline
    & $\log{\cal{L}}_{\text{max}}$& V & D & $\log{\cal{L}}_{\text{max}}$ & V & D & $\log{\cal{L}}_{\text{max}}$ & V & D & $\log{\cal{L}}_{\text{max}}$ & V & D  \\
%% Source (maxL, BF_V, BF_U): (Q6_U, Q4, Q4); (Q6_U, Q4, Q6); (q6_U, q4, Q4); (Q6_U, Q6, Q6)
%% 10/03/23 SL: All the above are wrong.
%% 10/03/23 Source (BF_V, BF_U): (q4, q4); (q4, q4); (q4, Q6); (Q6, Q6)
Black hole merger & 118.7 & 89.6  & 89.7    & 46.7 & 17.4 & 17.4 & 68.3 & 37.9 & 38.2    & 115.9 & 69.1 & 71.0 \\
Proca star merger & 121.0 & 90.7  & 93.4  & 36.7 & 13.4  & 15.5  & 62.5 & 29.5  & 32.4 & 107.4 & 71.1 & 76.3 \\
% $\cal{B}^{\text{PSM}}_{\text{BBH}}$ & & \cor{3.0}  & \cor{40.5}  &  &  & 0.02  & \cor{0.15} &  & $2\times 10 ^{-4}$  & \cor{$3\times 10 ^{-3}$} & \cor{7.2} & \cor{200.3} \\
$\log\cal{B}^{\text{PSM}}_{\text{BBH}}$ & & 1.1  & 3.7  &  &  -4.0  & -1.9 &  & -8.4  & -5.8 & & 2.0 & 5.3 \\
$\cal{B}^{\text{PSM}}_{\text{BBH}}$ & & 3.0  & 40.5  &  &  0.02  & 0.15 &  & $2\times 10 ^{-4}$  & $3\times 10 ^{-3}$ & & 7.2 & 200.3 \\

%Best Parameters  &   &  \\
%\hline
%\rule{0pt}{3ex}%
%Quasi-circular Binary Black Hole & 85.0 &  105.2 \\
%\rule{0pt}{3ex}%
%Head-on Equal-mass Proca Star & 85.4 &  106.7\\ 
%\rule{0pt}{3ex}%
%Head-on Unequal-mass Proca Star & 86.8 &  106.5 \\ 
%\rule{0pt}{3ex}%
%Head-on Binary Black Hole & 79.8 &  103.2 \\ \hline
\end{tabular}
%\end{tabularx}
\caption{\textbf{Summary of model selection on our selected GW events.} The three columns of each event are, in order, the maximum likelihood values, and the natural log Bayes factors (signal $vs.$ noise) obtained using either a standard prior uniform in co-moving volume (V) or a prior uniform in luminosity distance (D). The last row is the corresponding relative Bayes factors. For the BBH model, we report  the maximum values among the 2 mass ratio priors we tested using the \texttt{NRSur7dq4} model. The typical uncertainty in the log Bayes Factors is of order 0.1.}
\label{tab:logb1}
\end{center}
\end{table*}

\begin{table*}[t!]
\centering
\begin{center}
%\begin{ruledtabular}
%\begin{tabularx}{\columnwidth}{>{\raggedright\arraybackslash}Xrr}
\renewcommand{\arraystretch}{1.5}
\begin{tabular}{l|@{\hspace{1em}}c@{\hspace{1em}}|@{\hspace{1em}}c@{\hspace{1em}}|@{\hspace{1em}}c@{\hspace{1em}}|@{\hspace{1em}}c@{\hspace{1em}}}
%\hline
%\hline \\ 
Parameter  & GW190521 & GW200220 &  GW190426 & S200114f  \\ \hline
Primary mass $[M_\odot]$ & $126^{+13}_{-12}$ & $122^{+18}_{-19}$  & $129^{+35}_{-17}$   & $119^{+9}_{-14}$ \\
Secondary mass $[M_\odot]$ & $108^{+11}_{-15}$  & $105^{+15}_{-14}$ & $113^{+28}_{-13}$  & $88^{+16}_{-7}$
\\
Total / Final mass $[M_\odot]$ & $233^{+15}_{-16}$ & $228^{+24}_{-29}$  & $244^{+41}_{-26}$ & $207^{+16}_{-14}$
\\
Primary spin & $1.48^{+0.27}_{-0.14}$ & $1.44^{+0.21}_{-0.15}$  & $1.56^{+0.09}_{-0.12}$ & $1.56^{+0.04}_{-0.27}$
\\
Secondary spin & $1.28^{+0.14}_{-0.16}$ & $1.26^{+0.22}_{-0.12}$  & $1.37^{+0.19}_{-0.19}$ & $1.14^{+0.12}_{-0.01}$
\\
Final spin & $0.69^{+0.04}_{-0.04}$ & $0.66^{+0.10}_{-0.03}$    & $0.71^{+0.07}_{-0.04}$ & $0.66^{+0.03}_{-0.04}$
\\
Inclination $\pi/2-|\iota-\pi/2|$ [rad] & $0.68^{+0.35}_{-0.43}$ & $0.92^{+0.49}_{-0.23}$  & $0.65^{+0.54}_{-0.46}$  & $0.91^{+0.50}_{-0.24}$ 
\\
Luminosity distance [Mpc] & $568^{+356}_{-259}$  & $856^{+804}_{-421}$   &   $927^{+587}_{-591}$  &  $152^{+73}_{-61}$ 
\\
Right ascension & $3.55^{+2.66}_{-3.48}$  & $2.99^{+1.89}_{-0.49}$   &   $1.39^{+3.31}_{-0.53}$  &  $1.26^{+0.70}_{-0.13}$ 
\\
Declination & $0.48^{+0.47}_{-1.66}$  & $-0.06^{+0.51}_{-1.07}$   &   $0.08^{+0.58}_{-0.56}$  &  $-0.11^{+0.33}_{-0.34}$ 
\\
Polarization & $1.41^{+1.47}_{-1.16}$  & $1.51^{+1.47}_{-1.31}$   &   $1.60^{+1.52}_{-1.48}$  &  $0.31^{+1.30}_{-0.16}$ 
\\
Redshift $z$ & $0.12^{+0.06}_{{-0.05}}$ & $0.18^{+0.14}_{{-0.08}}$ & $0.18^{+0.11}_{{-0.12}}$ & $0.03^{+0.02}_{{-0.01}}$ 
\\
Total / Final redshifted mass $[M_\odot]$  & $260^{+9}_{-8}$ & $267^{+18}_{-16}$  & $289^{+26}_{-17}$ & $214^{+16}_{-14}$
\\
Primary field frequency $\omega_1/\mu_{\rm B}$  & $0.910^{+0.016}_{-0.023}$ &   $0.903^{+0.024}_{-0.032}$  & $0.920^{+0.007}_{-0.019}$ & $0.919^{+0.006}_{-0.043}$ 
\\
Secondary field frequency $\omega_2/\mu_{\rm B}$  & $0.867^{+0.035}_{-0.052}$  & $0.858^{+0.051}_{-0.050}$   & $0.888^{+0.032}_{-0.058}$  & $0.810^{+0.062}_{-0.010}$
\\
Boson mass $\mu_{\rm B}$ [$\times 10^{-13}$\,eV] & $8.69^{+0.61}_{-0.75}$ & $9.13^{+1.18}_{-1.30}$ &  $7.77^{+0.87}_{-0.96}$ & $10.20^{+0.68}_{-0.55}$ 
\\
Maximal boson star mass $[M_\odot]$  & $173^{+16}_{-11}$ & $165^{+27}_{-22}$ &  $193^{+28}_{-19}$ & $147^{+14}_{-9}$ \\

%ULogBayesFactor^{\text{Proca}}_{\text{BBH,Best}}  & $3.1$ (NRSur) & $2.5$ (NRSur)  & $1.0$ (XPHM)  & ---
%Evidence for $(2,0)$ mode & $\log{\cal B}  \simeq 0.6$ & ---
%\\[6pt]
%\hline
%\hline
%\end{tabularx}
\end{tabular}
%\end{ruledtabular}
\caption{\textbf{Parameters of the four events discussed in this work under a PSM scenario.}  We quote median values with symmetric $90\%$ credible intervals. Please see Table~\ref{tab:pe_bbh} in Appendix~\ref{sec:appBBHPE} for parameters obtained under the BBH scenario}.  
\label{tab:pe}
\end{center}
\end{table*}

\begin{figure*}[t!]
\begin{center}
\includegraphics[width=0.48\textwidth]{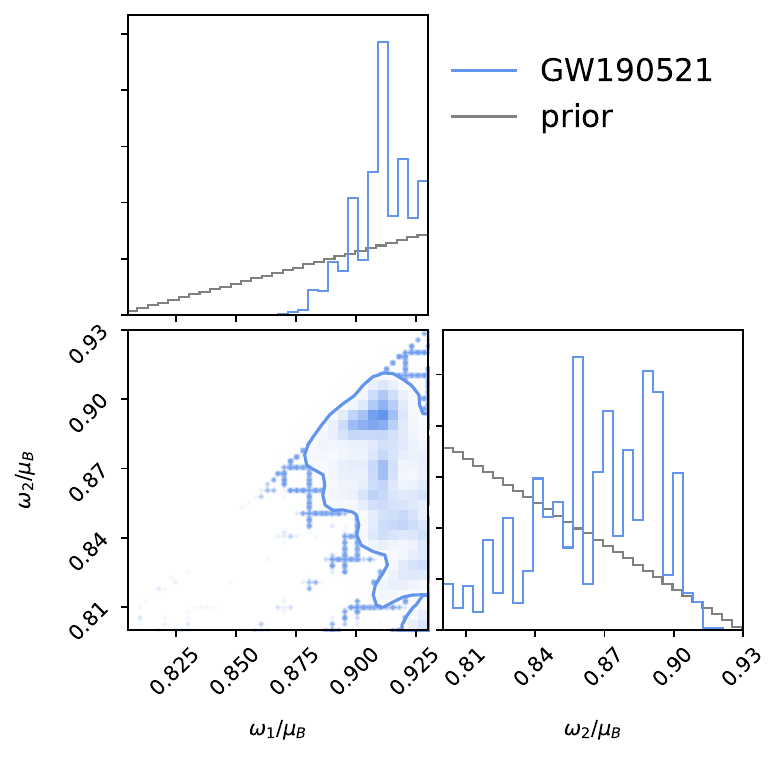}
\includegraphics[width=0.48\textwidth]{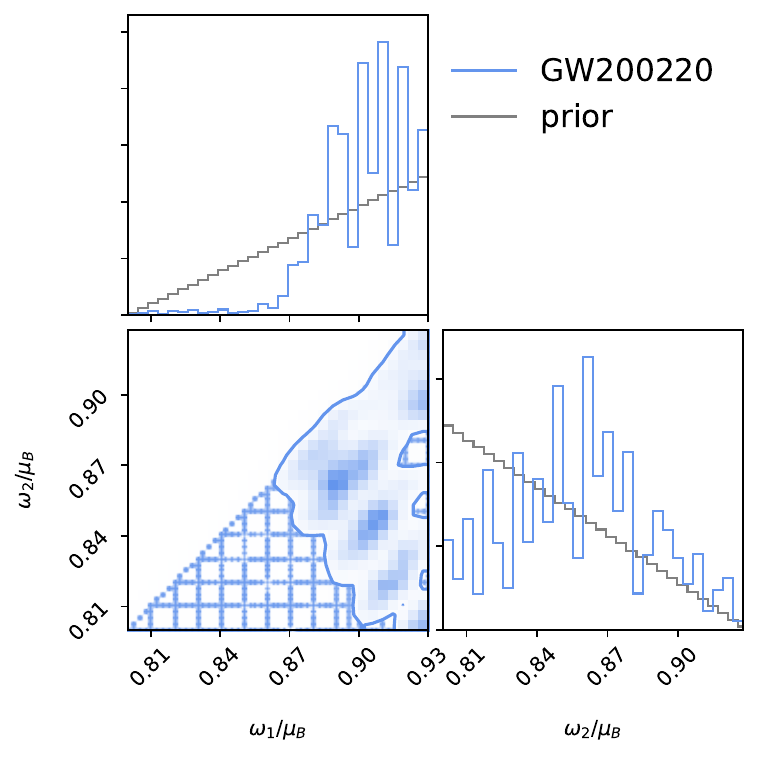}
\includegraphics[width=0.48\textwidth]{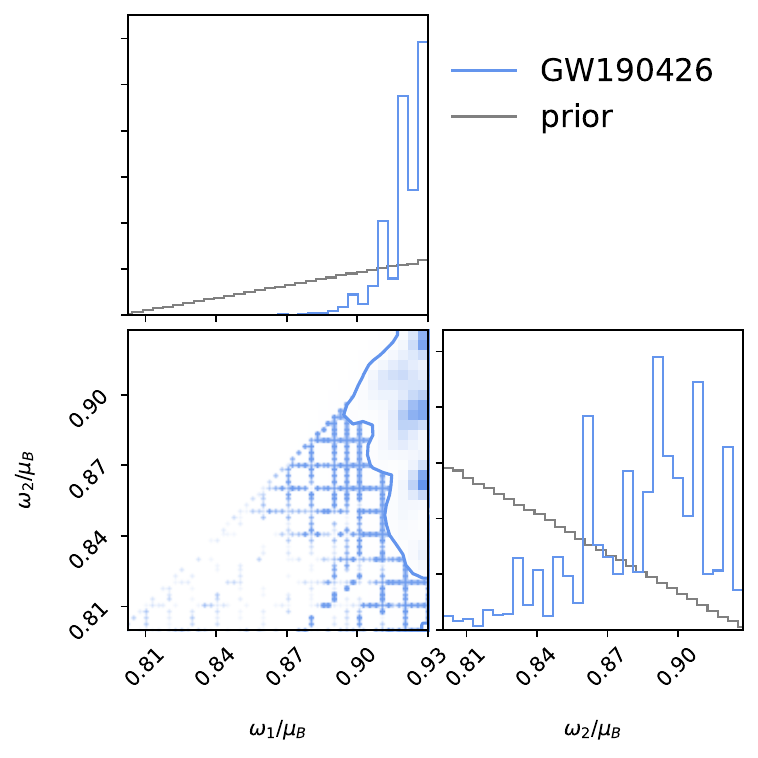}
\includegraphics[width=0.48\textwidth]{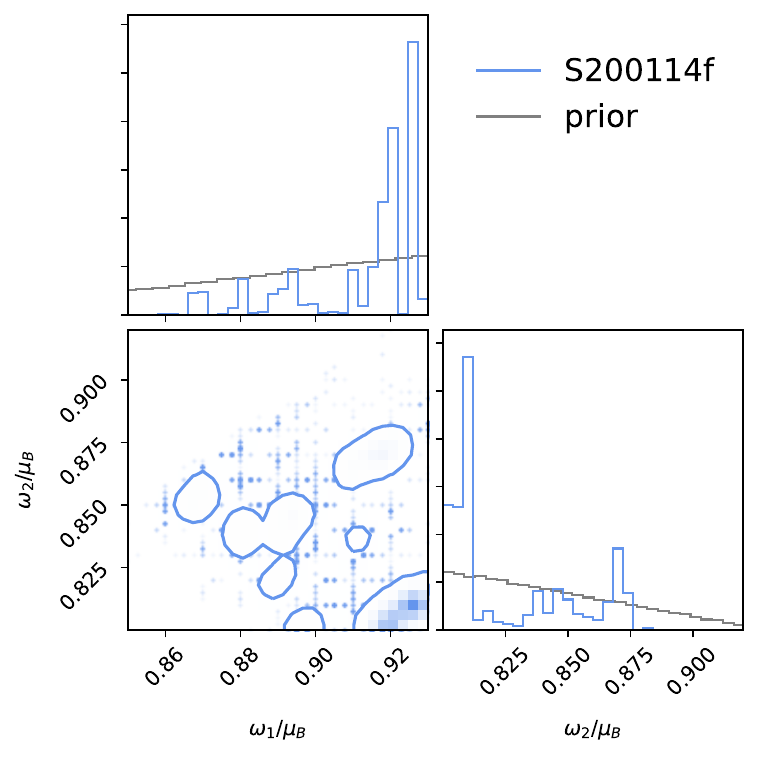}

\caption{\textbf{Posterior distributions for the field frequencies $\omega_{1,2}/\muB$ for the events analysed in this work}. We show the two-dimensional $90\%$ credible regions together with the corresponding one-dimensional posterior and prior distributions. The color darkness is proportional to the probability density. Note that while the contours appear to span continuous regions, the samples are actually placed on a discrete grid, as is most evident for the case of GW200220. For this reason, the probability density appears to be distributed in a fuzzy way within the contours and along discrete points outside them. Apparently ``missing'' points (e.g., in the bottom left region for GW190521) are actually white-coloured due to their very low probability density.} 
\label{fig:omega_posteriors_one}
\end{center}
\end{figure*}

\subsection{Parameter Estimation} 

We now discuss the properties we infer from each individual event. As mentioned above, our parameter inference results are summarised in Table~\ref{tab:pe}. In addition, Fig.~\ref{fig:omega_posteriors_one} shows two-dimensional credible regions for the field frequencies for the different events, together with the corresponding one-dimensional posterior distributions. Before diving into a detailed per-event discussion, we comment on two of the main \nncor{limitations} of our study, which are visible in the mentioned figure.\\ 

First, we note that our simulation catalogue is built as an expansion of that in~\cite{Proca}, mostly tailored to encompass GW190521. Consequently, for some events, the highest likelihood (best-fitting) points correspond to corner cases in our catalogue, making the most-probable regions of the parameter space to lay in such \nncor{corners} (see e.g. S200114f). On the one hand, this can lead to artificially small uncertainties in the frequencies $\omega_i/\mu_{\rm B}$ of the star fields and, therefore, to overly constrained boson-mass $\mu_{\rm B}$ estimates. On the other hand, the true best-fitting points may lay beyond the limits of our catalogue, making the evidences for the PSM model discussed above rather conservative.\\

Second, as also mentioned earlier, our simulation catalogue is limited to relative phases $\Delta\epsilon (t=0)=0$ at the start of our simulations, which leads to varying phase differences at merger $\Delta\epsilon (t=t_{\rm merger})$ for different combinations of $\omega_{1,2}/\muB$. For this reason, the different panels in Fig. \ref{fig:omega_posteriors_one} show a sort-of band structure that roughly corresponds similar values of $\Delta\epsilon (t=t_{\rm merger})$. While, again, this limits the physics present in our catalogue, this also means that a more complete catalog may better encompass the events we have analysed.\\

We now discuss individually the properties of each event. While in the following we will only focus on the parameters obtained under the PSM hypothesis, we provide a summary of those obtained under the BBH hypothesis in Appendix \ref{sec:appBBHPE}.

\subsubsection{GW190521} Our results are fully consistent with those reported in \cite{Proca}. For the final BH, we estimate a red-shifted final mass of $M_{z}=260^{+9}_{-8}\,M_\odot$ and a final spin of $a_f=0.69^{+0.04}_{-0.04}$, compatible with those reported by the LVK \cite{GW190521D}. We note that the final BH mass is essentially equal to the initial mass due to the negligible loss to GWs during head-on mergers, which also leads to a much lower source luminosity. We infer a luminosity distance around ten times closer than that estimated by the LVK at $d_{\rm L}=568^{+356}_{-259}$\,Mpc. Consequently, we obtain a much heavier source-frame mass of $M_{\rm src}=M_{z}/(1+z)=233^{+15}_{-16}\,M_\odot$. The individual source-frame mass estimates are $m_1 = 126^{+13}_{-12}\,M_\odot$ and $m_2 = 109^{+11}_{-15}\,M_\odot$. Remarkably, despite the significant growth of our simulation catalogue, these values are consistent with those reported in \cite{Proca} even though that study was limited to equal-mass PSMs.\\

As per the Proca-star specific parameters, Fig. \ref{fig:omega_posteriors_one} shows  informative posteriors for this and all the remaining events. We estimate star-frequencies $\omega_1/\mu_{\rm B}=0.910^{+0.016}_{-0.023}$ and $\omega_2/\mu_{\rm B}=0.867^{+0.035}_{-0.052}$, both consistent with those reported in \cite{Proca}. These, combined with the masses of the individual stars, allow us to estimate the mass of the underlying ultralight boson via
%\begin{equation}
%    \mu_{\rm B}=1.34\times10^{-10}\,\biggl(\frac{\mathcal{M}^{i}_{\text{PS}}}{M_{\rm BH}^{\rm final}\times (\mathcal{M}^{i}_{\text{PS}}/\mathcal{M}^T_{\text{PS}})}\biggl)\,\text{eV},
 %   \end{equation}
\begin{equation}
    \mu_{\rm B}=1.34\times10^{-10}\,\biggl(\frac{{\cal{M}}_1 + {\cal{M}}_2}{M^{\text{final} }_{\rm BH} / M_\odot}\biggl)\,\text{eV}.
    \label{eq:boson_mass}
\end{equation}
% As a remark to any future reader, the number 1.34e-10 comes from Planck mass squared, in a funny unit.
% 1.335e-10 = (M_Pl / eV) * (M_Pl / M_sun)
% So in a nutshell: M_Pl^2 = 1.335e-10 eV M_sun
%
Here, $\mathcal{M}_{1,2}=\mu_{\rm B}\, m_{1,2}/M^2_{\rm{Pl}}$ is a dimensionless mass parameter characterising each Proca star, $m_{1,2}$ denotes the source frame mass of each star, $M_{\rm{Pl}}$ the Planck mass and %and $\mathcal{M}^T_{\text{PS}}$ the total mass of the Proca binary system. 
$M_{\rm BH}^{\rm final}$ the mass of the final BH. %expressed in solar masses. 
We obtain  $\mu_{\rm B}^{\text{GW190521}}=8.69^{+0.61}_{-0.75}\times 10^{-13}$ eV. Finally, we make use of the boson mass and the individual star masses to infer the maximal mass $M^{\text{Proca}}_{\text{max}}$ that a Proca star could form by such bosons can have before collapsing to a black hole, given by
\begin{equation}
    M^{\rm Proca}_{\rm max} = %\qty(\frac{M_\text{max}}{M_{\odot}}) %=
    1.125\,\biggl(\frac{1.34\times10^{-10}\,\text{eV}}{\mu_{\rm B}}\biggl)M_{\odot}.
\end{equation}

This expression comes from the formula of the maximum mass of a given bosonic star model~\cite{liebling2017dynamical}. The factor $1.125$ depends on the model and here we use the corresponding value of the maximum mass of a $m=1$ spinning Proca star in the fundamental state~\cite{Herdeiro:2019mbz}. We obtain $M^{\rm Proca}_{\rm max}=173^{+16}_{-12}\,M_\odot$.

\begin{figure}[t!]
\begin{center}
\includegraphics[width=0.47\textwidth]{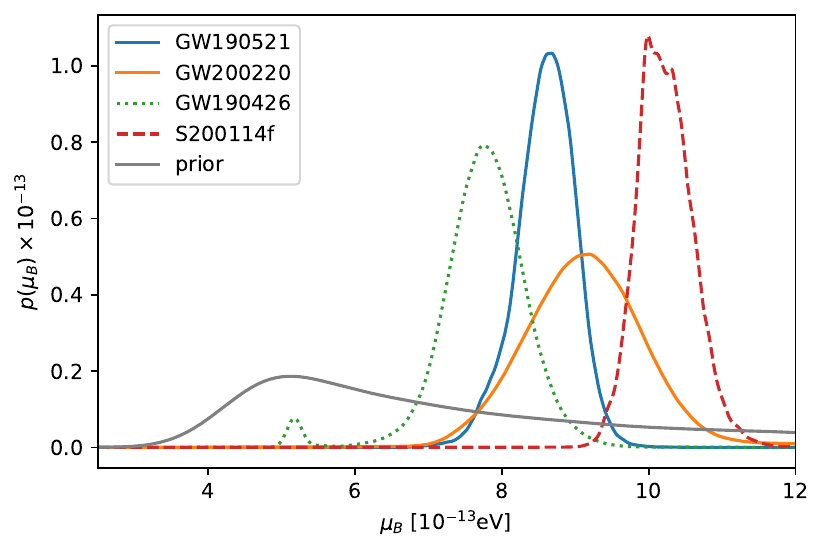}
\caption{\textbf{Boson-mass estimates for the events analysed in this work.} The coloured curves show the posterior distributions for the boson mass. The grey curve denotes the boson-mass prior, which is determined by our priors in the star frequencies, total red-shifted mass and luminosity distance.% \ncor{Change the xlabel to $\mu_{\rm B}$.}
}
\label{fig:mub}
\end{center}
\end{figure}

\subsubsection{GW200220} 

GW200220 was detected by the matched-filter search \texttt{PyCBC}~\cite{Usman:2015kfa} during the third observing run of Advanced LIGO and Virgo with an inverse-false-alarm rate (IFAR) of 0.15\,yr and a probability of astrophysical origin $p_{\text{astro}}= 0.62$ \cite{abbott2021gwtc3}. This is significantly lower than that of GW190521, which was found with an IFAR of 4900\, yr \cite{GW190521D} \footnote{This IFAR was obtained through the search for generic transients \texttt{coherent WaveBurst} (\texttt{cWB}). GW190521 was later associated $p_{\text{astro}}$ values of 0.93 and 1 \cite{abbott2021gwtc2} by the matched-filter search algorithms \texttt{PyCBC}~\cite{Usman:2015kfa} and \texttt{gstlal} \cite{Messick:2016aqy}, which are sub-optimal for this kind of event due to e.g., the omission of orbital precession in the search templates \cite{CalderonBustillo:2017skv}}.. Despite its low significance, GW200220 outstands as the third-heaviest BBH reported to date. While in the previous section we showed that this event is more consistent with a BBH, it is still interesting to discuss the properties we obtain under the PSM scenario.\\

We find that GW200220 is essentially a more distant copy of GW190521, with consistent individual masses of $m_1 = 122^{+18}_{-19}M_\odot$ and $m_2 = 105^{+15}_{-14}$; and a similar final spin of $a_f=0.66^{+0.10}_{-0.03}$ but located at a slightly larger $d_{\rm L}=856^{+804}_{-421}$\,Mpc. More interestingly, while we stress again the marginal character of this event, we obtain very similar field frequencies of $\omega_1/\mu_{\rm B}=0.903^{+0.024}_{-0.032}$ and $\omega_2/\mu_{\rm B}=0.858^{+0.051}_{-0.050}$. The frequency posteriors for GW200220 differ from those of GW190521 in two main aspects. First, the larger loudness of GW190521 makes the likelihood to be more peaked, discarding the low $\omega_{1,2}/\muB$ region, as is clear in Fig. \ref{fig:omega_posteriors_one}. Second, the posterior for GW200220 clearly shows the ``band-structure'' caused by the varying value of the relative phase at merger across our catalogue. Again, we understand that this is less obvious for GW190521 due to its larger loudness.\\

As expected from the above frequency and mass values, we estimate boson-mass of $\mu_{\rm B}^{\text{GW200220}}=9.13^{+1.18}_{-1.30}\times 10^{-13}$\,eV completely consistent with that of GW190521. Using the formalism in \cite{Ashton2021_overlap_int}, we can test the hypothesis that the two events are sourced by the same ultra-light boson, i.e., that they share the same boson mass. For two events A and B, we can compute the odds-ratio\footnote{The odds-ratio is defined as $\mathcal{O}_{C/U} = \mathcal{I}_{\mu_{\rm B}}\,\pi_{C/U}$, where $\pi_{C/U}$ is the prior odds of the two hypotheses, common v.s. uncorrelated, and we have implicitly assumed equal prior probabilities, i.e.: $\pi_{C/U} = 1$.} for the common v.s. uncorrelated mass through the overlap integral
\begin{equation}
    {\cal{I}}_{\mu_{\rm B}}^{\rm AB} = \int \frac{p(\mu_{\rm B} | \mathrm{A})\,p(\mu_{\rm B} | \mathrm{B})}{\pi(\mu_{\rm B})}\,\dd\mu_{\rm B},
\label{eq:overlap}
\end{equation}
where $\pi(\mu_{\rm B})$ denotes our prior on the boson mass, represented by the grey curve in Fig.~\ref{fig:mub}. For the pair GW190521-GW200220 we obtain a odds-ratio ${\cal{I}}_{\mu_{\rm B}}=5.5$ favouring a common $\mu_{\rm B}$. This means that, if we consider that the two events share the same boson, the relative evidence for the PSM vs. BBH scenarios rises by a factor of 5.3. Later, we will showcase how this result can be exploited in the context of population studies in section \ref{sec:pop}.\\

Finally, we infer a maximal Proca star mass $M^{\text{Proca}}_{\text{max}}=165^{+27}_{-22}\,M_\odot$. On the one hand, this is consistent with the one inferred from GW190521. On the other, the total masses of both events are consistently larger than the  estimated maximal Proca star masses. This implies that, in both cases, the remnant hyper-massive boson star formed at the end of the two mergers has enough mass to collapse into a black-hole and yield the corresponding characteristic ringdown signal expected by current gravitational-wave searches.

\subsubsection{GW190426} 

GW190426 was detected by a version of the matched-filter search \texttt{PyCBC} specifically targeting BBH signals \cite{Nitz2020_PyCBCBBH} with an IFAR of $0.25$\,yr and a $p_{\text{astro}}=0.75$ \cite{GWTC2.1}, again significantly lower than that of GW190521. While, under our current catalogue, this event is strongly discarded as a PSM, it is still interesting to look at some of the properties that are inferred under such a scenario. First, we note that the primary field frequency $\omega_1/\mu_{\rm B}=0.919^{+0.006}_{-0.043}$ clearly rails against the upper limit of our catalogue (see also Fig. \ref{fig:omega_posteriors_one}). This evidences that we need to enlarge our catalogue to correctly encompass this event. Nevertheless, at the same time, it is interesting to note that we obtain a boson-mass of $\mu_{\rm B}^{\text{GW190426}}=7.77^{+0.87}_{-0.96}\times 10^{-13}$\,eV lower than (despite consistent with) those inferred from the previous two events. In particular, we find a overlap integrals ${\cal{I}}_{\mu_{\rm B}}=1.8$ and ${\cal{I}}_{\mu_{\rm B}}=3.7$ favouring the common-boson hypothesis when comparing this event with GW190521 and GW200220 respectively.

\subsubsection{S200114f} 
S200114f is a short-duration transient observed during the second half of the third observing run of Advanced LIGO and Virgo \cite{abbott2021gwtc3,o3_imbh}. This intriguing trigger was missed by matched-filter searches targeting black hole mergers (which omit orbital precession \cite{Harry:2016ijz,Bustillo:2016gid,Chandra2020_Nuria} and higher-order harmonics \cite{Capano:2013raa,Harry:2017weg,CalderonBustillo:2017skv,Chandra2022_IMBHHMsearch}) but was observed by the model-agnostic search coherent Wave Burst \cite{Klimenko:2015ypf} with an IFAR of 34\,yr \cite{o3_imbh} \footnote{This is reduced to 17\,yr after applying a trials factor accounting for the fact that this trigger was searched for using both Hanford-Livingston and Hanford-Livingston-Virgo data \cite{Klimenko:2015ypf,o3_imbh}}. Due to the lack of detection by matched-filter searches, S200114f has not been labelled as a confirmed detection, but, nevertheless, nor has it been conclusively classified as background noise either. Remarkably, parameter estimation was performed on this trigger with three different state-of-the-art waveform models \cite{NRSur7dq4,SEOBNRv4PHM,XPHM_Pratten}, with all results across different models returning values for the individual masses. Rather than revealing that this trigger is not a black hole merger, or even not of astrophysical origin, these results showcase the inconsistencies between these BBH approximants at the regions of the parameter space that best fit the signal. Additionally, while the morphology of this trigger is consistent with that of a family of noise transients known as Tomte glitches \cite{Merritt2021_tomtes}, it was not possible to conclusively rule out an astrophysical origin. We therefore consider it interesting to analyse this event from the perspective of further waveform models and, in particular, under our PSM catalogue.

In terms of its masses, we find that S200114f is essentially a lighter and more nearby version of GW190521 with a much larger inclination. We estimate a final total red-shifted mass of $M_{z}=214^{+16}_{-14}\ M_\odot$ and a distance of $d_{\rm L}=152^{+73}_{-61}$\,Mpc. Owing to the standard distance prior, the louder BBH scenario should be implicitly favoured by our analysis. Despite this, we obtain ${\cal{B}}^{\text{PSM}}_{\text{BBH}}{\sim} 7.2$, slightly preferring the PSM scenario. Moreover, removing the effect of such prior yields a ${\cal{B}}^{\text{PSM}}_{\text{BBH}}\simeq 200$, strongly preferring the PSM model.

The above combination of red-shifted mass and distance results in a source-frame mass of $M_{\rm src}=207^{+16}_{-14}\ M_\odot$. The final black hole would have a spin of $a_{\rm f}=0.66^{+0.03}_{-0.04}$. The main difference in the intrinsic properties of S200114f w.r.t. GW190521 arises from the frequency of their bosonic fields. We estimate $\omega_1/\mu_{\rm B}=0.919^{+0.006}_{-0.043}$ and $\omega_2/\mu_{\rm B}=0.810^{+0.062}_{-0.010}$ for this event. We note that the extremely small uncertainties of $\delta\omega_i=0.01$ in the lower and upper ends of the respective frequency ranges are solely due to the fact that this event lies on the edge of our simulation catalogue, which makes our posterior distributions rail against such limits (see Fig.\ref{fig:omega_posteriors_one}. On the one hand, this means that all the provided results are over-constrained even within the head-on paradigm. On the other hand, this reveals that there is room for improvement in fitting this event within the scenario we propose. Altogether, we obtain a value for the boson mass $\mu_{\rm B}^{\text{S200114f}}=10.20^{+0.68}_{-0.55}\times 10^{-13}$\,eV, larger than for the previous events. While we find that the common-boson hypothesis is favoured with ${\cal{I}}_{\mu_{\rm B}}=3.7$ w.r.t. GW200220, it is rejected w.r.t. GW190521 with ${\cal{I}}_{\mu_{\rm B}}=0.1$. Finally, the boson masses of S200114f and GW190426 are highly inconsistent with ${\cal{I}}_{\mu_{\rm B}}=0.02$.\\

Even though some of the studied pairs of events yield rather inconsistent boson masses, we stress that we are imposing the very restrictive scenario of a head-on merger. Recall that the frequency of the bosonic field -- which determines the boson masses -- fixes the spins of the individual stars and consequently the spin of the final BH. Therefore, the preferred star frequencies for the merging stars are those that can lead to the correct final BH spin. Expanding our numerical relativity catalogue to less eccentric configurations would provide an extra contribution from the orbital angular momentum to the final spin, therefore allowing for a wider range of star frequencies and, consequently, boson masses. The expectation is that without the head-on restriction the true boson mass posteriors should be significantly broader, which would lead to a much better consistency for the mass across events. For these reasons, we think it is quite remarkable that the analysed events yield the slightest consistency.

\begin{table}
\centering
\renewcommand{\arraystretch}{1.35}
\begin{tabular}{l | c c c c }
        & GW190521 & GW200220 & GW190426 &  S200114f  \\[4pt]
Triplet & 0.1 (0.2)    & 0.02 (0.05)    & 0.6  (1.3)   &  12.7 (5.6)  \\ 
\hline  
GW200220             & 5.3 (6.3) &    --     &   --        &  --   \\ 
GW190526             & 3.3 (1.1) & 1.8 (0.9) &   --        &  --   \\ 
S200114f             & 0.1 (0.2) & 3.7 (2.9) & 0.02 (0.04) &  --   \\
\hline
\end{tabular}
\caption{\textbf{Mass-overlap integrals for pairs and triplets.} The three bottom rows show the overlap integrals $\mathcal{I}^{AB}_{\mu_{\rm B}}$ for each pair of the events we study. The top row shows the overlap integral for each of the possible triplets, excluding the event on the top of the corresponding column. Values within parentheses correspond to analyses using a uniform distance prior while the rest correspond to a standard prior uniform in co-moving volume. Values larger than one favour the common-boson hypothesis over the uncorrelated one. The overlap integral values for GW190521 with itself are 9.62 (9.92).}
\label{tab:overlap}
\end{table}

\section{Population} 
\label{sec:pop}

The existence of multiple events that can be compared to our PSM model invites the question of whether statistical evidence for these objects can be accumulated across the observed events, even in the absence of conclusive evidence coming from a single one (see e.g., \cite{Saleem2022} for a similar application). In other words, we can estimate whether the observed set contains a fraction $\zeta$ of PSMs. Starting from our observational data set of four events $\{ d_i\}$, we consider a population of compact objects consisting of a fraction $\zeta$ of PSMs and a fraction $1-\zeta$ of BBHs. With this, we can compute the likelihood of our data set given $\zeta$ as
\begin{equation}
    \begin{aligned}
        p(\{d_i\}|\zeta) &= \prod_{i=1}^{N=4} \big{[}p(d_i|\text{PSM})\,\zeta + p(d_i|\text{BBH})\,(1-\zeta) \big{]} \\ 
                         &\propto \prod_{i=1}^{N=4} \big{[}{\cal{B}}^{\text{PSM}}_{\text{BBH,i}}\,\zeta + (1-\zeta )\big{]},
    \end{aligned}
    \label{eq:pop}
\end{equation}
where ${\cal{B}}^{\text{PSM}}_{\text{BBH,i}} = {\cal{B}}^{\text{PSM}}_{\text{i}}/{\cal{B}}^{\text{BBH}}_{\text{i}} $ denotes the relative Bayes factor between PSM and BBH models of the $i$-th event.\\

\subsection{Boson-mass agnostic calculation}

The left panel of Fig. \ref{fig:pop} shows the posterior distribution of $\zeta$, where we use the Bayes factors reported in Table \ref{tab:logb1}. We note that in all the cases we will discuss, we impose an uniform prior on $\zeta \in [0,1]$. Solid curves include S200114f as a real event while dashed ones exclude it. Blue curves correspond to a uniform prior in co-moving volume. In these cases, we see that ignoring S200114f returns a posterior that peaks near $\zeta=0$ and, at the same time, shows support all the way to $\zeta=1$. In particular, we obtain $\zeta = 0.27^{+0.43}_{-0.25}$, with $\zeta > 0.05$ at the $90\%$ credible level. The inclusion of S200114f as a true event raises this to $\zeta = 0.39^{+0.38}_{-0.33}$ with $\zeta > 0.11$ at the $90\%$ credible level, with a peak at $\zeta \simeq 0.3$. Red curves correspond to our uniform distance prior. In this case, ignoring S200114f we obtain a posterior peaking at $\zeta \simeq 0.3$ with a $90\%$ lower bound of $\zeta = 0.13$. Including S200114f as a real event raises the latter to $\zeta = 0.25$ (i.e, at least event should be a PSM instead of a BBH) with a peak at $\zeta \simeq 0.6$.\\

\subsection{Exploiting boson-mass consistencies}

The fact that some of the events show consistent boson masses further invites the question of whether these can be analysed assuming a common mass value. In such a case, the evidence for the PSM model would rise due to the reduction of the number of parameters and the consequent reduction of the Occam penalty (see e.g., \cite{Bustillo2021,Thrane2019}). Given the original prior for the boson-mass $\pi(\mu_{\rm B})$, the original posterior for each event $p_i(\mu_{\rm B})$ and a new prior for the boson-mass $\pi^{*}(\mu_{\rm B})$, the new value of the Bayesian evidence ${\cal{Z}}^*_i$ for each event can be obtained through
\begin{equation}
    {\cal{Z}}^{*}_i = \mathcal{Z}_i\int \pi^{*}(\mu_{\rm B}) \frac{p(\mu_{\rm B})}{\pi(\mu_{\rm B})} \,\dd\mu_{\rm B} = {\cal{I}}^{\pi^{*}}_{\mu_{\rm B}}{\cal{Z}}_i,
\end{equation}
where ${\cal{Z}}_i$ denotes the Bayesian evidence obtained under the original prior $\pi(\mu_{\rm B})$. While in principle a mass-prior assuming a unique ultra-light boson should be given by a delta function centred at a given mass, here we set a prior equal to the posterior for the most significant of our four events. This is, we choose $\pi^{*}(\mu_{\rm B}) = p^{\text{GW190521}}(\mu_{\rm B})$. The updated evidence for the remaining events under the PSM hypothesis is therefore given by 
\begin{equation}
    {\cal{Z}}^{*}_i = \mathcal{Z}_i \int p^{\text{GW190521}}(\mu_{\rm B})\frac{p(\mu_{\rm B})}{\pi(\mu_{\rm B})}\, \dd\mu_{\rm B} = {\cal{I}}^{\text{GW190521},i}_{\mu_{\rm B}}{\cal{Z}}_{i}.
\end{equation}
The factor ${\cal{I}}^{\text{GW190521},i}_{\mu_{\rm B}}$ is known as the overlap integral and, as previously shown in \cite{Ashton2021_overlap_int}, is equivalent to the relative Bayes factor between the common-source v.s. uncorrelated source hypotheses for the two compared events. In Table~\ref{tab:overlap}, we display these values for all signal pairs, together with the corresponding three-event integrals\footnote{The ``triple'' integral is computed through ${\cal{I}}_{\mu_{\rm B}}^{ABC} = \int \frac{p(\mu_{\rm B} | \mathrm{A})\,p(\mu_{\rm B} | \mathrm{B})\,p(\mu_{\rm B} | \mathrm{C})}{\pi(\mu_{\rm B})\pi(\mu_{\rm B})}\,\dd\mu_{\rm B}$. We note, however, that ${\cal{I}}_{\mu_{\rm B}}^{ABC}$ does not enter any of our calculations and it is only provided for comparison} purposes. The new PSM v.s. noise Bayes factor is then given by ${\cal{B}}_i^{*}={\cal{I}}^{\text{GW190521}, i}_{\mu_{\rm B}}\,{\cal{B}}_i$. Finally, by replacing $\mathcal{B}^{\rm PSM}_{\rm i}$ in Eq.~\eqref{eq:pop} with these, we can recompute the posterior distribution of the fraction of PSMs $\zeta$ under the assumption that all events share the same boson as GW190521.\\

\begin{figure*}[t!]
\begin{center}
\includegraphics[width=0.49\textwidth]{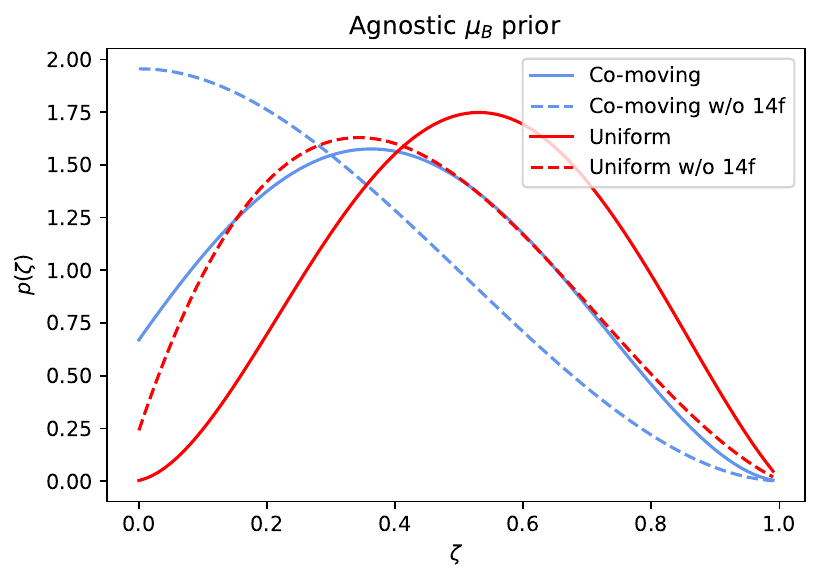}
\includegraphics[width=0.49\textwidth]{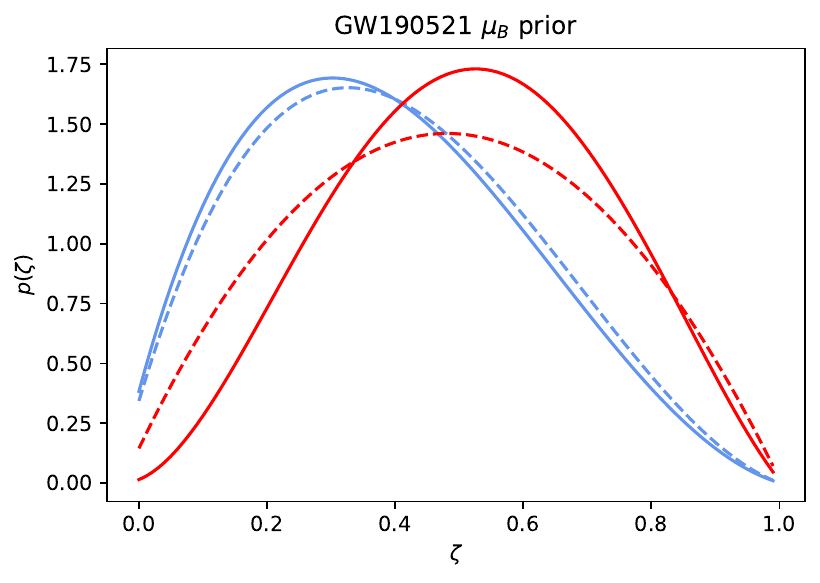}
\caption{\textbf{Population fractions of boson-star mergers $\zeta$ within our data set for two different priors}. In the left panel, we ignore any correlations between the boson mass obtained for our events. In the right panel, we impose a boson-mass prior given by the posterior for GW190521. Blue curves make use of a distance prior uniform in co-moving volume while red ones impose a uniform-in-distance prior. Finally, solid (dashed) curves include (exclude) S200114f as a real gravitational-wave event.
}
\label{fig:pop}
\end{center}
\end{figure*}

The right panel of Fig.~\ref{fig:pop} shows the new posteriors of $\zeta$. Exploiting common masses has dramatic consequences when S200114f is not considered as a true event. This is expected as the overlap integrals of the remaining two events support the common boson hypothesis, therefore increasing their evidence as PSMs. In particular, for each of our two distance priors, we now obtain posteriors peaked at $\zeta=0.3$ and $\zeta=0.5$ and $90\%$ lower bounds of $\zeta=0.11$ and $\zeta=0.24$. While a similar qualitative effect is observed when including S200114f, this is quantitatively less dramatic. The reason is that the raised $\cal{B}^{\text{PSM}}_{\text{BBH}}$ for the other events are now accompanied by a reduction of that for S200114f due to its highly inconsistent boson mass with respect to GW190521.\\

The above should be considered as a proof-of-principle calculation with relevant shortcomings that can artificially favour each of the PSM and BBH hypotheses. First, we have ignored the prior on the relative abundance of BHs and Proca stars in the Universe. Additionally, we have ignored other kinds of possible exotic compact binaries as, for instance, mixed BH-PS mergers. Second, because at the moment no simulations for circular PSM exist, we ignore the fact that highly eccentric (let alone head-on) mergers are highly astrophysically suppressed. Finally, we also note that the black-hole merger model~\cite{NRSur7dq4} is limited to non-eccentric binaries with mass-ratio $q\leq 6$ and that some of these events may be better reproduced when adding the effect of orbital eccentricity, as it is the case for GW190521~\cite{Bustillo2021,RomeroShaw2020_ecc_apjl,Gayathri2022_ecc_natastro,Gamba2022_ecc_natastro}, or even by mass-ratios larger than those allowed by the model. On the other hand, we also note that our PSM model is also incomplete and constrained to a narrow number of cases, which causes some of the analysed events to lay on the edges of our parameter space. Increasing our parameter coverage would most likely lead to improved fits and, therefore, increased evidence of these events.

\section{Discussion \\}
 %{\it \textbf{Discussion.}} 
Despite their canonical interpretation as black-hole mergers, short GW transients displaying barely any pre-merger emission merit further exploration of their possible origin. We have compared four such events to a catalogue of 759 numerical-relativity simulations of PSMs. Performing model selection with respect to vanilla quasi-circular BBH mergers, we find that the most significant of these events (GW190521) and the loud trigger S200114f favour the PSM hypothesis. The weaker events GW200220 and GW190426 respectively weakly and strongly reject the hypothesis. Remarkably, we find that two of the \textit{catalogued} GW events which are not strongly discarded as PSMs, namely GW190521 and GW200220, yield consistent boson masses around $9\times 10^{-13}$\,eV. Next, we have performed the first population study of compact binaries -- restricted to the intermediate-mass black-hole range here treated -- considering a mixed black hole-Proca star merger population. We note that the latter is a rather proof-of-concept exercise that, moreover, provides conclusions only about the observation set as opposed to the underlying population; and ignores any (unknown) priors on the relative abundance of BBHs and PSMs. In addition, turning this into a proper population study, would also require the usage of selection effects. Nevertheless we note that our study was still enough to showcase the potential benefit of exploiting boson-mass consistencies across events. 
 
This is the first extensive and systematic analysis of GW events under an exotic compact-merger scenario alternative to BBHs. Although our new simulation catalogue has been significantly expanded since our initial study \cite{Proca}, it still suffers from important limitations. These are mainly the range of parameters covered by our numerical simulations and the fact that all of these correspond to the unrealistic astrophysical configuration of a head-on merger. The latter limits the type of morphologies we can possibly fit due to the shortness of the templates, significantly over-constrains our parameter estimates; and also intrinsically disfavors the PSM model due to its weak luminosity. On the other hand, our limited range of field frequencies coverage may prevent us from correctly fitting some of the events we analyse. For instance, we know that the numerical simulations best fitting S200114f and GW190426 lay in the edges of our catalogue. This implies, e.g., that an analysis under an enhanced simulation catalogue may return not only better fits to the data but also modify our parameter estimates, e.g., those of the boson masses.

While progress is made towards numerical simulations of more realistic and less eccentric configurations, we highlight that our results are highly promising and should strongly motivate the pursuit of such extended catalogues. First, these simple configurations suffice to fit the data as well as the most developed BBH models, if not better. Second, even though the standard prior in typical GW parameter estimation is by-default designed to prefer loud circular configurations for which GW detectors have a much larger reach, our analysis shows that in some cases the Proca scenario is marginally preferred. In fact, when removing such ``bias'' to foresee what results would be obtained considering louder and circular configurations, two events show a comparable preference to both scenarios and the other two, GW190521 and S200114f, show stronger preferences for PSM.

The existence of an ultralight bosonic field would have profound implications. It could at least account for part of dark matter, since it would give rise to a remarkable energy extraction mechanism from astrophysical spinning BHs, which eventually form new sorts of ``hairy" BHs~\cite{herdeiro2014kerr,herdeiro2016kerr}. In addition, such a field could serve as a guide toward beyond-standard-model physics, possibly pointing to the stringy axiverse. From an astrophysical perspective, the existence of massive bosonic stars could also have an impact on black-hole populations, if these objects merge and collapse frequently, contributing to the formation of intermediate-mass black holes.

\section*{Acknowledgements \\}

We thank Tom Callister and Kaze Wong for enlightening discussions about population studies and Xisco Jimenez Forteza for comments on the manuscript. The corner plots in Fig. \ref{fig:omega_posteriors_one} have been generated  with the \texttt{corner} package~\cite{corner}. The analysed LIGO-Virgo data and the corresponding power spectral densities, in their strain versions, are publicly available at the online Gravitational-Wave Open Science Center \cite{SoftwareX,OpenDataArxiv}. This research has made use of data or software obtained from the Gravitational Wave Open Science Center (gwosc.org), a service of LIGO Laboratory, the LIGO Scientific Collaboration, the Virgo Collaboration, and KAGRA. LIGO Laboratory and Advanced LIGO are funded by the United States National Science Foundation (NSF) as well as the Science and Technology Facilities Council (STFC) of the United Kingdom, the Max-Planck-Society (MPS), and the State of Niedersachsen/Germany for support of the construction of Advanced LIGO and construction and operation of the GEO600 detector. Additional support for Advanced LIGO was provided by the Australian Research Council. Virgo is funded, through the European Gravitational Observatory (EGO), by the French Centre National de Recherche Scientifique (CNRS), the Italian Istituto Nazionale di Fisica Nucleare (INFN) and the Dutch Nikhef, with contributions by institutions from Belgium, Germany, Greece, Hungary, Ireland, Japan, Monaco, Poland, Portugal, Spain. KAGRA is supported by Ministry of Education, Culture, Sports, Science and Technology (MEXT), Japan Society for the Promotion of Science (JSPS) in Japan; National Research Foundation (NRF) and Ministry of Science and ICT (MSIT) in Korea; Academia Sinica (AS) and National Science and Technology Council (NSTC) in Taiwan. JCB received the support of a fellowship from ``la Caixa'' Foundation (ID 100010434) and from the European Union’s Horizon 2020 research and innovation programme under the Marie Skłodowska-Curie grant agreement No 847648. The fellowship code is LCF/BQ/PI20/11760016. JCB is also supported by the research grant PID2020-118635GB-I00 from the Spain-Ministerio de Ciencia e Innovaci\'{o}n. JAF is supported by the Spanish Agencia Estatal de Investigaci\'on  (PGC2018-095984-B-I00, PID2021-125485NB-C21) and by the  Generalitat  Valenciana (PROMETEO/2019/071). This work is supported by the Center for Research and Development in Mathematics and Applications (CIDMA) 
through the Portuguese Foundation for Science and Technology (FCT - Funda\c {c}\~ao para a Ci\^encia e a Tecnologia), reference UIDB/04106/2020, and by national funds (OE), through FCT, I.P., in the scope of the framework contract foreseen in the numbers 4, 5 and 6 of the article 23, of the Decree-Law 57/2016, of August 29, changed by Law 57/2017, of July 19. We also acknowledge support  from  the  projects  PTDC/FIS-OUT/28407/2017,  CERN/FIS-PAR/0027/2019, PTDC/FIS-AST/3041/2020,  CERN/FIS-PAR/0024/2021 and 2022.04560.PTDC. NSG is supported by the Spanish Ministerio de Universidades, through a María Zambrano grant (ZA21-031) with reference UP2021-044, funded within the European Union-Next Generation EU.   
This work has further been supported by the European Union’s Horizon 2020 research and innovation (RISE) programme H2020-MSCA-RISE-2017 Grant No. Fu\text{NF}iCO-777740 and by the European
Horizon Europe staff exchange (SE) programme HORIZON-
MSCA-2021-SE-01 Grant No. NewFunFiCO-101086251. 
We acknowledge the use of IUCAA LDG cluster Sarathi for the computational/numerical work. The authors acknowledge computational resources provided by the CIT cluster of the LIGO Laboratory and supported by National Science Foundation Grants PHY-0757058 and PHY0823459; and the support of the NSF CIT cluster for the provision of computational resources for our parameter inference runs. This material is based upon work supported by NSF's LIGO Laboratory which is a major facility fully funded by the National Science Foundation. This manuscript has LIGO DCC number P2200169. 

\appendix
\renewcommand\thesection{\Roman{section}}
\renewcommand\thesubsection{\thesection.\alph{subsection}}
\counterwithout{equation}{section}
\addtocounter{equation}{-1}

\section{Parameter inference and model selection with discrete waveform models}\label{sec:appendix1}

Common gravitational-wave data parameter inference is carried out making use of semi-analytical waveform models that span a continuous parameter space as, e.g., phenomenological \cite{XPHM_Pratten,Khan:2015jqa}, effective-one-body \cite{SEOBNRv4PHM} or numerical-relativity surrogates \cite{NRSur7dq4}. On the one hand, this enables the exploration of a continuous set of parameters. On the other, this facilitates to impose any desired Bayesian priors on the intrinsic source parameters, like the individual masses and spins.\\

The above is in contrast with the situation encountered when the ``waveform model'' consists on a finite and discrete set of numerical relativity simulations, characterised by parameters $\theta_{s}$. In the following we describe our procedure to extract parameter posterior distributions and Bayesian evidences using such simulation set and, in particular, we describe in detail our procedure to impose given priors on the parameters $\theta_{s}$, which in our case correspond to the two star frequencies $\omega_{1,2}/\mu_{\rm B}$. For completeness, we will denote the parameters that we can continuously sample by $\theta_c$, namely the total mass, source orientation, signal polarization, sky-location, luminosity distance and time of arrival.

\subsection*{Quantities of interest}

The posterior probability $p(d\mid\theta)$ for given source parameters $\theta=\{\theta_s,\theta_c\}$, according to a signal model $M_i$, given detector data $d$, is given by 

\begin{equation}\tag{\ref{eq:bayesian_prob}}
    p_{M_i}(\theta\,|\,d) = \frac{\pi(\theta)\,{\cal{L}}_{M_i}(d \,|\, \theta)}{{\cal{Z}}_{M_i}},
\end{equation}

where the Bayesian evidence for the model $M_i$ is given by  
\begin{equation}\tag{\ref{eq:evidence}}
    {\cal{Z}}_{M_{i}}=\int_{\Theta} \pi(\theta) {\cal{L}}_{M_{i}}(d\,|\,\theta) \,\dd\theta.
\end{equation}

The marginal posterior distribution for the parameters $\theta_k$ is obtained through 

\begin{equation}
    p^{\rm marg}_{M_i}(\theta_k\,|\,d) = \int \pi(\theta) {\cal{L}}_{M_i}(d\mid \theta) \Pi_{m\neq k} \,\dd\theta_m .
    \label{eq:marg_prob}
\end{equation}

Finally, given two waveform models $M_1$ and $M_2$, the relative probability for the data given the models, or relative Bayes Factor ${\cal{B}}^{M_1}_{M_2}$, is given by 
\begin{equation}\tag{\ref{eq:bayesfactor}}
{\cal{B}}^{M_1}_{M_2} = \frac{{\cal{Z}}_{M_1}}{{\cal{Z}}_{M_2}}.
\end{equation}

\subsection*{Discrete waveform models}
Continuous waveform models allow to sample the parameter space in a continuous manner. First, this virtually allows to perform integrals \eqref{eq:evidence} and \eqref{eq:marg_prob} in a continuous way. Second, and more important, it allows to impose any desired prior $\pi(\theta)$ on the explored parameters.\\ 

Because of the high computational cost of numerical relativity simulations, it is not possible to generate waveforms in a continuous manner. Instead, we are forced to work with a discrete set of points in the parameter space spanned by the parameters on which our simulations depend, namely, the frequencies of the two boson stars $\omega_{1,2}/\mu_{\rm B} \equiv \theta_s$, while the rest of parameters $\theta_c$ can be sampled in a continuous way. In practice, this means that integrals over $\omega_{1,2}/\mu_{\rm B}$ become discrete sums, yielding:

\begin{eqnarray}
    {\cal{Z}}_{\text{Proca}}&=&\sum_{i,j} \pi(\omega_{1}^{i}/\mu_{\rm B},\omega_{2}^{j}/\mu_{\rm B}) \nonumber\\
   &&\times \Delta \omega_{1}^{i}/\mu_{\rm B} \Delta \omega_{2}^{j}/\mu_{\rm B} {\cal{L}}_{\rm marg}(\omega_{1}^{i}/\mu_{\rm B},\omega_{2}^{j}/\mu_{\rm B})
\label{eq:z_sum}
\end{eqnarray}  

 where ${\cal{L}}_{\rm marg}(\omega_{1}^{i}/\mu_{\rm B},\omega_{2}^{j}/\mu_{\rm B})$ denotes the marginalised likelihood for pair of frequencies, i.e., for each of our numerical simulations, and is given by:

 \begin{equation}
    {\cal{L}}(\omega_{1}^{i}/\mu_{\rm B},\omega_{2}^{j}/\mu_{\rm B}) = \int_{\theta_c} \pi(\theta_c){\cal{L}}_{M_{i}}(d\,|\,\theta_c,\omega_{1}^{i}/\mu_{\rm B},\omega_{2}^{j}/\mu_{\rm B}) \,\dd\theta_c.
\label{eq:margl_sum}
\end{equation}
\\

\subsection*{Bayesian priors}

\begin{figure}[t!]
\begin{center}
\includegraphics[width=0.49\textwidth]{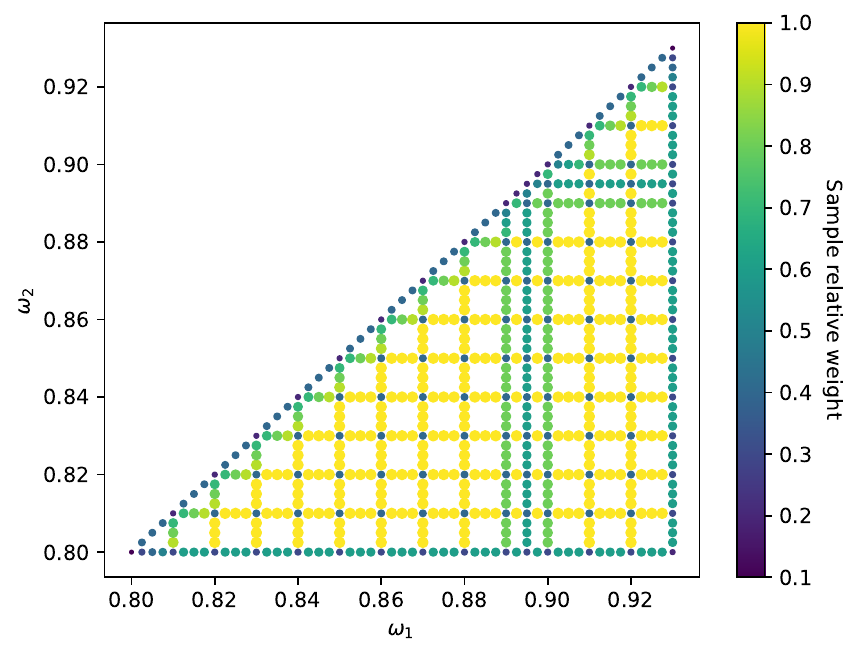}
\caption{\textbf{Catalogue of numerical simulations of proca-star merger simulations and relative weights}. We show our catalog of numerical simulations labeled by the values of the star-frequencies $\omega_1/\muB$ and $\omega_2/\muB$, with $\omega_1 /\muB \geq \omega_2/\muB$. Note these do not span an uniform grid. The colour code and relative size of the points indicates the weight each simulation is given (relative to the maximum weight) to impose an uniform prior in $(\omega_1/\muB,\omega_2/\muB)$. This is inversely proportional to the density of simulations at the corresponding point.
}
\label{fig:weights}
\end{center}
\end{figure}

The discreteness of our numerical simulations set makes it, in principle, difficult to set physically sensible Bayesian priors on $\pi(\omega_i/\mu_{\rm B},\omega_j/\mu_{\rm B})$. To exemplify what we mean by this, consider Eq.~\ref{eq:bayesian_prob} and recall that, by definition of priors $\sum_{i,j} \pi(\omega_{1}^{i}/\mu_{\rm B},\omega_{2}^{j}/\mu_{\rm B}) \,\Delta \omega_{1}^{i}/\mu_{\rm B} \,\Delta \omega_{2}^{j}/\mu_{\rm B} = \sum_{i,j} w_{i,j} = 1$. That is, the Bayesian evidence ${\cal{Z}}_{\text{Proca}} = \sum_{i,j} w_{i,j} {\cal{L}}_{\rm marg}(w_1^{i}/\mu_{\rm B},w_2^{j})$ is just the weighted average of the individual marginal evidence for each numerical simulation in our catalog.\\ 

If no relative weights are assigned to each of the simulations, then $w_{i,j} = 1/N_{\rm sim}\, \forall i,j$, with $N_{\rm sim}$ denoting the total number of simulations. This is therefore equivalent to imposing a prior $\pi(\omega_1^{i}/\mu_{\rm B},\omega_2^{j}/\mu_{\rm B})$ proportional to the density of points in our simulation grid, which would be different from a ``more physically reasonable'' uniform prior in the two star frequencies.\\

\subsection*{Imposing \nncor{a} uniform prior in the star frequencies}

We note that our previous work~\cite{Proca}, made use of a set of simulations spanning an uniform grid in $\omega_i/\mu_{\rm B}=\omega_j/\mu_{\rm B}=\omega/\mu_{\rm B}$ with step $\Delta w$, which corresponds to the diagonal in Fig.~\ref{fig:weights}. This trivially enabled us to impose a flat prior in $\omega/\mu_{\rm B}$ by simply applying equal weights $w_i$ to all of our simulations. Similarly, the same was used when using a secondary set of simulations consisting on a fixed $\omega_1/\mu_{\rm B} = 0.895$ and uniformly varying $\omega_2/\mu_{\rm B}$.\\

Fig. \ref{fig:weights} shows our simulation catalogue, which does not span an uniform grid in $(\omega_1/\mu_{\rm B},\omega_2/\mu_{\rm B})$.
If the simulations did span an uniform grid, then the evidence $\cal{Z}_{\text{Proca}}$ under a uniform prior in $\omega_{1,2}/\mu_{\rm B}$, would simply be given by the plain average of ${\cal{L}}_{i,j}$, with uniform weights.

In contrast, however our simulations \nncor{cover} certain regions with larger density than others, rather leaving uniformly distributed holes in such space. This means that, in principle, taking the plain average of ${\cal{L}}_{i,j}$ would over-weight certain regions of the parameter space, corresponding to a prior over-favouring certain (evenly distributed) regions of the parameter space. We note, however, that this will only have visible effects if the typical range of variation of the likelihood as a function of $\omega_{1,2}/\muB$ is much shorter than the typical separation between our simulations. In fact, we have checked that this produces results indistinguishable from those obtained after imposing a strictly uniform prior \footnote{For instance we obtain a $\log {\cal{B}}=90.8$ for the PSM model without strictly imposing \nncor{a} uniform prior while we obtain a value of $90.7$ using a strictly uniform prior (see Table~\ref{tab:logb1} in the main text).}\\

In order to impose a strictly uniform prior, we can simply weight each of the individual evidences ${\cal{L}}_{i,j}$ the inverse of the local density at such point $\rho_{i,j}$, yielding $w_{i,j}\propto \rho^{-1}_{i,j}$. Equivalently, one can interpret this as the intuitive process of associating area elements $\Delta \omega_1^{i}/\mu_{\rm B}\Delta \omega_2^{j}/\mu_{\rm B} \propto 1/\rho_{i,j}$, keeping $\pi(\omega_1^{i}/\mu_{\rm B},\omega_2^{j}/\mu_{\rm B})$ uniform to each of the simulations in our grid. The weights $w_{i,j}$ are shown in Fig.~\ref{fig:weights}. As it is expected, simulations placed at regions of high density as, e.g., those at the borders of the $(\omega_1,\omega_2)/\mu_{\rm B}$ triangle and those at the intersection ``nodes'' are significantly down-weighted.
We note that this procedure is completely equivalent to simply interpolating ${\cal{L}}_{\rm marg}(\omega_1/\mu_{\rm B},\omega_2/\mu_{\rm B})$ across our catalog and re-computing its values in an uniform grid (as we did in the Supplementary material of \cite{Proca} for the case of head-on BH mergers), with the exception that, in this case, we ``explicitly define the interpolation scheme'' and, therefore, know the respective weight of each of our simulations. This then allows us to compute posterior probability distributions on the different parameters.\\

Finally, marginal posterior probabilities for $\omega_1/\mu_{\rm B}$ (and similarly for $\omega_2/\mu_{\rm B}$) can be simply obtained as 

\begin{equation}
    p^{\rm marg}_{M_i}(\omega_{1}/\mu_{\rm B}|\,d) = \pi(\omega_1/\mu_{\rm B}) \,{\cal{L}}_{\rm marg}(\omega_{1}/\mu_{\rm B}),
\label{eq:p_marg_w1}
\end{equation}

where 

\begin{equation}
\begin{aligned}
 &{\cal{L}}_{\rm marg}(\omega_1 / \muB) = \\ & \sum_{i} \pi (\omega_1/\muB,\nncor{\omega_{2}^i}/\muB) {\cal{L}}(\nncor{\omega_1/\muB,\omega_{2}^i}/\muB) \Delta\nncor{\omega_{2}^i}/\muB
%{\cal{L}}_{\rm marg}\qty(\frac{\omega_1}{\mu_{\rm B}}) = \sum_{i} \pi\qty(\frac{\omega_1}{\mu_{\rm B}},\frac{\omega_{2,i}}{\mu_{\rm B}}) {\cal{L}}\qty(\frac{w_1}{\muB},\frac{w_{2,i}}{\muB}) \Delta\frac{\omega_{2,i}}{\muB}
\label{eq:L_marg_w1}
\end{aligned}
\end{equation}.

\subsection*{Calculation of the weights $\omega_{i,j}$}

While there are a plethora of methods to estimate the local density of points in a two-dimensional space, here we describe our approach to estimate the local density of the grid shown in Fig. \ref{fig:weights}. Such grid can be constructed in two main ways, which we will refer to as ``vertical'' and ``horizontal''.\\ 

The ``vertical method'' consists on initially placing an uniformly grid along the $x$-axis, whose points have separations $\Delta\omega_1/\muB=0.0025$. This way, we can associate to the i'th element this grid 1-dimensional volume element $\Delta x^{i}_{v} = \Delta\omega_1^{i}/\muB=(\omega_1^{i-1}/\muB-\omega_1^{i+1}/\muB)/2$. Note that this value is equal to 0.0025 for all points except for the end points of the grid, for which $\Delta x^{v} = 0.00125$. Next, in order to build the two-dimensional grid, one just places points along the vertical direction $\omega_2/\muB$, using variable steps $(\Delta\omega_2/\muB)(\omega_1/\muB)$ that depend on $\omega_1/\muB$, as is obvious in Fig~\ref{fig:weights}. Analogously to the $x$-axis discussion, each point is now associated \nncor{with} $\Delta y^{\nncor{j}}_{v} = \Delta\omega_2^{\nncor{j}}/\muB=(\omega_2^{\nncor{j}-1}/\muB-\omega_2^{\nncor{j}+1}/\muB)/2$. With this, each point of the \nncor{grid} is associated \nncor{to} an area element $\Delta{A}^{i,j}_{v} = \Delta x^{i}_{v}\Delta y^{\nncor{j}}_{v}$ \nncor{which} is equal to the inverse of the local density $1/\rho^{v}_{i,j} \propto \omega^{v}_{i,j}$.\\

We note that, while perfectly \nncor{legitimate}, the above calculation leads to an asymmetric weighting of the yellow points in Fig.~\ref{fig:weights} placed along vertical and horizontal lines, as points placed along the verticals would be associated much higher local densities. In order to symmetrise this, notice that the grid can also be build by inverting the above procedure, in what we call ``horizontal'' construction. This is, one first builds a vertical grid with steps $\Delta\omega_2/\muB=0.0025$. This way, now each element is associated \nncor{to} a 1-dimensional volume element $\Delta y^{j}_{h} = \Delta\omega_2^{\nncor{j}}/\muB=(\omega_2^{j-1}/\muB-\omega_2^{j+1}/\muB)/2 = 0.0025$, except for the end points that are associated $\Delta y^{j}_{v} = 0.00125$. Analogously to the previous case, one then places points along the horizontal direction using steps that depend on $\omega_2\nncor{/}\muB$. With this, each point of the is associated an area element $\Delta{A}^{i,j}_{h} = \Delta x^{i}_{h}\Delta y^{\nncor{j}}_{h}$, which is equal to the inverse of the local density $1/\rho^{h}_{i,j} \propto \omega^{h}_{i,j}$.\\

Finally, we obtain symmetric area elements and weights $\Delta{A}^{i,j}=(\Delta{A}^{i,j}_{v}+\Delta{A}^{i,j}_{h})/2$ and $\omega_{i,j} \propto A_{i,j}$, which we represent in Fig. \ref{fig:weights}.

\section{Assessment of error systematics of numerical waveforms}

In this Appendix we provide details on our numerical waveforms for Proca-star mergers. In particular, we discuss the possible impact of our waveform extraction method, initial data, and numerical grid resolution. In addition, we briefly discuss the impact of possible variations of the initial separation of the Proca stars in our simulations. We note that while the former two aspects have to do with the accuracy of our simulations, the latter implies variations of the physical properties of the system \cite{sanchis2022impact}.

\subsection*{Extraction radius}

Our waveforms are extracted at a finite radius $r_{\rm{GW}}\mu=100$, where $\mu$ denotes a characteristic scale that ranges in $M^{*}\mu \in [0.622,0.946]$ across our catalog and $M^{*}$ denotes the mass of a single star. In typical NR units, where the extraction is expressed in units of the total mass $M$ of the binary, this corresponds to extraction radii $r_{\rm{GW}} = 100M/(M_1^{*}\mu+M_2^{*}\mu) \in [53,80]\,M$. We note that this is in contrast with typical extrapolation to null infinity \cite{Nakano2015_extrapolation,Taylor2013_extrapolation}, which is done for most numerical simulations of BBHs used in GW data analyses e.g. \cite{GW150914_NR}. Extraction at finite radius can lead to systematic errors, specially when more than one GW mode is present in the signal \cite{Bustillo:2015ova}. We have performed a series of tests to ensure that our extraction strategy does not induce systematic errors that can influence our analysis, given the SNR of our signals.\\

\begin{figure}[t!]
\begin{center}
\includegraphics[width=0.45\textwidth]{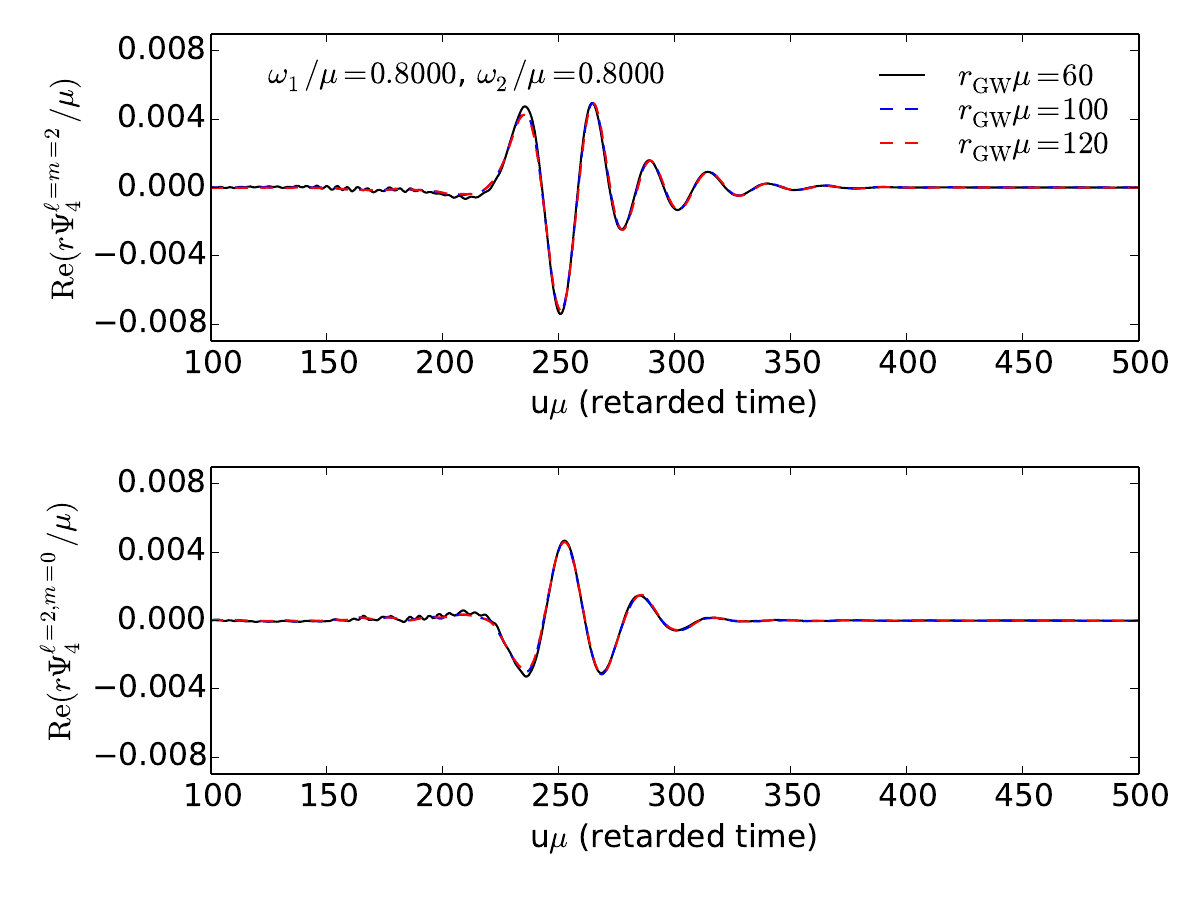}
\includegraphics[width=0.45\textwidth]{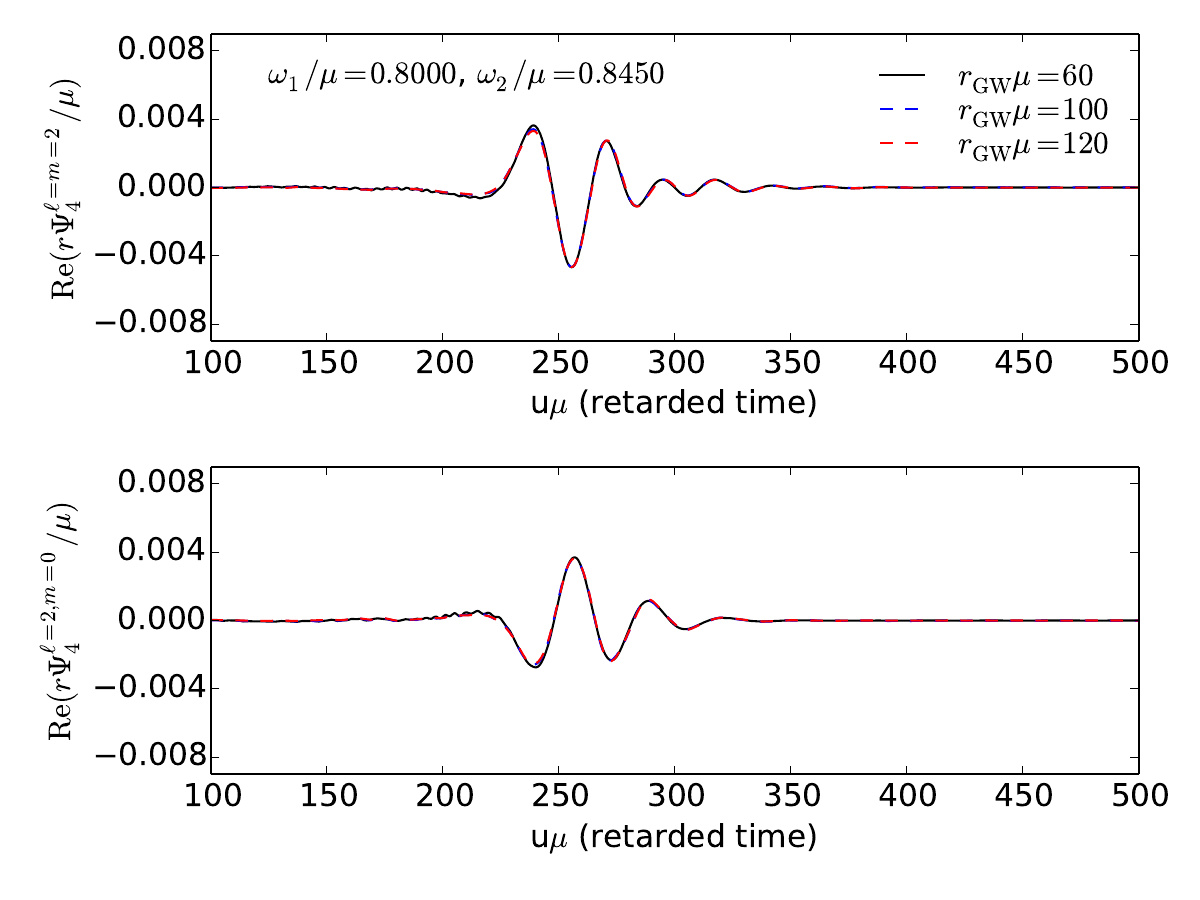}
\includegraphics[width=0.45\textwidth]{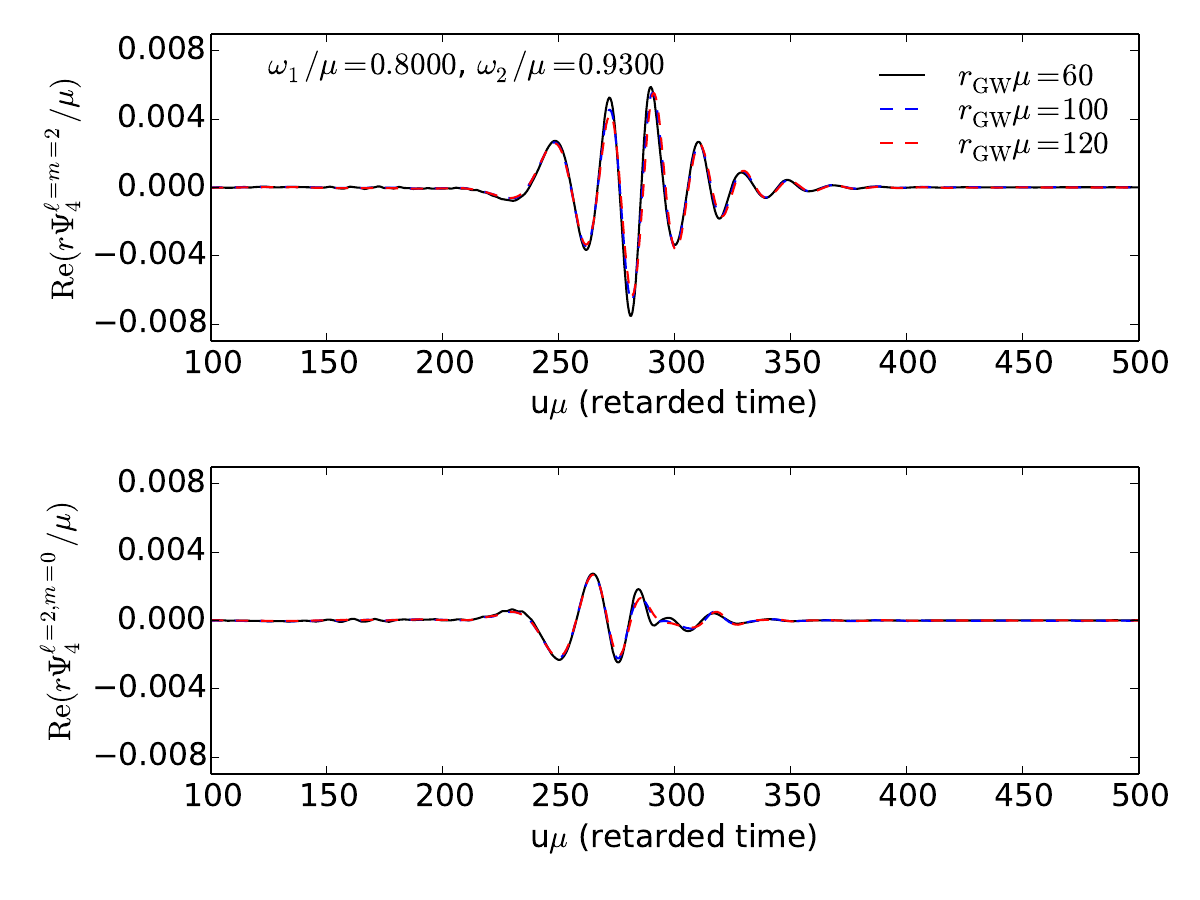}
\caption{\textbf{Gravitational waveforms extracted at different radii}. We show that the waveforms from different Proca star collisions overlap for different extraction radii.
}
\label{fig:extraction-radius}
\end{center}
\end{figure}

\begin{figure}[t!]
\begin{center}
\includegraphics[width=0.45\textwidth]{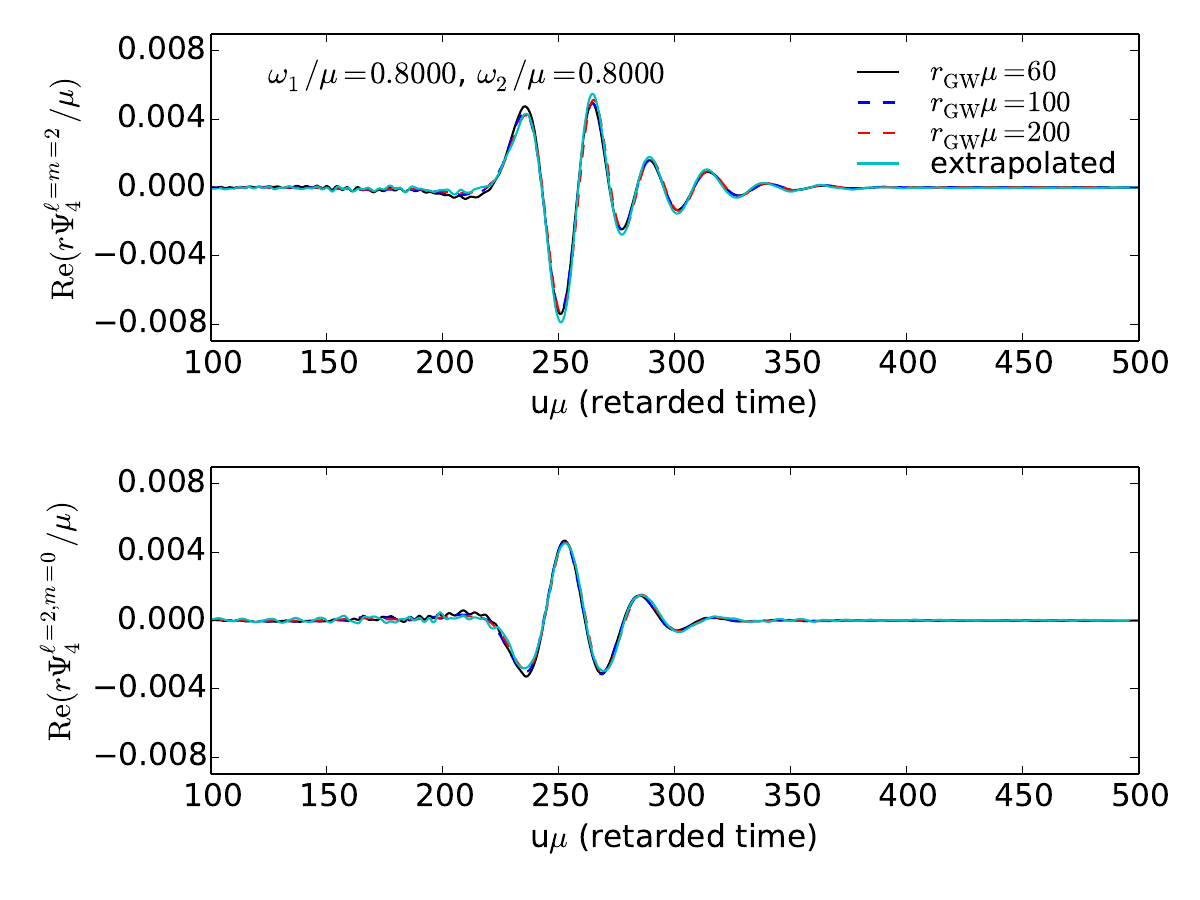}
\includegraphics[width=0.45\textwidth]{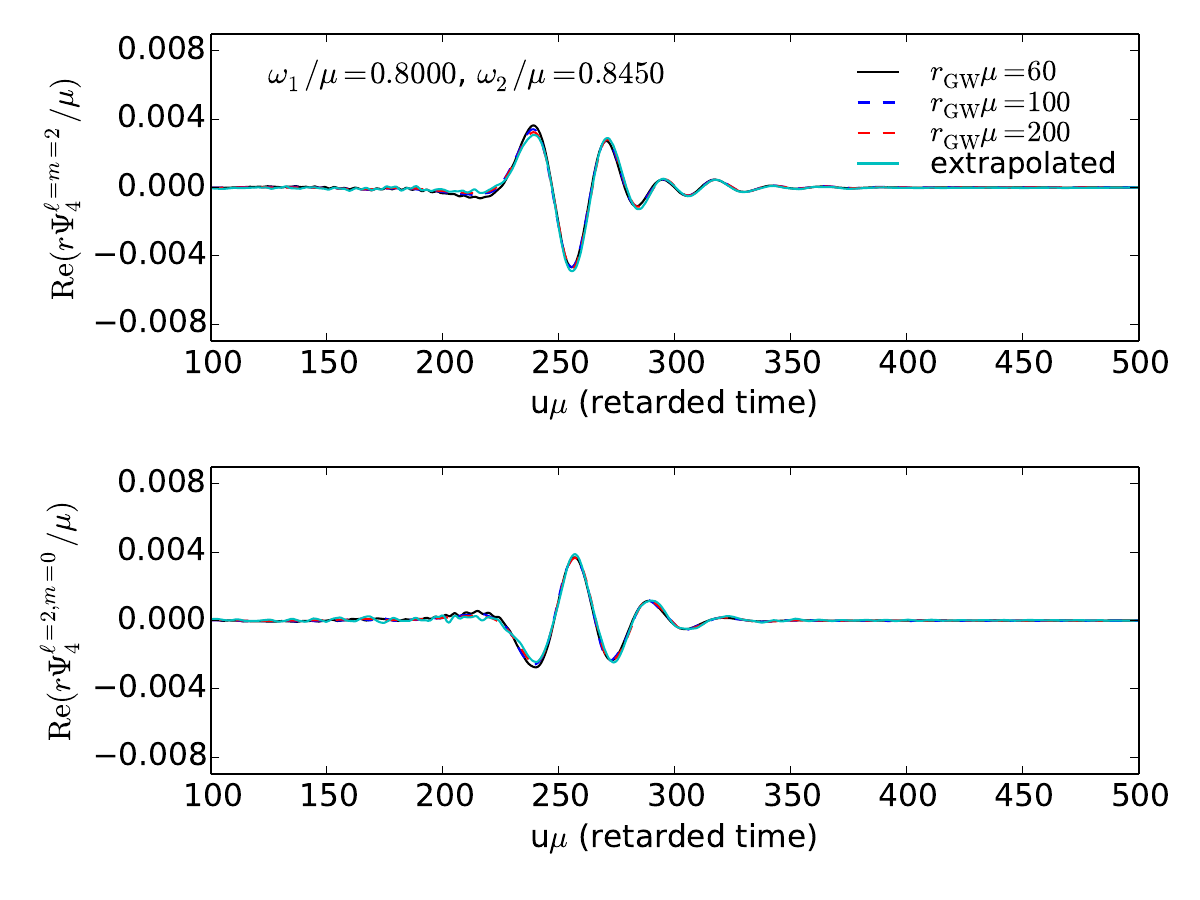}
\includegraphics[width=0.45\textwidth]{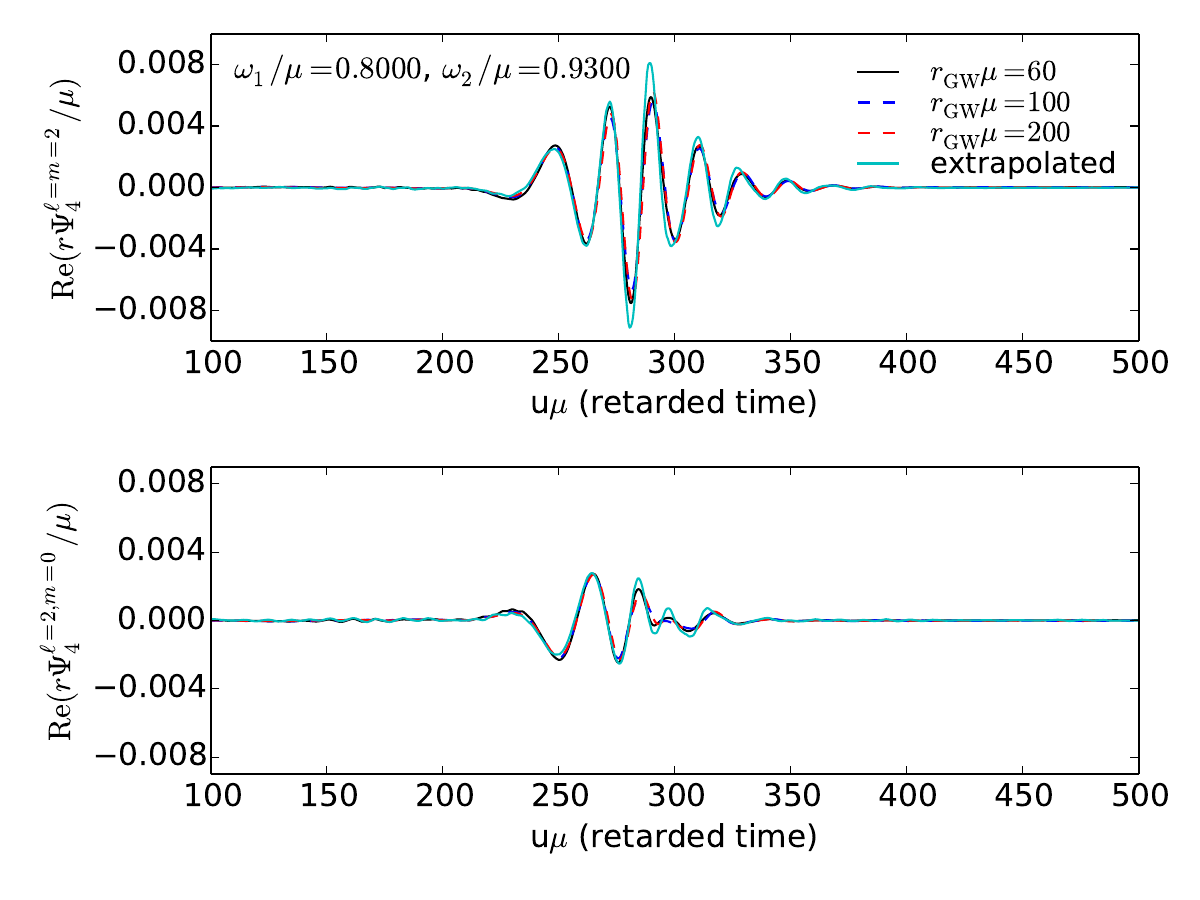}
\caption{\textbf{Gravitational waveforms extracted at different radii and extrapolated to scri+}. We show that extrapolation leads to waveforms that overlap well with those extracted at our largest initial radius.
}
\label{fig:extrapolation}
\end{center}
\end{figure}

\begin{figure*}[t!]
\begin{center}
\includegraphics[width=0.508\textwidth]{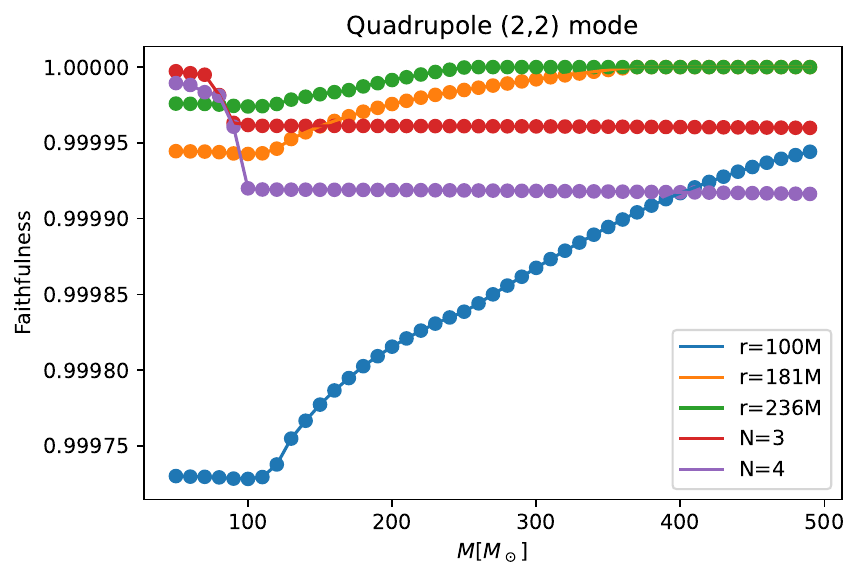}
\includegraphics[width=0.482\textwidth]{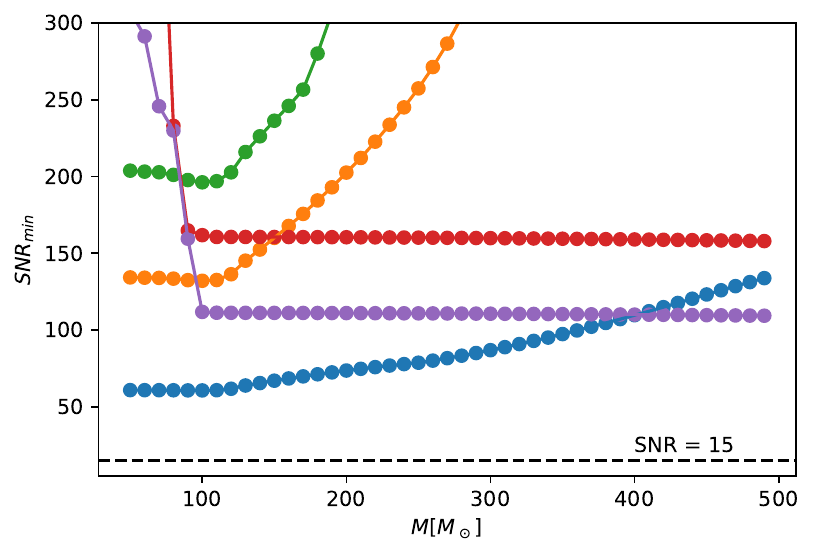}
\caption{\textbf{Assessment of systematic errors due to finite extraction radius in numerical simulations: Quadrupole modes}. The left panel shows the overlap between the dominant quadrupole modes of numerically simulated waveforms of a mass-ratio $q=3$ non-spinning BBH, obtained by the SXS collaboration. We compare simulations extracted at several finite radii, extrapolated to null infinity at $N=3,4$ to a reference $N=2$ waveform. The right panel shows the minimum SNR needed to distinguish the corresponding pair of waveforms.
}
\label{fig:radius}
\end{center}
\end{figure*}

\begin{figure*}[t!]
\begin{center}
\includegraphics[width=0.32\textwidth]{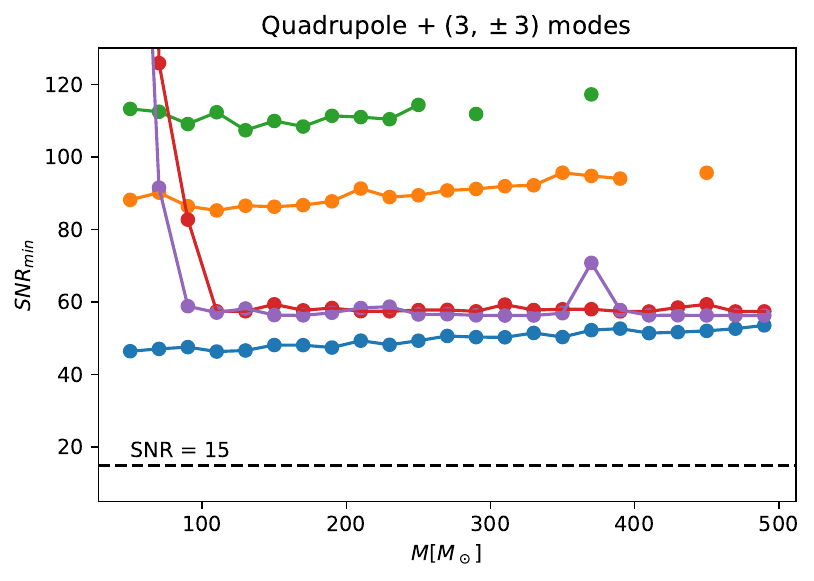}
\includegraphics[width=0.32\textwidth]{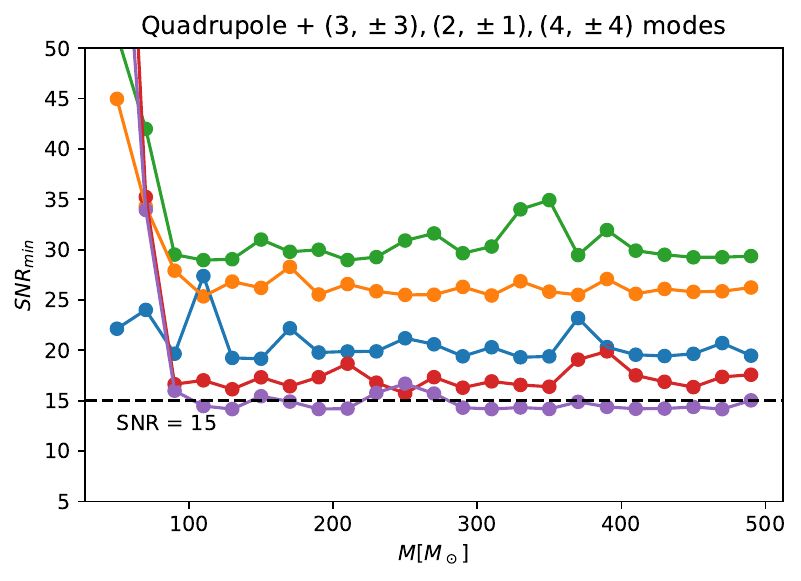}
\includegraphics[width=0.32\textwidth]{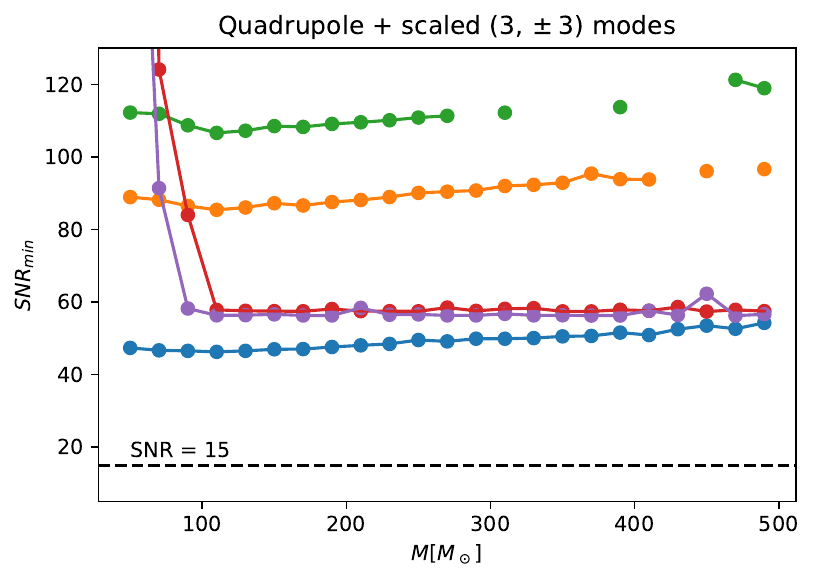}
\caption{\textbf{Assessment of systematic errors due to finite extraction radius in numerical simulations: several modes}. Same as Fig. \ref{fig:radius} but for the case of edge-on signals, i.e., observed on the orbital plane of the source. We minimize the faithfulness as a fuction of the azimuthal angle of the observer. The left panel shows the result of including only the strongest sub-dominant mode $(3,\pm 3)$. The central panel further includes the $(2,\pm 1)$ and $(4,\pm 4)$ modes. Finally, the right panel is equal to the left one, but with the $(3,\pm 3)$ scaled to have the same amplitude as the quadrupole one. The goal of this is to ``mimic'' the situation of our head-on mergers, where two modes are co-dominant. We note that apparently missing points are due to very large values of the SNR$_{min}$ which, unlike in Fig. \ref{fig:radius}, does not evolve monotonically as a function of the total mass due to the presence of higher-order modes}.

\label{fig:radius_hm}
\end{center}
\end{figure*}

\subsubsection{Visual inspection}

First, for illustrative purposes, we show in Fig.~\ref{fig:extraction-radius} that the waves extracted at different extraction radii for some selected cases, once appropriately shifted and re-scaled, overlap in the wave zone, as expected. The retarded time $u_\mu$ is defined as the difference between the coordinate time and the tortoise
coordinate: $r^*=r+2M\,\nncor{\log}(r/2M-1)$, where $M$ is the total mass of the system~\cite{boyle2009extrapolating,SXSCatalog,hamilton2023catalogue}. The overlap between waveforms is excellent for the more compact stars, although it decreases as we increase the value of $\omega_2/\mu$ for fixed $\omega_1/\mu$ . The re-scaled maximum peaks differ at radii $r_{\rm{GW}}\mu=60$ and 120 by $\sim2\%$ in the most compact case (equal-mass with $\omega/\mu=0.8000)$ to $\sim15\%$ for $\omega/\mu=0.9300$, showing that the extraction radius $r_{\rm{GW}}\mu=60$ is too close to the source. In Fig.~\ref{fig:extrapolation} we plot the waveform resulting from extrapolating the waveform to null infinity through a third-order polynomial fit, using the waveforms
from the three different extraction radii, namely $r_{\rm GW}\mu=60, 100, 200$, to obtain $r\Psi_4^{2;m}$. 

\subsubsection{Quantitative analysis using SXS waveforms}

Second, in order to understand the quantitative impact of the extraction at finite radius, we have also considered BBH waveforms from a $q=3$ non-spinning BBH \footnote{We choose this source with the goal of having several prominent sub-dominant emission modes.} from the SXS catalog \cite{SXS,SXSCatalog} (namely \texttt{SXS:BBH:0030}), both extracted at different finite radii and extrapolated to null infinity. We compute the overlap $O$ of these waveforms, as a function of the total mass, to a reference extrapolated $N=2$ waveform, where $N$ is the order of the polynomial expansion used to extrapolate the modes~\cite{SXSCatalog} \footnote{$N=2$ waveforms are recommended as reference waveforms in the SXS catalog paper \cite{SXSCatalog}.} with the exact same parameters (also known as ``faithfulness'' $F$). \nncor{We note that, as it is common practice, we maximise this over the time of arrival and global relative phases}. In addition, we compute the corresponding minimum SNR needed to distinguish the compared waveforms. This is given by SNR$_{\text{min}}=N_{\text{dim}}/(\sqrt{2(1-F)})$, where $N_{\text{dim}}$ denotes the number of parameters whose measurability can be affected by model innacuracies. For the case of the quadrupole modes alone these are $\{\omega_1,\omega_2,M\}$, so that $N_{\text{dim}}=3$, while for full waveforms we add the orientation angles $(\theta_{JN},\varphi)$, so that $N_{\text{dim}}=5$ (see \cite{Lindblom:2008cm,Chatziioannou2017_Accuracy} and e.g. \cite{PINN,Bustillo:2015qty,Hannam_nature_precession} for applications).
We do this for both the quadrupole modes alone and for the waveform observed at an edge-on location ($\theta_{JN}=\pi/2$) at random azimuths $\varphi \in [0,2\pi]$, including a varying number of GW modes. We assume a flat power-spectral density with a lower frequency cutoff of 11\,Hz. In principle, we would like to use the results of this analysis to draw conclusions about the impact of finite extraction radius in our Proca-star merger waveforms. We note, however, that while lowest extraction radius included in the SXS catalog is $r_{\rm{GW}}=100\,M$, this is typically larger that that in our catalogue $r_{\rm{GW}}\mu=100$, which as mentioned earlier corresponds to $r_{\rm{GW}}\in [53,80]\,M$. For this reason, we will later show comparisons between our waveforms extracted at $r_{\rm{GW}}\mu=100$ to waveforms extracted at $r_{\rm{GW}}\mu=200$, which we specifically obtained to perform these tests.\\

The left panel of Fig.~\ref{fig:radius} shows the corresponding overlaps for the case where we only include the quadrupole mode while the right panel shows the corresponding SNR$_{\rm min}$. In all cases such are above 70, which is four times the loudness of the GW events we consider. Moreover, we note that extrapolation to null infinity induces well-known systematics in the late ringdown part of the waveform. As an example, the $N=4$ quadrupole mode differs more from the $N=2$ than the $r_{\rm{GW}}=100\,M$ one for masses beyond $400\,M_\odot$. 

Figure \ref{fig:radius_hm} shows our results for edge-on cases where we include several modes. We illustrate the progressive degradation of the faithfulness as modes are included. To this end, the right and central panels show, respectively, the SNR$_{\rm min}$ obtained when including only the $(3,\pm 3)$ modes and when further adding the $(2,\pm 1)$ and $(4,\pm 4)$ modes. First, we note that the minimum SNRs we obtain for the $r_{\rm{GW}}=100\,M$ cases fall to $\simeq 40$ (way beyond those of our signals) and $20$ (above, but near our typical SNR of 15). Moreover, we highlight that \textit{extrapolated waveforms $N=3,4$ are unsuitable for GW analyses}, if $N=2$ is taken as a reference, as in that case SNR$_{\rm min}$ consistently hits the value of 15. 

Finally, in the right panel of Fig. \ref{fig:radius_hm} we try to mimic the situation in our PSM waveforms, where two modes are co-dominant. To this end, we include a $(3,\pm 3)$ mode re-scaled so that its amplitude at merger is equal to that of the $(2,\pm 2)$. We note despite a slight (not visible in the figure) degradation of the faithfulness with respect to the left panel, SNRs of $\simeq 50$ would be needed for the finite radius systematics to impact our analysis.

\subsubsection{Quantitative analysis for our Proca-star waveforms}

\begin{figure}[t!]
\begin{center}
\includegraphics[width=0.49\textwidth]{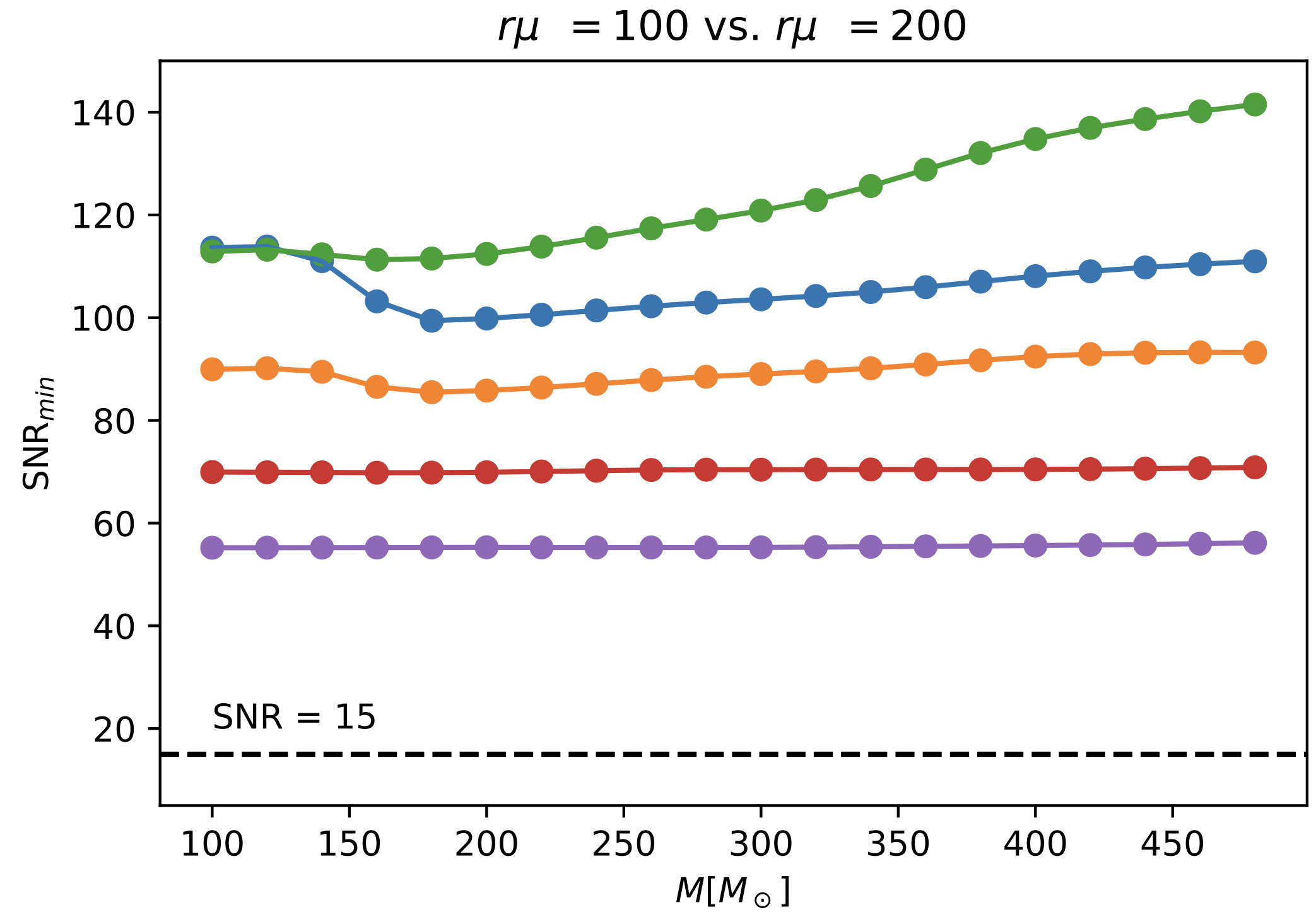}
\includegraphics[width=0.49\textwidth]{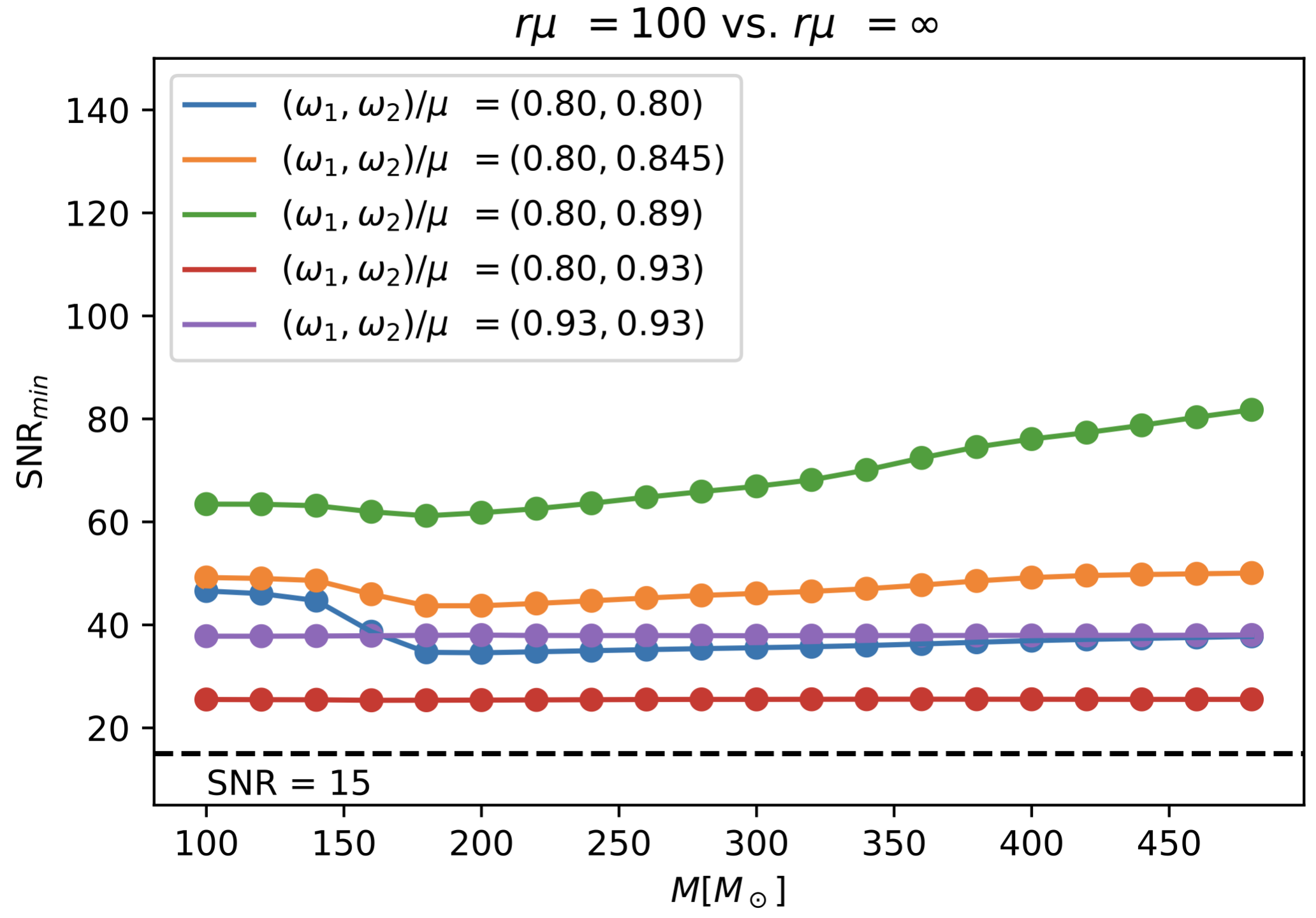}
\caption{\textbf{Impact of extraction radius in our Proca-star merger simulations}. We show the same as in the right panel of Fig. \ref{fig:radius} but for the case of head-on Proca-star mergers. The top panel compares the quadrupole modes of waveforms extracted at $r_{\rm{GW}}\mu=100$, which we used throughout our work, and $r_{\rm{GW}}\mu=200$, consistent with the $r_{\rm{GW}}=100M$ cases shown in Fig. \ref{fig:radius}. Differences between these waveforms are only detectable at SNRs of $50$ in the worst case, which is beyond the SNR $\simeq 15$ of the signals we study. The bottom panel compares waveforms extracted at $r_{\rm{GW}}\mu=100$ to those extrapolated to null infinity. Differences are only detectable, in the worst case, for SNRs of $25$.}
\label{fig:radius_proca}
\end{center}
\end{figure}

Finally, Fig. \ref{fig:radius_proca} shows the aforementioned comparison between Proca-star waveforms extracted at $r_{\rm{GW}}\mu=100$ and $r_{\rm{GW}}\mu=200$, where the latter are now beyond the minimal radius of $r_{\rm{GW}}=100\,M$ studied for the SXS case, in particular ranging in $r_{\rm{GW}}\in[106,160]M$ We consider five cases where the primary star frequency is fixed to $\omega_1/\muB=0.80$ and the secondary one is varied between the minimal and maximal frequencies in our catalog, which respectively correspond to cases of maximal and minimal compactness. We add a case where both stars are minimally compact. The figure shows that SNRs beyond $\simeq 25$, in the most pessimistic case (corresponding to the least compact secondary star), are needed for the differences between the two waveforms to be detectable within the mass range we explore.

\subsection*{Initial data}
\begin{figure}[t!]
\begin{center}
\includegraphics[width=0.45\textwidth]{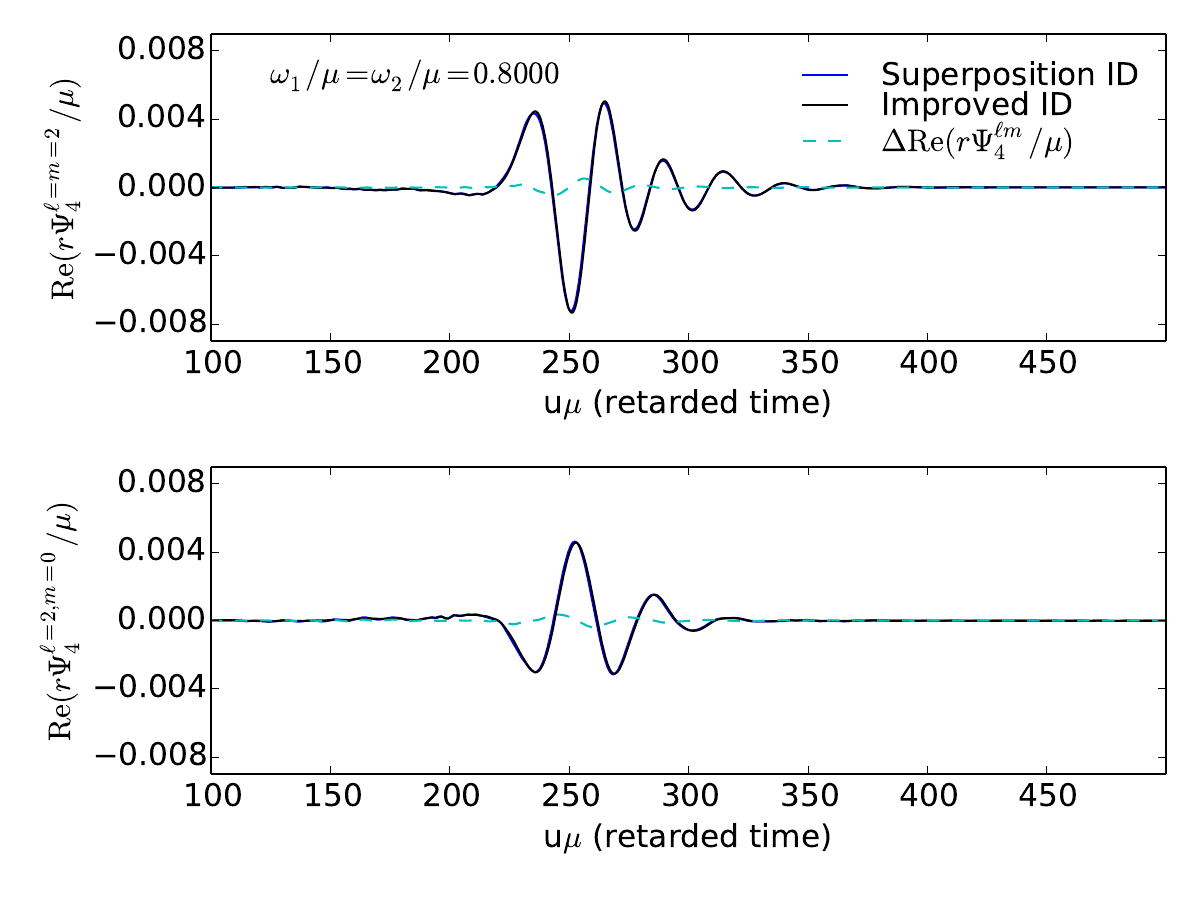}
\caption{\textbf{Gravitational waveforms obtained using different initial data}. We consider two equal-mass collisions of Proca stars with $\omega/\mu=0.8000$ for two different initial data: plain superposition and the improved method described in~\cite{helfer2022malaise}.
}
\label{fig:gw-initial-data}
\end{center}
\end{figure}

\begin{figure}[t!]
\begin{center}
\includegraphics[width=0.49\textwidth]{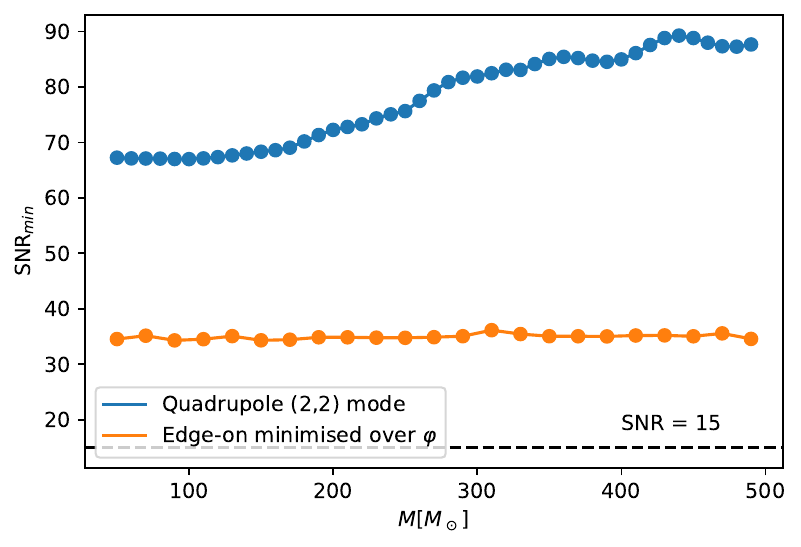}
\caption{\textbf{Assessment of systematic errors due to initial data}. Same as the right panels of Fig.~\ref{fig:radius}, but comparing waveforms extracted from simulations using our initial data and that computed using the methods in ~\cite{helfer2022malaise,evstafyeva2022unequal}. The simulated system has equal-star frequencies of $\omega/\mu=0.8000$.
}
\label{fig:initialdata}
\end{center}
\end{figure}

Until recently, including the period during which this work was developed, state-of-the-art bosonic-star mergers were performed using a plain superposition initial data~\cite{palenzuela2007head,bezares2017final,sanchis2019head,jaramillo2022head}. This is known to lead to constraint violations which can result in artificial effects, including a typical initial burst of spurious GWs known as ``junk radiation''. %\cite{Higginbotham2019}. 
%Obtaining 
Accurate, constraint-satisfying initial data 
%requires elliptic solvers, which have been developed 
has been obtained 
only very recently ~\cite{aurrekoetxea2023cttk,siemonsen2023binary}. In the future we will update our waveform catalogue with numerical simulations that use appropriate constraint-satisfying initial data. While such developments are being pursued, we have adopted the intermediate step proposed in~\cite{helfer2022malaise,evstafyeva2022unequal} to improve the plain superposition initial data. We have implemented this method for the equal-mass case and compared it with our waveforms from the equal-mass collision of our most massive and compact star configuration $\omega/\muB=0.8000$ $\omega/\mu=0.8000$. This comparison is displayed in Fig.~\ref{fig:gw-initial-data}. The waveform corresponding to the improved initial data is slightly time shifted but the difference between the two methods at the peak of the waveform is around 2.1\%. Once again, to assess the impact of our initial data in GW analyses, we computed the faithfulness and limiting SNRs shown in Fig. \ref{fig:radius} for the case of one of our Proca-star mergers using our initial data and the methods from~\cite{helfer2022malaise,evstafyeva2022unequal}. In this comparison we remove by hand the junk radiation of the simulations, which we also do in our main analysis. The results, shown in Fig. \ref{fig:initialdata}, reveal that our initial data would only impact our analyses for SNRs above $\simeq 35$.

\subsection*{Initial star separation}

In our simulations, the two Proca stars are released from rest at an initial distance of $D\mu=40$. We note that the choice of $D$ is somewhat similar to that of the reference frequency at which eccentricity is defined for eccentric systems, similarly leading to varying phenomenology. In particular, starting our simulations at different distances would change the intrinsic luminosity of the system, the ``impact parameter'' of the two stars due to the different amount of frame dragging, and, as explained in \cite{sanchis2022impact}, it would introduce a varying relative phase of the complex field of the two stars at merger. Therefore, our catalogue is rather sub-optimal in covering the possible physics of head-on mergers.\\

\begin{figure}[t!]
\begin{center}
\includegraphics[width=0.45\textwidth]{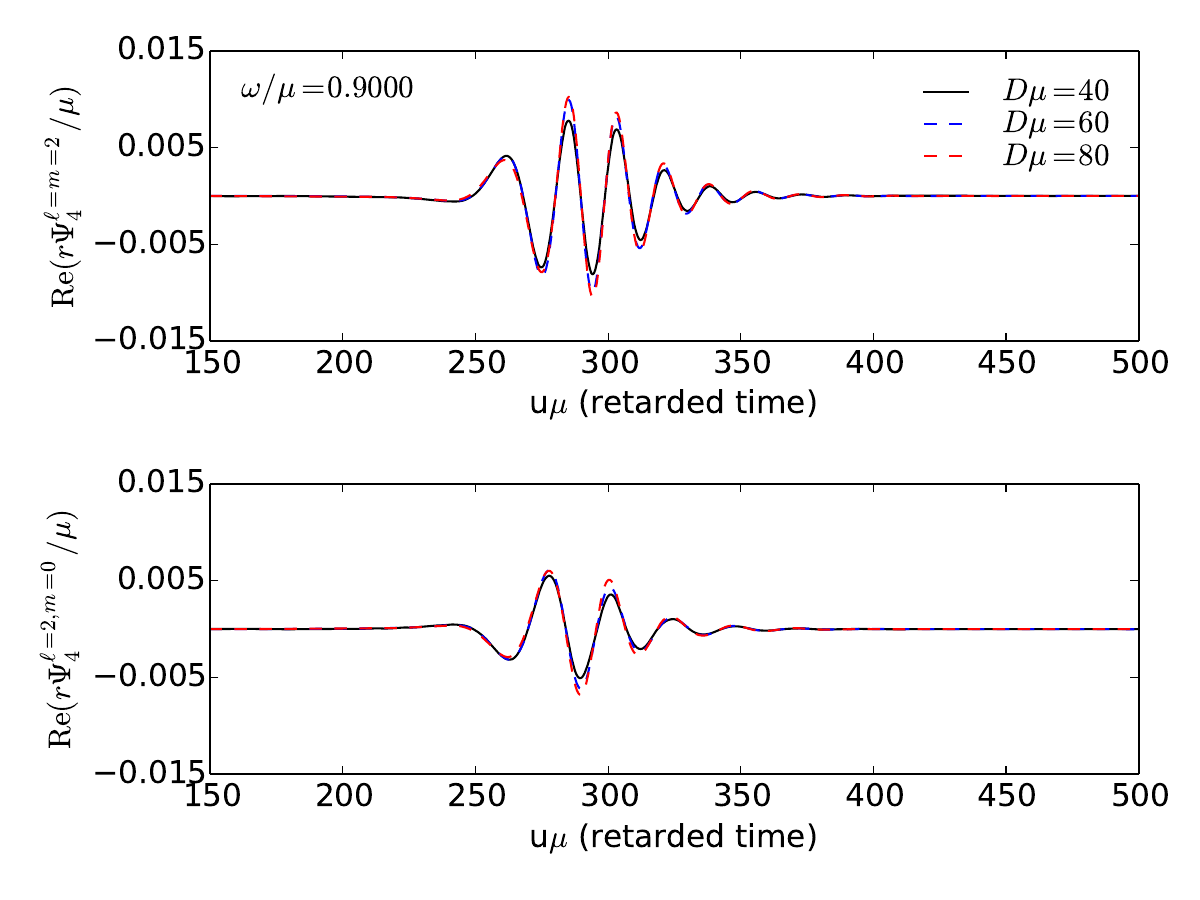}
\includegraphics[width=0.45\textwidth]{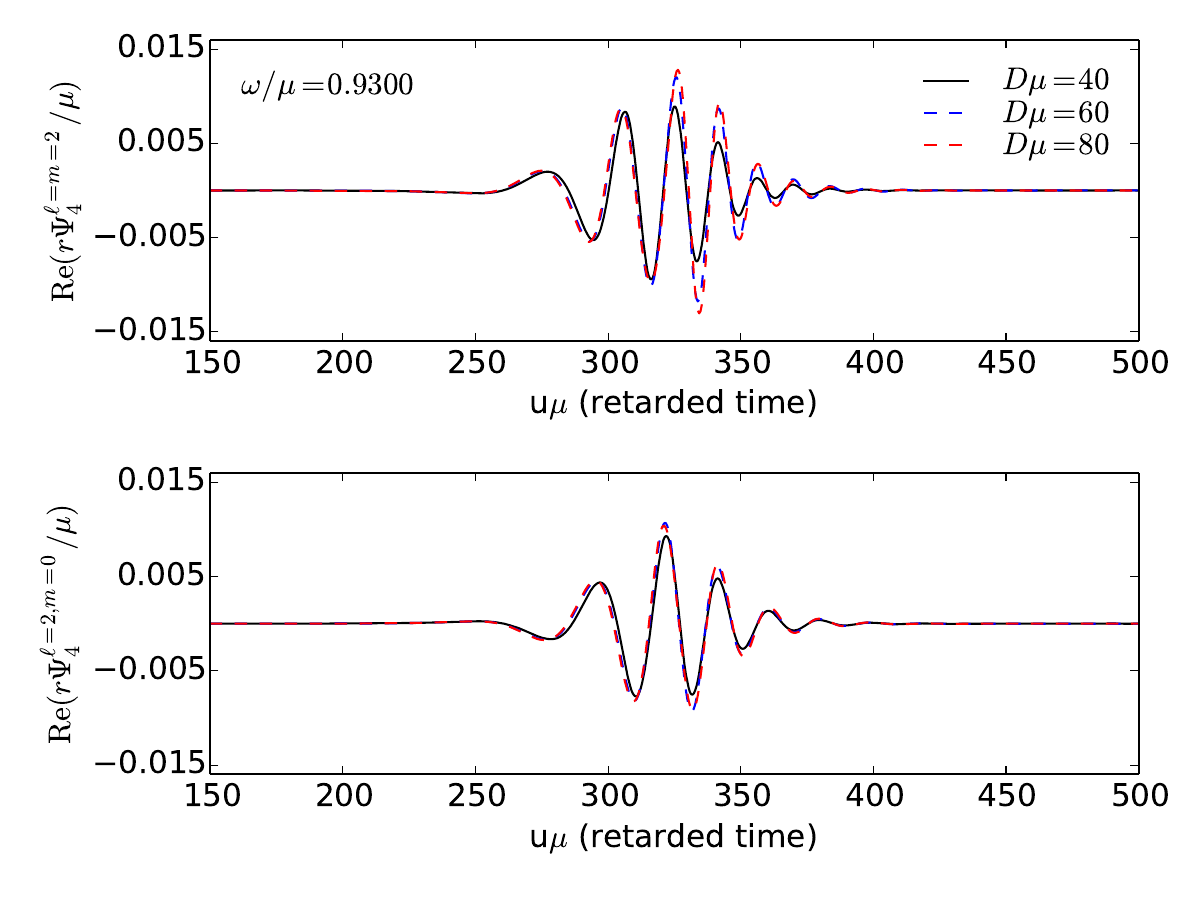}
\caption{\textbf{Gravitational waveforms from different initial separation distances}. Increasing the initial separation between the stars leads to slightly larger amplitudes.}

\label{fig:distance-radius}
\end{center}
\end{figure}

In order to assess the differences in the waveforms, we have also performed two equal-mass collisions (with $\omega/\mu=0.9000$ and $\omega/\mu=0.9300$) at different initial distances. We choose equal-mass collisions to avoid the impact of relative phases at merger from ``affecting'' the results we show here (for details on that issue, see \cite{sanchis2022impact}). The results are displayed in Fig.~\ref{fig:distance-radius}. We find that the waveforms are similar in all cases. However, as expected, larger initial separation distances induce a slightly larger amplitude, in particular in the $\ell=m=2$ mode (see Fig.~\ref{fig:distance-radius}). Moreover, since increasing the initial separation leads to intrinsically louder sources, it also leads to larger estimated distances, which would be less penalized by the distance prior uniform in co-moving volume, therefore increasing our preference for the Proca-star merger model. In this sense, our Bayes Factors are rather conservative.

An ideal catalogue would make use of all possible separations, which is however unrealistic, as these can be infinitely many. As said above, this situation is similar to the choice of the infinitely many reference frequencies at which eccentricity can be defined in eccentric systems. Therefore, the ability of our current catalogue to reproduce gravitational-wave signals is still limited.

\subsection*{Numerical grid resolution}

For a convergence study on the gravitational waveforms we refer the interested reader to the appendix section of~\cite{sanchis2022impact}. \nncor{In addition, we have checked that the mismatch between the waveforms we use in this study, which we label in ~\cite{sanchis2022impact} as ``high'' and those of ``very high'' resolution is of order $10^{-4}$, so that SNRs of order $\sim 80$ are needed for differences to have an impact.}

\section{Maximum likelihood parameters}

We report in Tables \ref{tab:maxl_proca} and \ref{tab:maxl_bbh} the parameters of the waveforms yielding the maximum likelihood values, reported in Table \ref{tab:logb1} and plotted in Figures \ref{fig:21g}-\ref{fig:14f}. We note that for the BBH cases, we report the mass ratio in terms of $q=m_2/m_1 \leq 1$. The orientation is reported in terms of the inclination angle between the total angular momentum and the line-of-sight $\theta_{JN}$ and the azimuthal angle of the observer around the source $\varphi$, i.e., understood as angle formed by the projection of the line-of-sight onto the orbital plane and the line separating the two BHs. Finally, spins are represented through the magnitudes $a_{1,2}$, the tilt angles between the spins and the total angular momentum $\theta_{1,2}$, the relative azimuthal angle between the two spins $\phi_{12}$ and the angle between the total and the orbital angular momentum $\phi_{JN}$. These are the typical parameters sampled in the parameter estimation code \texttt{Bilby} \cite{Ashton:2018jfp}. All parameters are estimated at a reference frequency of 11Hz.\\

\begin{table*}[t!]
\centering
\begin{center}
%\begin{ruledtabular}
%\begin{tabularx}{\columnwidth}{>{\raggedright\arraybackslash}Xrr}
\renewcommand{\arraystretch}{1.5}
\begin{tabular}{l|@{\hspace{1em}}c@{\hspace{1em}}|@{\hspace{1em}}c@{\hspace{1em}}|@{\hspace{1em}}c@{\hspace{1em}}|@{\hspace{1em}}c@{\hspace{1em}}}
%\hline
%\hline \\ 
Parameter  & GW190521 & GW200220  & GW190426 & S200114f \\ \hline

Total red-shifted mass $[M_\odot]$  & $267.88$ & $383.12$ & $404.20$ & $236.20$ 
\\
Inclination $\theta_{JN}$ [rad] & $2.41$ & $1.34$  & $1.85$  & $1.07$\\
Azimuth $\varphi$ & $5.22$ & $5.61$  & $3.72$ & $3.28$ 
\\
Luminosity distance [Mpc] & $267.83$  & $55.52$ &  $48.78$   &   $71.07$   
\\
Polarization $\psi$ & $1.29$ & $1.68$  & $0.92$ & $1.52$ 
\\
Right ascension $\alpha$ & $3.94$ & $4.82$  & $1.92$ & $1.94$ 
\\
Declination $\delta$ & $0.91$ & $-0.95$  & $-0.44$ & $0.11$ 
\\
Primary field frequency $\omega_1/\mu_{\rm B}$  & $0.9000$ &   $0.8800$ & $0.9000$  & $0.8800$ 
\\
Secondary field frequency $\omega_2/\mu_{\rm B}$  & $0.8550$  & $0.8075$   & $0.8500$ & $0.8325$

\end{tabular}
\end{center}
\caption{\textbf{Maximum likelihood values for our analysed events, under the Proca-star merger hypothesis}. We quote the inclination in terms of the angle between the line-of-sight and the total angular momentum $\theta_{JN}$ as well as the azimuthal angle of the observer $\varphi$ (see Appendix I in \cite{Proca}).}
\label{tab:maxl_proca}
\end{table*}

\begin{table*}[t!]
\centering
\begin{center}
%\begin{ruledtabular}
%\begin{tabularx}{\columnwidth}{>{\raggedright\arraybackslash}Xrr}
\renewcommand{\arraystretch}{1.5}
\begin{tabular}{l|@{\hspace{1em}}c@{\hspace{1em}}|@{\hspace{1em}}c@{\hspace{1em}}|@{\hspace{1em}}c@{\hspace{1em}}|@{\hspace{1em}}c@{\hspace{1em}}}
%\hline
%\hline \\ 
Parameter  & GW190521 & GW200220  & GW190426 & S200114f \\ \hline

Total red-shifted mass $[M_\odot]$  & 254.44 & 308.78 & 303.82 & 280.15 
\\
Mass ratio   & 0.75 & 0.64 & 0.88 & 0.17 
\\
Primary spin $a_1$  & 0.93 & 0.93 & 0.86 & 0.98 
\\
Secondary spin $a_2$  & 0.95 & 0.96 & 0.26 & 0.99 
\\
Primary tilt $\theta_1$  & 1.72 & 1.07 & 0.81 & 2.77 
\\
Secondary tilt $\theta_2$  & 2.73 & 2.13 & 0.84 & 0.51 
\\
Spin-spin azimuth $\phi_{12}$  & 4.47 & 6.16 & 1.14 & 4.54 
\\
Total-orbital momentum azimuth $\phi_{JL}$  & 5.82 & 383.12 & 5.76 & 4.04 
\\
Inclination $\theta_{JN}$ [rad] & 1.99 & 1.47  & 1.60  & 2.15
\\
Azimuth $\varphi$ & 5.85 & 2.00  & 0.23 & 1.82 
\\
Luminosity distance [Mpc] & 1509.35  & 2257.87 &  343.44   &   355.28   
\\
Polarization $\psi$ & 1.13 & 1.69  & 2.31 & 3.06 
\\
Right ascension $\alpha$ & 4.37 & 3.44  & 0.63 & 1.93 
\\
Declination $\delta$ & 0.84 & 0.49  & -0.60 & 0.02 
\end{tabular}
\end{center}
\caption{\textbf{Maximum likelihood values for our analysed events, under the BBH hypothesis}. Mass-ratios are quoted as $q=m_2/m_1 \leq 1$.}
\label{tab:maxl_bbh}
\end{table*}

\label{sec:appmaxL}

\section{Parameter estimates under the black-hole merger hypothesis}
\label{sec:appBBHPE}

In this section we report the parameter estimates for our studied events under the analysis with the BBH model \texttt{NRSur7dq4}. As stated in the main text, we effectively use 8 different priors for our runs, which consist on different combinations:
\begin{itemize}
    \item Mass ratio: uniform in $1/q \in [1,Q_{\rm max}]$ and $q \in [1/Q_{\rm max},1]$.
    \item Mass-ratio limit: $Q_{\rm max}\in\{4,6\}$.
    \item Luminosity distance: uniform in co-moving volume and uniform in luminosity distance.\\
\end{itemize}

Table \ref{tab:pe_bbh} reports the parameter estimates for the four events in terms of median and symmetric $90\%$ credible intervals. These are obtained under a distance prior uniform in co-moving volume, using the mass-ratio prior that maximises the Bayesian evidence. In other words, these correspond to the column ``V'' for the BBH model quoted in Table \ref{tab:logb1}. For S200114f, this corresponds to the mass-ratio prior uniform in $1/q\in[1,6]$ while for the rest this corresponds to the prior uniform in $q\in[1/4,1]$. We highlight that the parameters obtained for GW190521 are completely consistent with those in \cite{GW190521D} and that those for S200114f clearly rail against the limits of the parameter space covered by \texttt{NRSur7dq4}. In particular, the posterior for the mass ratio rails against the $q=1/6$ limit. This could motivate the usage of waveform models allowing for larger mass ratios like \texttt{SEOBNRv4PHM}~\cite{SEOBNRv4PHM} or \texttt{IMRPhenomXPHM}~\cite{XPHM_Pratten} may be in order. However, while spin estimates indicate that significant spin magnitudes and orbital precession are needed to reproduce this event, the mentioned models model precession through post-newtonian or effective-one body approximations that break down during the merger-ringdown inspiral, damaging their accuracy \cite{Hannam_nature_precession,SEOBNRv4PHM}. \nncor{However, see} \cite{PhenomNRPrecession} for a phenomenological model calibrated using precessing NR simulations. 

\begin{table*}[t!]
\centering
\begin{center}
%\begin{ruledtabular}
%\begin{tabularx}{\columnwidth}{>{\raggedright\arraybackslash}Xrr}
\renewcommand{\arraystretch}{1.5}
\begin{tabular}{l|@{\hspace{1em}}c@{\hspace{1em}}|@{\hspace{1em}}c@{\hspace{1em}}|@{\hspace{1em}}c@{\hspace{1em}}|@{\hspace{1em}}c@{\hspace{1em}}}
%\hline
%\hline \\ 
Parameter  & GW190521 & GW200220  & GW190426 & S200114f \\ \hline

Primary mass $[M_\odot]$  & $87^{+22}_{-15}$ & $86^{+31}_{-21}$ & $121^{+38}_{-29}$ & $209^{+69}_{-17}$
\\
Secondary mass   $[M_\odot]$  & $72^{+20}_{-19}$ & $68^{+27}_{-24}$& $86^{+25}_{-36}$ & $37^{+11}_{-4}$ 
\\
Total mass  $[M_\odot]$  & $158^{+23}_{-14}$ & $153^{+32}_{-24}$& $205^{+25}_{-21}$ & $246^{+15}_{-12}$ 
\\
Total red-shifted mass  $[M_\odot]$  & $271^{+14}_{-15}$ & $280^{+23}_{-21}$& $329^{+27}_{-31}$ & $280^{+14}_{-9}$ 
\\
Final mass  $[M_\odot]$  & $150^{+22}_{-14}$ & $145^{+30}_{-22}$& $194^{+23}_{-20}$ & $243^{+15}_{-12}$ 
\\
Final red-shifted mass  $[M_\odot]$  & $258^{+12}_{-13}$ & $266^{+20}_{-18}$& $310^{+23}_{-27}$ & $277^{+14}_{-9}$ 
\\
Final spin   & $0.71^{+0.05}_{-0.06}$ & $0.71^{+0.08}_{-0.07}$& $0.80^{+0.07}_{-0.09}$ & $0.37^{+0.06}_{-0.06}$ 
\\
Primary spin $a_1$  & $0.72^{+0.22}_{-0.64}$ & $0.55^{+0.40}_{-0.49}$ & $0.73^{+0.24}_{-0.60}$  & $0.96^{+0.20}_{-0.23}$
\\
Secondary spin $a_2$  & $0.76^{+0.21}_{-0.65}$ & $0.54^{+0.41}_{-0.48}$ & $0.56^{+0.36}_{-0.52}$ & $0.88^{+0.10}_{-0.28}$ 
\\
Primary tilt $\theta_1$  & $1.57^{+0.94}_{-0.96}$ & $0.41^{+0.17}_{-1.00}$ & $0.84^{+1.25}_{-0.62}$ & $2.58^{+0.28}_{-0.27}$ 
\\
Secondary tilt $\theta_2$  & $1.57^{+0.92}_{-1.02}$ & $1.46^{+1.16}_{-1.04}$ & $1.10^{+1.30}_{-0.81}$  & $1.48^{+0.53}_{-0.54}$ 
\\
Effective spin $\chi_{\rm eff}$  & $0.01^{+0.29}_{-0.35}$ & $0.06^{+0.37}_{-0.36}$ & $0.35^{+0.35}_{-1.46}$ & $-0.66^{+0.14}_{-0.10}$
\\
Effective precessing spin $\chi_{\rm p}$ & $0.74^{+0.21}_{-0.36}$ & $0.51^{+0.37}_{-0.38}$ & $0.49^{+0.36}_{-0.32}$ & $0.51^{+0.20}_{-0.30}$
\\
Inclination $\theta_{JN}$ [rad] & $1.02^{+1.75}_{-0.73}$ & $1.42^{+1.30}_{-1.05}$  & $1.82^{+0.96}_{-1.29}$ & $2.30^{+0.33}_{-0.35}$\\
Azimuth $\varphi$ & $3.49^{+2.53}_{-0.22}$ & $2.99^{+2.67}_{-2.68}$  & $3.24^{+2.59}_{-2.73}$ & $1.00^{+4.90}_{-0.78}$ 
\\
Luminosity distance [Mpc] & $4381^{+2605}_{-2424}$  & $5366^{+3801}_{-3023}$ &  $3477^{+2753}_{-1696}$  &   $665^{+371}_{-289}$   
\\
Polarization $\psi$ & $1.77^{+1.13}_{-1.47}$ & $1.50^{+1.50}_{-1.23}$  &$1.58^{+1.16}_{-1.16}$ & $0.34^{+2.75}_{-0.29}$ 
\\
Right ascension $\alpha$ & $-0.82^{+1.63}_{-0.38}$ & $3.01^{+1.89}_{-0.54}$  & $4.33^{+0.94}_{-3.44}$ & $1.96^{+0.03}_{-0.80}$ 
\\
Declination $\delta$ &  $1.77^{+1.63}_{-0.38}$ & $-0.41^{+0.44}_{-0.63}$  & $0.22^{+0.44}_{-0.63}$ & $0.07^{+2.75}_{-0.29}$ 

\end{tabular}
\end{center}
\caption{\textbf{Posterior parameter distributions under the black-hole merger hypothesis}. We quote medians together with $90\%$ credible intervals for the runs yielding the largest Bayesian evidence, with distance prior  uniform in co-moving prior. The first three columns use a prior uniform in $q\in [1/4,1]$ while the last column uses a prior flat in $1/q \in [1,6]$.}
\label{tab:pe_bbh}
\end{table*}

\bibliography{psi4_observation.bib}

\end{document}